\documentclass[12pt]{iopart}
\usepackage{iopams}

\usepackage{graphicx}  
\usepackage{dcolumn}   
\usepackage{bm}        
\usepackage{xspace}
\usepackage{xcolor}
\usepackage{float}
\usepackage{subcaption}
\usepackage [T1]{fontenc}
\usepackage [utf8]{inputenc}
\usepackage[paperwidth=210mm,paperheight=297mm,centering,hmargin=2cm,vmargin=2.5cm]{geometry}
\usepackage{multirow}
\usepackage[square,numbers,sort&compress]{natbib}
\usepackage[colorlinks]{hyperref}
\newcommand{\text}{\mathrm}
\newcommand{\eqref}[1]{(\ref{#1})}

\newcommand{\tw}{\ensuremath{t_\mathrm{w}}\xspace}
\newcommand{\dd}{\mathrm{d}}

\newcommand{\Tg}{\ensuremath{T_\mathrm{g}}\xspace}
\newcommand{\Tc}{\ensuremath{T_\mathrm{c}}\xspace}
\newcommand{\Tm}{\ensuremath{T_\mathrm{m}}\xspace}
\newcommand{\teff}{\ensuremath{t^\mathrm{eff}}\xspace}
\newcommand{\tnew}{\ensuremath{t_\mathrm{new}}\xspace}
\newcommand{\itDelta}{\mathit{\Delta}}
 
\DeclareUnicodeCharacter{2212}{-}

\begin{document}
\title[Spin-glass dynamics in the presence of a magnetic field]{Spin-glass dynamics in the presence of a magnetic field: exploration of microscopic properties}

\author{I.~Paga$^{1,2,*,\ddagger}$, Q. Zhai$^{3,*}$, M. Baity-Jesi$^4$, E. Calore$^5$, A. Cruz$^{6,7}$,
L.~A.~Fernandez$^{2,7}$, J. M.~Gil-Narvion$^7$, I.~Gonzalez-Adalid~Pemartin$^2$,
A Gordillo-Guerrero$^{8,9,7}$, D.~I\~niguez$^{6,7,10}$, A.~Maiorano$^{11,12,7}$, E.~Marinari$^{13,12}$,
V. Martin-Mayor$^{2,7}$, J. Moreno-Gordo$^{7,6}$, A.~Mu\~noz-Sudupe$^{2,7}$, D. Navarro$^{14}$,
R. L. Orbach$^3$, G. Parisi$^{13,12}$, S.~Perez-Gaviro$^{15,7,6}$, F. Ricci-Tersenghi$^{13,12}$,
J. J. Ruiz-Lorenzo$^{16,9,7}$, S.~F.~Schifano$^{17}$, D. L. Schlagel$^{18}$, B. Seoane$^{2,7}$,
A. Tarancon$^{6,7}$, R.~Tripiccione$^5$, D. Yllanes$^{19,7}$} 

\address{$^1$ Dipartimento di Fisica, Sapienza Universit\`a di Roma, INFN, Sezione di Roma 1, I-00185 Rome, Italy} 
\address{$^2$ Departamento de   F\'\i{}sica Te\'orica, Universidad Complutense, 28040 Madrid, Spain}
\address{$^3$ Texas Materials Institute, The University of Texas at Austin, Austin,
  Texas 78712, USA}
\address{$^4$ Eawag,  Überlandstrasse 133, CH-8600 Dübendorf, Switzerland}
\address{$^5$ Dipartimento di Fisica e Scienze della
  Terra, Universit\`a di Ferrara e INFN, Sezione di Ferrara, I-44122
  Ferrara, Italy}
\address{$^6$ Departamento de F\'\i{}sica Te\'orica,
  Universidad de Zaragoza, 50009 Zaragoza,
  Spain}
\address{$^7$ Instituto de Biocomputaci\'on y F\'{\i}sica de
  Sistemas Complejos (BIFI), 50018 Zaragoza, Spain}
\address{$^8$ Departamento de
  Ingenier\'{\i}a El\'ectrica, Electr\'onica y Autom\'atica, U. de
  Extremadura, 10003, C\'aceres, Spain}
\address{$^9$ Instituto de
  Computaci\'on Cient\'{\i}fica Avanzada (ICCAEx), Universidad de
  Extremadura, 06006 Badajoz, Spain}
\address{$^{10}$ Fundaci\'on ARAID, Diputaci\'on General de
  Arag\'on, 50018 Zaragoza, Spain}
\address{$^{11}$ Dipartimento di Biotecnologie, Chimica e
  Farmacia, Università degli Studi di Siena, 53100, Siena,
  Italy}
\address{$^{12}$ Dipartimento di Fisica, Sapienza
  Universit\`a di Roma, and CNR-Nanotec, I-00185 Rome,
  Italy}
\address{$^{13}$ INFN, Sezione di Roma 1, I-00185 Rome,
  Italy}
\address{$^{14}$ Departamento de Ingenier\'{\i}a,
  Electr\'onica y Comunicaciones and I3A, U. de Zaragoza, 50018
  Zaragoza, Spain}
\address{$^{15}$ Centro Universitario de la Defensa, 50090
  Zaragoza, Spain}
\address{$^{16}$ Departamento de F\'{\i}sica,
  Universidad de Extremadura, 06006 Badajoz,
  Spain}
\address{$^{17}$ Dipartimento di Scienze Chimiche e Farmaceutiche, Università di Ferrara e INFN  Sezione di Ferrara, I-44122 Ferrara, Italy}
\address{$^{18}$ Division of Materials
  Science and Engineering, Ames Laboratory, Ames, Iowa 50011, USA}
\address{$^{19}$ Chan Zuckerberg Biohub, San Francisco, CA 94158, USA}
\address{$^*$ These authors contributed equally to this work.}
\address{$^\ddagger$  ilaria.paga@uniroma1.it}
\date{\today}
\maketitle
\begin{abstract}
The synergy between experiment, theory, and simulations enables a microscopic
analysis of spin-glass dynamics in a magnetic field in the vicinity of and below
the spin-glass transition temperature $\Tg$. The spin-glass correlation length,
$\xi(t,\tw;T)$, is analysed both in experiments and in simulations
in terms of the waiting time \tw\ after the spin glass has been cooled
down to a stabilised measuring temperature $T<\Tg$
and of the time $t$ after the magnetic field is changed.
This correlation length is extracted experimentally for a
CuMn 6 at. $\%$ single crystal, as well as for simulations on the
Janus~II special-purpose supercomputer, the latter with time and length scales comparable to
experiment. The non-linear magnetic susceptibility is reported from experiment
and simulations, using $\xi(t,\tw;T)$ as the scaling variable. Previous
experiments are reanalysed, and disagreements about the nature of the Zeeman
energy are resolved. The growth of the spin-glass magnetisation in zero-field
magnetisation experiments, $M_\mathrm{ZFC}(t,\tw;T)$, is measured from
simulations, verifying the scaling relationships in the dynamical or
non-equilibrium regime. Our preliminary search for the de Almeida-Thouless
line in $D=3$ is discussed.
\end{abstract}
\submitto{J. Stat. Mech.: Theor. and Exp.}

\newpage
\tableofcontents
\newpage
\section{Introduction}
\label{Sec:intro}
This paper examines in detail the dynamics of spin glasses in the
vicinity of and below their condensation temperature $\Tg$ in the presence of a
magnetic field, an exploration relevant to 
many other condensed-matter glassy systems: fragile molecular glasses,
polymers, colloids, super-cooled liquids, and now even social science through
``Resource dynamics on species packing in diverse ecosystems'' \cite{cui:20}, to
name a few.  The advent of realistic time and length scales on the Janus~II dedicated
supercomputer generates a synergy between theory, experiment, and
simulations that encompasses a thorough examination of 
spin-glass dynamics reaching from the low-temperature regime to
the vicinity of \Tg~\cite{janus:18,zhai:19,zhai-janus:20a}.
 
The special nature of this approach is the use of the spin-glass correlation
length, $\xi(t,\tw;T)$ as the primary factor in the analysis
\cite{zhai-janus:20a}, where $t$ is the time after a magnetic field
change, when measurements of the magnetisation take place; $\tw$ is the waiting
time after the system is quenched from above \Tg\ to temperatures
within the condensation regime, and before the change in magnetic field; and
$T$ is the temperature. This correlation length can be
extracted from experiment and simulations, under dynamical or non-equilibrium
configurations~\cite{joh:99,zhai:17b,janus:08b,janus:09b}.  As such, it will be
used for a detailed study of a new powerful scaling law for the non-linear
magnetisation near to and below $\Tg$.  We shall show that this law not only accounts
for the experiments presented in this paper, but also clears up some historical
issues about the nature of the Zeeman, or magnetic-field, energy in the
spin-glass state~\cite{joh:99,bert:04}.
 
To be specific about the temperature and magnetic field protocol, we present an
analysis of the zero-field-cooled magnetisation, $M_{\mathrm {ZFC}}(t,\tw;H)$, as a
function of $t,\tw$ and magnetic field $H$ at prescribed temperatures $T\le
\Tg$.  The protocol is one where the ``sample'' is quenched from a temperature
$T>\Tg$ to a measuring temperature $T\le \Tg$ in zero magnetic field.  The word
\textit{quench} means different things experimentally and in simulations.  In
the former, there is a finite \textit{cooling rate} as the system is brought
from above $\Tg$ to the measurement temperature $T$.  It is typically of the
order of one to tens of seconds per degree of cooling.  In the case of simulations,
it is instantaneous. Though on the surface this would seem a difficult issue,
in fact temperature chaos~\cite{bray:87b,banavar:87} (which we now know to be
present in non-equilibrium dynamics as well~\cite{janus:20}) makes the two
approaches similar if not identical. Experimentally, though the cooling rate is
finite, lowering the temperature sufficiently ($\delta T$ larger than
milli-Kelvins) creates new spin-glass states without knowledge of previous
history (a process termed rejuvenation~\cite{jonason:98}). This is the reason that the
magnetic susceptibility is reproducible from one experiment to another, without
recourse to the cooling rate. Thus, the \emph{final} state reached upon an
experimental temperature quench is as fresh as the state arrived at in
simulations upon instantaneous quenching.
 
After the measurement temperature $T$ is reached, the system is held for a time
$\tw$, the \textit{waiting time}, after which a magnetic field $H$ is turned
on.  The resulting magnetisation, $M_{\mathrm {ZFC}}(t,\tw;H)$, is then measured
over a time interval $t$.  The response consists of two terms: an instantaneous
increase in magnetisation (the so-called \textit{reversible magnetisation}),
and a slowly increasing part termed the \textit{irreversible
magnetisation}.  The latter is found to depend upon all of the factors $t,\tw,
H$.  The rise of the irreversible term is typically very slow, taking literally
times of the order of the age of the universe to reach equilibrium.  For this
reason, a spin glass, once perturbed from a quasi-equilibrium state, never
reaches equilibrium, and an experiment is always in a dynamical or non-equilibrium
regime.

The sum of the reversible and irreversible magnetisations grows towards a
well-defined ``target'', the so-called \textit{field-cooled} magnetisation,
$M_{\mathrm {FC}}$, for which the measuring protocol is the opposite of the
zero-field magnetisation.  Namely, at $T>\Tg$, a magnetic field $H$ is turned
on, and then the temperature is quenched to $T\le \Tg$.  Typically, $M_{\text
{FC}}$ is relatively constant, but not without its own dynamics.  If the
magnetic field is suddenly removed, the magnetisation immediately decays by its
reversible part (the same as in the zero-field case), followed by a slow decay
termed the irreversible part or $M_{\mathrm {TRM}}(t,\tw;H)$, the
thermo-remanent magnetisation {\it dependent upon the waiting time $\tw$}.  In
general, it is found that
\begin{equation}
\label{eq:superpositionM}
M_{\mathrm {FC}} = M_{\mathrm {ZFC}}(t,\tw;H) + M_{\mathrm {TRM}}(t,\tw;H)\,.
\end{equation}
This is known as the extended principle of superposition \cite{nordblad:97}.  There is
an immense literature covering both $M_{\mathrm {ZFC}}(t,\tw;H)$ and
$M_{\mathrm {TRM}}(t,\tw;H)$ measurements, and the physical insights gained
from them~\cite{vincent:97,nordblad:97,vincent:07}.
 
Our approach is rather different, in that we choose to represent the dynamics in
terms of the spin-glass correlation length $\xi(t,\tw;H)$.  This quantity
was first extracted from experiment in the work of Joh \emph{et al.}~\cite{joh:99}, who
developed a protocol based on the relaxation function
$S(t,\tw;H)$ defined by
\begin{equation}
\label{eq:S_definition}
S(t,\tw;H) = \dd\bigg[{\frac {-M_{\mathrm {TRM}}(t,\tw;H)}{H}}\bigg]\bigg/\dd\ln\,t\,.
\end{equation}
It is known that $S(t,\tw;H)$ peaks at what is termed an effective waiting
time, $t_H^{\text {eff}}$, which is usually of the order of
$\tw$~\cite{nordblad:86}.  This time is characteristic of the decay of
$M_{\text {TRM}}(t,\tw;H)$, or, through Eq.~\eqref{eq:superpositionM}, of the
increase of $M_{\text {ZFC}}(t,\tw;H)$ with time $t$.  As noted by Lederman \textit{et al.}~\cite{lederman:91} and Hammann \textit{et al.}~\cite{hammann:92}, for states distributed according
to ultrametric symmetry, the dynamics is controlled by a largest free-energy barrier
height, $\itDelta_{\text {max}}$, associated with the state that has the
smallest overlap with the initial state, $q_{\text {min}}$.  Thus, $t_H^{\text
{eff}}$ can be associated with $\itDelta_{\text {max}}$ through the usual
Arrhenius law:
\begin{equation}
\label{eq:Delta_max_definition}
\itDelta_{\text {max}}=k_\text{B}T\big(\ln t_H^{\text {eff}}-\ln \tau_0\big)\,,
\end{equation}
where $\tau_0$ is a characteristic exchange time, $ \tau_0 \sim \hbar/k_\text{B}\Tg$.
 
\begin{figure}[t]
        \centering 
        \includegraphics[width = 0.6\columnwidth]{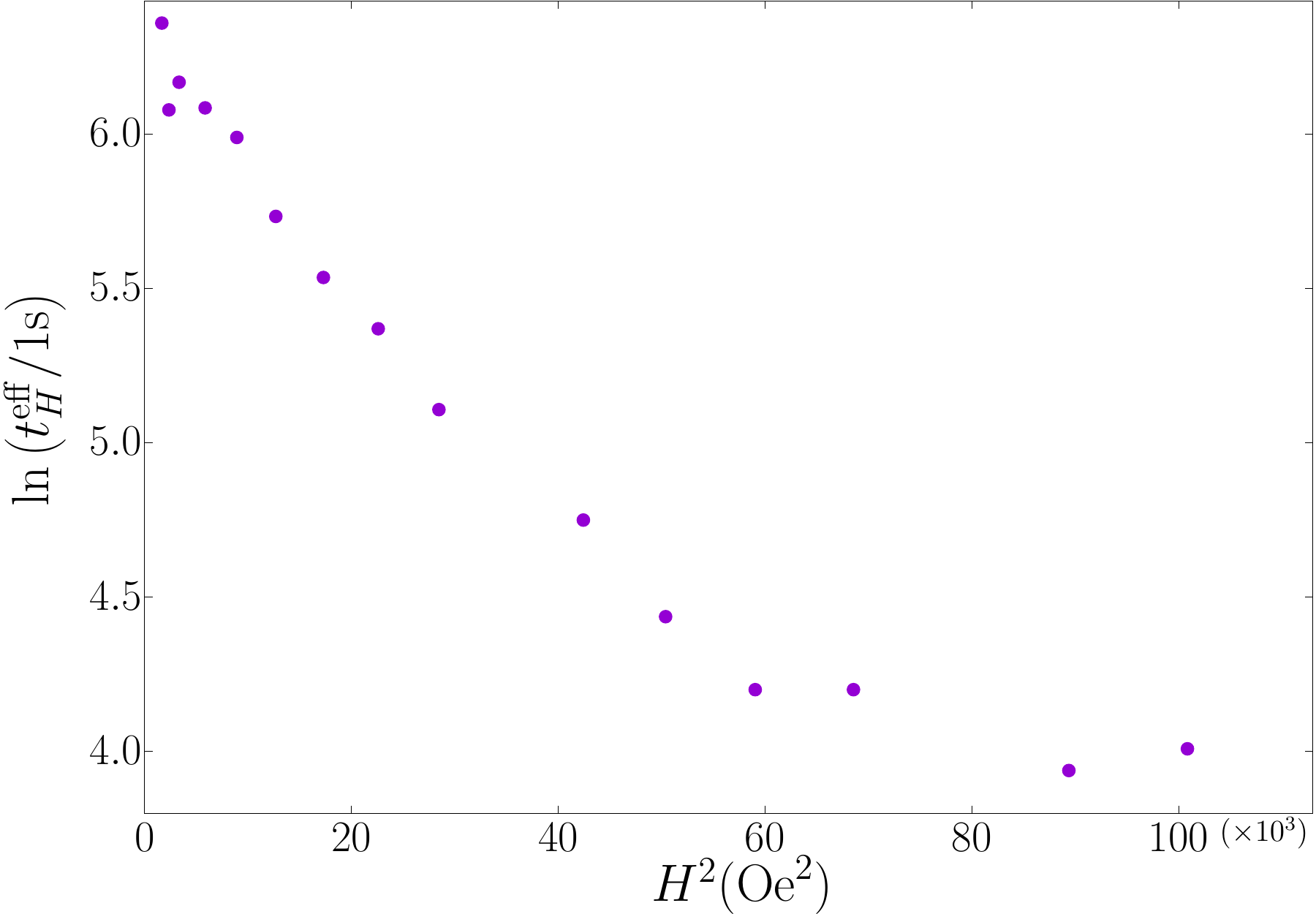}
	\caption{A plot of $\ln t_H^{\text {eff}}$, extracted from the Zeeman energy $E_\text{Z}$ as
in Eq.~\eqref{eq:zeeman_Ez_definition}, vs $H^2$ for Cu:Mn 6 at. \% ($T/\Tg =
0.83, \, \tw = 480$ s) at fixed $\tw$ and $T$. Data taken from Ref.
\cite{joh:99}.}
\label{fig:joh99_data}
\end{figure}
In order to extract a correlation length, Joh et \textit{al.}~\cite{joh:99}
used the notion that the free-energy barrier heights were reduced in the
presence of a magnetic field by the Zeeman energy, $E_\text{Z}$,
\cite{joh:99,guchhait:14,bouchaud:92,vincent:95,janus:17b} and that, for small
magnetic fields $H$,
\begin{equation}
\label{eq:zeeman_Ez_definition}
E_\text{Z} = \big(V_{\text {corr}}/a_0^3\big)\chi_{\text {FC}}H^2\,,
\end{equation}
where $\chi_{\text {FC}}$ is the field-cooled magnetic susceptibility per spin,
$V_{\text {corr}}$ is the correlated volume, and $a_0$ the average spatial
separation of the magnetic ions, so that the number of correlated spins
is $N_\mathrm{corr}=V_{\text {corr}}/a_0^3$.  Joh \emph{et al.} took
\begin{equation}
\label{eq:Ncorr_definition}
N_\text{corr}=V_{\text{corr}}/a_0^3\approx {\frac {4\pi}{3}}\xi^3\,,
\end{equation}
where $\xi$ is in units of $a_0$.  We now know that a more appropriate relationship would be
\begin{equation}
\label{eq:Ncorr_new_def}
N_\text{corr}=V_{\text {corr}}/a_0^3={\frac {4\pi}{3}}\xi^{3-\theta(\tilde x)/2}\equiv b\,\xi^{3-\theta(\tilde x)/2}\,,
\end{equation}
where $b$ is a geometrical factor, and $\theta(\tilde x)$ is the replicon exponent
\cite{janus:17b}. Using Eqs.~\eqref{eq:S_definition} through
\eqref{eq:zeeman_Ez_definition}, the experiments in~\cite{joh:99} 
produced the plot in Fig.~\ref{fig:joh99_data}.
The data in the limit of small magnetic fields $H$ were clearly linear in
$H^2$, allowing Eqs.~\eqref{eq:zeeman_Ez_definition} through~\eqref{eq:Ncorr_new_def} 
to set a value for $\xi$.  

The deviation from
linearity in $H^2$ was puzzling, leading to the authors' stating: ``We do not
have a satisfactory explanation for this change in slope.  A different
description predicts a linear dependence of $E_\text{Z}$ upon $H$, which can be made
to fit the data (\ldots) but with a significant deviation at small field changes''.
It is the purpose of this paper to analyse the entirety of the data considering 
non-linear terms in the spin-glass magnetisation according to a new scaling
law.  In addition to analysing magnetisation data from new experiments, 
we shall also show that the data of Fig.~\ref{fig:joh99_data}, and
subsequent experiments of Bert {\it et al.}~\cite{bert:04} on the Ising spin
glass Fe$_{0.5}$Mn$_{0.5}$TiO$_3$, fit the new scaling law well, obviating the
need to question Eq.~\eqref{eq:zeeman_Ez_definition}, and putting to rest
the controversy over the nature of the Zeeman energy in spin glasses.
 
This paper brings together a complete set of magnetisation measurements of a
single crystal of the prototypical spin glass, CuMn 6 at. \%.  The beauty of
the experimental results is that the correlated volume is most certainly
spherical (as opposed to thin films where the correlated volume is of pancake
geometry \cite{zhai:17b}), and unlimited by finite-size crystallites separated
by grain boundaries~\cite{rodriguez:04,rodriguez:13}.  Accompanying these
measurements are the remarkable simulations of the Janus~II dedicated
supercomputer, which, for the first time, yield spin-glass correlations that
approach experimental time and length scales. Indeed,  the range of correlation
lengths that we are able to achieve both experimentally and numerically is
itself a breakthrough. The current manuscript reports results up to $\xi
\approx 23.6 \, a_0$ ($a_0$ is the typical Mn-Mn distance), which represents a
step forward by a factor of three from previous work~\cite{janus:17b}. On
the experimental side, we reach a correlation length four times larger than in
Ref.~\cite{joh:99}.

The synergy between these two approaches, combined with theory, opens up a new
vista for spin-glass dynamics.  A direct outgrowth of this collaboration is the
introduction of the new magnetisation scaling law, which encompasses the full
range of magnetic fields for temperatures in the vicinity of the condensation
temperature $\Tg$~\cite{zhai-janus:20a}.  This scaling law successfully
describes both experimental and simulation results and, as noted above, will
resolve a nearly three-decade-old controversy concerning the nature of the
magnetic state.

Although our simulations were not designed to that end, we take the occasion as
well to attempt a preliminary search for the de Almeida-Thouless (dAT)
line in the phase diagram of the 3D spin glass.

The paper is organised as follows.  Section \ref{Sec:exp_details} details the
experimental measurements of the non-linear magnetisation in the CuMn spin
glass.  Section \ref{Sec:Janus_details} describes the nature of our numerical
simulations.  Section \ref{Sec:S_of_t} introduces the response function,
Eq.~\eqref{eq:S_definition}, and its extraction from experiment and
simulations.  Section \ref{Sec:scaling_law_all} develops the new scaling law,
and applies it to both experimental and simulation results. In addition,
Section \ref{Sec:overshoot_phenomena} shows the nature of the growth of the
numerical correlation length $\xi$ in the presence of a magnetic field at
temperatures close to the critical temperature $\Tg$. We observed interesting
overshoot phenomena that we prove to be general, as they are observed even in
ferromagnetic systems. Section \ref{Sec:xi_t_tw_dAT} investigates the 
dAT phase boundary in $D = 3$. Important technical details are
provided in the appendices. Finally, Section \ref{Sec:conlusion} summarises our
results, and points to future opportunities stemming from the synergy expressed
in this paper between theory, experiment, and simulations.

\section{Experimental details}
\label{Sec:exp_details}
\noindent
The experimental measurements were made with a CuMn $\sim$ 6 at. \%
single-crystal sample, prepared using a Bridgman method.  The Cu and Mn were
arc melted several times in an argon environment and cast in a copper mold.
The ingot was then processed in a Bridgman furnace.  Both X-ray fluorescence
and optical observation showed that the beginning of the growth is a single
phase.  More details can be found in Appendix~A of Ref.~\cite{zhai:19}.  The
transition temperature, $\Tg = 31.5$~K, was determined from the temperature at
which $M_{\text {ZFC}}(T)$ first began to depart from $M_{\text {FC}}(T)$.
 
The magnetisation measurements were made using a commercial DC SQUID.  The
sample was quenched from 40~K at 10~K/min to the measuring temperature $\Tm$ in
zero magnetic field.  After stabilisation of the temperature, the system was
aged for a waiting time $\tw$ before a magnetic field was applied, and the
magnetisation $M_{\text {ZFC}}(t,\tw;\Tm)$ recorded as a function of time $t$.
The temperatures $\Tm$ were chosen as 28.5~K, 28.75~K, and 29~K, so that
$\Tm\geq 0.9\Tg$.  The magnetic fields ranged from 16~Oe to 59~Oe.  Table
\ref{tab:exp_details} displays the relevant experimental parameters, including
the effective replicon exponent $\theta(\tilde x)$.
 
\begin{table}[tb]
	\caption{The values of the measuring temperatures $\Tm$ and waiting
times $\tw$ for the four experimental regimes, the respective correlation
lengths at times $\tw$ (in units of the average Mn-Mn spacing $a_0$), and the
effective replicon exponent $\theta (\tilde x)$ obtained from Eqs.
\eqref{eq:x_equiv_lJxi_def} and \eqref{eq:Ncorr_vs_xi} below (see also
Appendix~\ref{Appendix:josephson_lJ}).}
\begin{indented}
\item[]\begin{tabular}{@{}c  c  c  c  c}
\br
                & $\Tm$ (K) & $t_{\mathrm{w}}$ (s) & $\xi(\tw)/a_0$ & $\theta (\tilde x)$\\
\mr
                {Exp. 1} & 28.50 & 10\,000 & 320.36 &  0.337\\
                {Exp. 2} & 28.75 & 10\,000 & 341.76 &  0.344\\
                {Exp. 3} & 28.75 & 20\,000 & 359.18 &  0.342\\
                {Exp. 4} & 29.00 & 10\,000 & 391.27 &  0.349\\
\br
        \end{tabular} 
        \label{tab:exp_details}
\end{indented}
\end{table}
 
\begin{table}[tb]
   \caption{Main parameters for each of our numerical simulations:
	     $T$, $\tw$, $\xi(\tw)$, the longest simulation time
$t_\mathrm{max}$, the replicon exponent $\theta(\tilde x)$ (see Appendix
\ref{Appendix:josephson_lJ}) and the value of $C_\mathrm{peak}(\tw)$ employed
in Eq.~\eqref{eq:C_teff_Cpeak}. Here and in the rest of the paper, error
bars are one standard deviation.}
\begin{indented}
\item[] \begin{tabular}{@{}c c  c l D{.}{.}{-1} c  l c c}
\br
                        & & $T$ &\multicolumn{1}{c}{$\tw$} & \multicolumn{1}{c}{$\xi(\tw;H=0)$} & & \multicolumn{1}{c}{$t_\mathrm{max}$}  & $\theta(\tilde{x})$ & $C_\text{peak}$ \\
                        \mr
                        {Run 1} & & 0.9 & $2^{22}$     & 8.294(7)  && $2^{30}$     & 0.455  & 0.533(3)  \\ 
                        {Run 2} & & 0.9 & $2^{26.5}$   & 11.72(2)  && $2^{30.5}$   & 0.436 & 0.515(2)  \\ 
                        {Run 3} & & 0.9 & $2^{31.25}$  & 16.63(5)  && $2^{33}$     & 0.415 & 0.493(3)  \\
                        {Run 4} & & 1.0 & $2^{23.75}$  & 11.79(2)  && $2^{28}$     & 0.512 & 0.422(2)  \\
                        {Run 5} & & 1.0 & $2^{27.625}$ & 16.56(5)  && $2^{30}$     & 0.498 & 0.400(1)  \\
                        {Run 6} & & 1.0 & $2^{31.75}$  & 23.63(14) && $2^{35}$     & 0.484 & 0.386(4)  \\
                        \br
       \end{tabular}
   \label{tab:details_NUM}
\end{indented}
\end{table}

\section{Some details of the simulations}
\label{Sec:Janus_details}
We carried out massive simulations on the  Janus~II supercomputer
\cite{janus:14} to study the Ising-Edwards-Anderson (IEA) model in a cubic
lattice with periodic boundary conditions and size $L=160 \, a_0$, where $a_0$
is the average distance between magnetic moments, see Table~\ref{tab:details_NUM} for the simulation details. The $N=L^D$ Ising spins,
$s_{\boldsymbol{x}} = \pm 1$, interact with their lattice nearest neighbours through
the Hamiltonian:
\begin{equation}
\mathcal H= - \sum_{\langle \boldsymbol x, \boldsymbol y\rangle} J_{\boldsymbol x \boldsymbol y} s_{\boldsymbol{x}} s_{\boldsymbol{x}} - H \sum_{\boldsymbol{x}} s_{\boldsymbol{x}} \; ,
\end{equation}
where the quenched disordered couplings are $J_{\boldsymbol x \boldsymbol y}= \pm 1$ with
$50\%$ probability. We name a particular choice of the couplings a
\emph{sample}. In the absence of an external magnetic field, $H=0$, this model
undergoes a spin-glass transition at the critical temperature $\Tg=1.102(3)$~\cite{janus:13}.
 
As we explained in the Introduction, in order to simulate the experimental
zero-field-cooling (ZFC) protocol the following
procedure was performed: We  place the initial random spin configuration 
instantaneously at the working temperature $T$ and let relax for a time
$\tw$ at $H=0$. At time $\tw$, we turn on the external magnetic field 
and we start recording the magnetic density,
\begin{equation}
  M_\mathrm{ZFC}(t,\tw;H)=\frac{1}{160^3} \sum_{\boldsymbol x}\ s_{\boldsymbol x} (t+\tw;H)\,,
\end{equation}
as well as the correlation function,
\begin{equation}
\label{eq:Ctw_def}
  C(t,\tw;H)=\frac{1}{160^3} \sum_{\boldsymbol x}\  s_{\boldsymbol x}(\tw;0)\, s_{\boldsymbol x} (t+\tw;H)\, .
\end{equation}
The non-equilibrium dynamics was simulated with a Metropolis
algorithm; the numerical time unit being the lattice sweep, roughly
corresponding to one picosecond of physical time \cite{mydosh:93}. For each
temperature and waiting time, see Table~\ref{tab:details_NUM}, several magnetic
fields were simulated. For computational reasons, one single independent sample
was simulated for each case. We checked, however, the robustness and the
sample independence of our results in a single case, studied in detail in
Appendix~\ref{Appendix:Cpeak_samples}.
 
According to Ref.~\cite{janus:17b}, the value of the dimensionless magnetic
field $H$ used in the numerical simulation can be matched to the physical
field.  This relation was estimated from experimental
$\mathrm{Fe}_{0.5}\mathrm{Mn}_{0.5}\mathrm{TiO}_3$ data \cite{aruga_katori:94}.
We found that $H=1$ in the IEA model  corresponded to $\approx5 \times
10^4$~Oe physically.  Hence, our experimental range (16~Oe to 59~Oe)
corresponds to magnetic field $0.0003 \lesssim H \lesssim 0.0012$ in the IEA
model. However, the signal-to-noise ratio, which scales linearly in $H$ for
small fields~\cite{mezard:87}, limited our simulation to $H \geq 0.005$,
equivalent to a physical $H=250$~G.
 
In order to match the experimental and numerical scales, we exploited
dimensional analysis~\cite{fisher:85} to relate $H$ and the reduced temperature
$\hat{t} = ( \Tg -T)/\Tg$ through the scaling
\begin{equation}
\label{eq:F-S_relation}
\hat{t}_\mathrm{num} \approx \hat{t}_\mathrm{exp} \left( \frac{H_\mathrm{num}}{H_\mathrm{exp}}\right)^\frac{4}{\nu(5-\eta)} \; ,
\end{equation}
where $\nu=2.56(4)$ and $\eta=-0.39(4)$ are $H=0$ critical exponents
\cite{janus:13}, while subscripts exp and num stand for experiment and
simulation, respectively. According to Eq.~\eqref{eq:F-S_relation}, and
minding signal-to-noise limitations, we can match the experimental and numerical scales
by increasing $\hat{t}_\mathrm{num}$, resulting in $0.89 \lesssim
T_\mathrm{num} \lesssim 0.99$. Given our pre-existing database of long
simulations at $H=0$ \cite{janus:18}, it has been convenient to work at
temperatures $T_\mathrm{num}=0.9$ and $T_\mathrm{num}=1.0$. Table
\ref{tab:details_NUM} displays the relevant numerical parameters, including the
effective replicon exponent $\theta(\tilde{x})$ and the $C_\mathrm{peak}(\tw)$
values that will be introduced and explained in Section~\ref{Sec:C_peak}.

Let us finally remark that Section~\ref{Sec:overshoot_phenomena} uses a different
set of simulations from the rest of the  paper.

\section{Measurements and computations of the relaxation rate}
\label{Sec:S_of_t}

In this section we describe the  relaxation function $S(t,\tw;H)$ (Section~\ref{Sec:St_extracting_num})
and explain how $\teff_H$ is extracted from simulations (Section~\ref{Sec:C_peak}).

\subsection{Extracting the relaxation function $S(t,\tw;H)$}
\label{Sec:St_extracting_num}
\begin{figure}[t]

        \centering
        \includegraphics[width = 0.7\columnwidth]{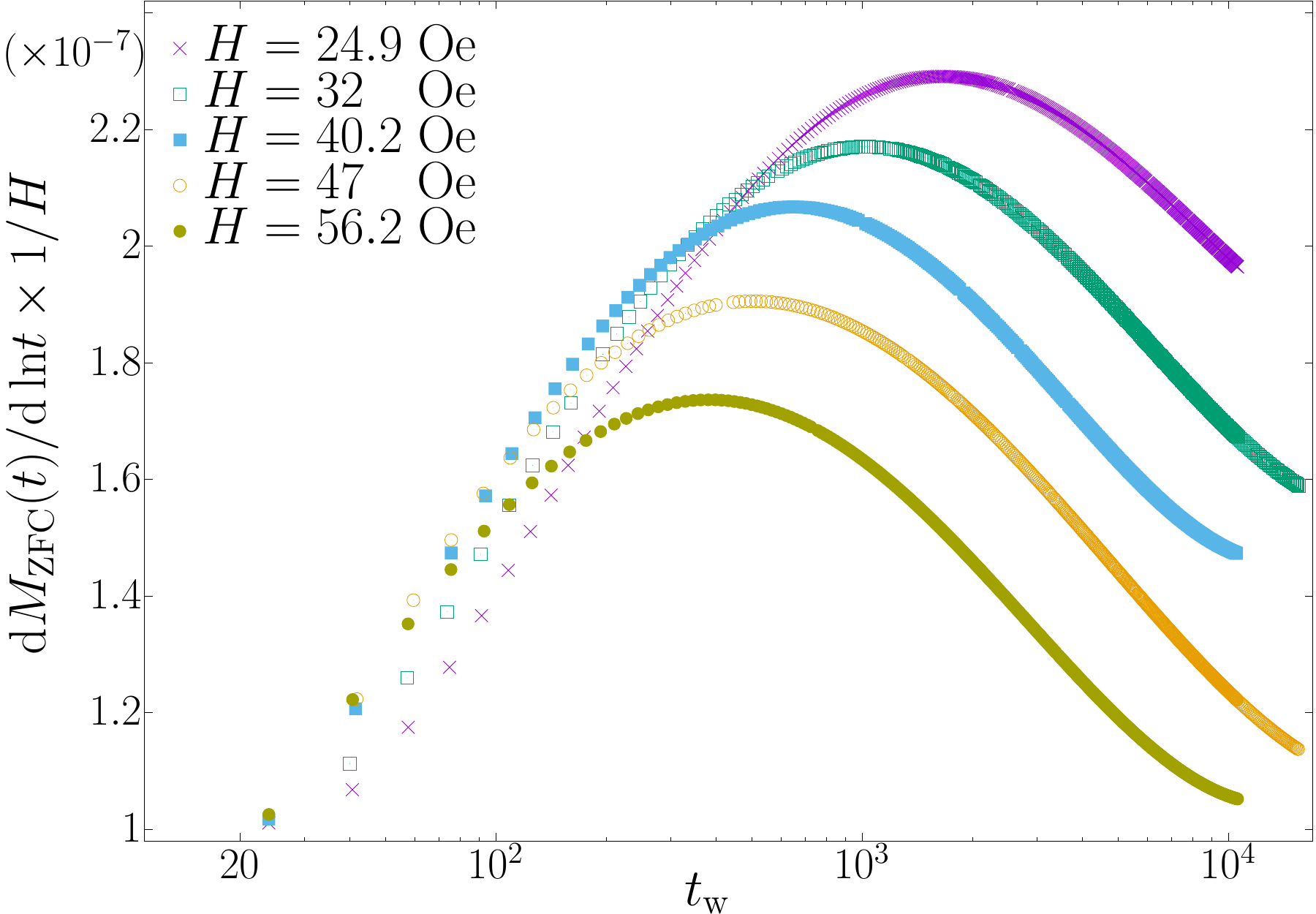}
	\caption{Example of $S(t,\tw;H)$ measurements for different magnetic
fields.  The sample is a single crystal CuMn 6 at. \%, and the measurements
were taken at a waiting time of $\tw=10\,000$~s and at $T=28.5$~K.  The
time at which $S(t,\tw;H)$ peaks defines $t_H^{\text {eff}}$, the effective
response time.  The shift to shorter times as $H$ increases is the measure of
the reduction of $\itDelta_{\text {max}}$ with increasing Zeeman interaction, and
is used to extract the linear and non-linear terms in the magnetic
susceptibility.}
        \label{fig:exp_St_behavior}
\end{figure}

The main quantity used in our experiments of \cite{joh:99} is the relaxation function $S(t,\tw;H)$:
\begin{equation}
  S(t,\tw ;H)=\frac{\mathrm{d} M_{\mathrm{ZFC}}(t,\tw;H)}{\mathrm{d}\ln t}\,,
  \label{eq:S_oft}
\end{equation}
which exhibits a local maximum at time  $\teff_H$.
Experimentally, measurements of $M_\mathrm{ZFC}(t,\tw;H)$ enable the evaluation
of the relaxation function $S(t,\tw;H)$ directly. A representative set of data
for $\Tm = 28.5$~K and $\tw=10^4$~s is displayed in Fig.~\ref{fig:exp_St_behavior}. 
Numerically, the calculation of $S(t,\tw;H)$ is sensitive to the relative errors of 
the magnetisation density, which increase as
\begin{equation}
\frac{\delta M_\mathrm{ZFC}(t,\tw;H)}{M_\mathrm{ZFC}} \propto \frac1H \, .
\end{equation} 

\begin{figure}[t]
 \includegraphics[width=\linewidth]{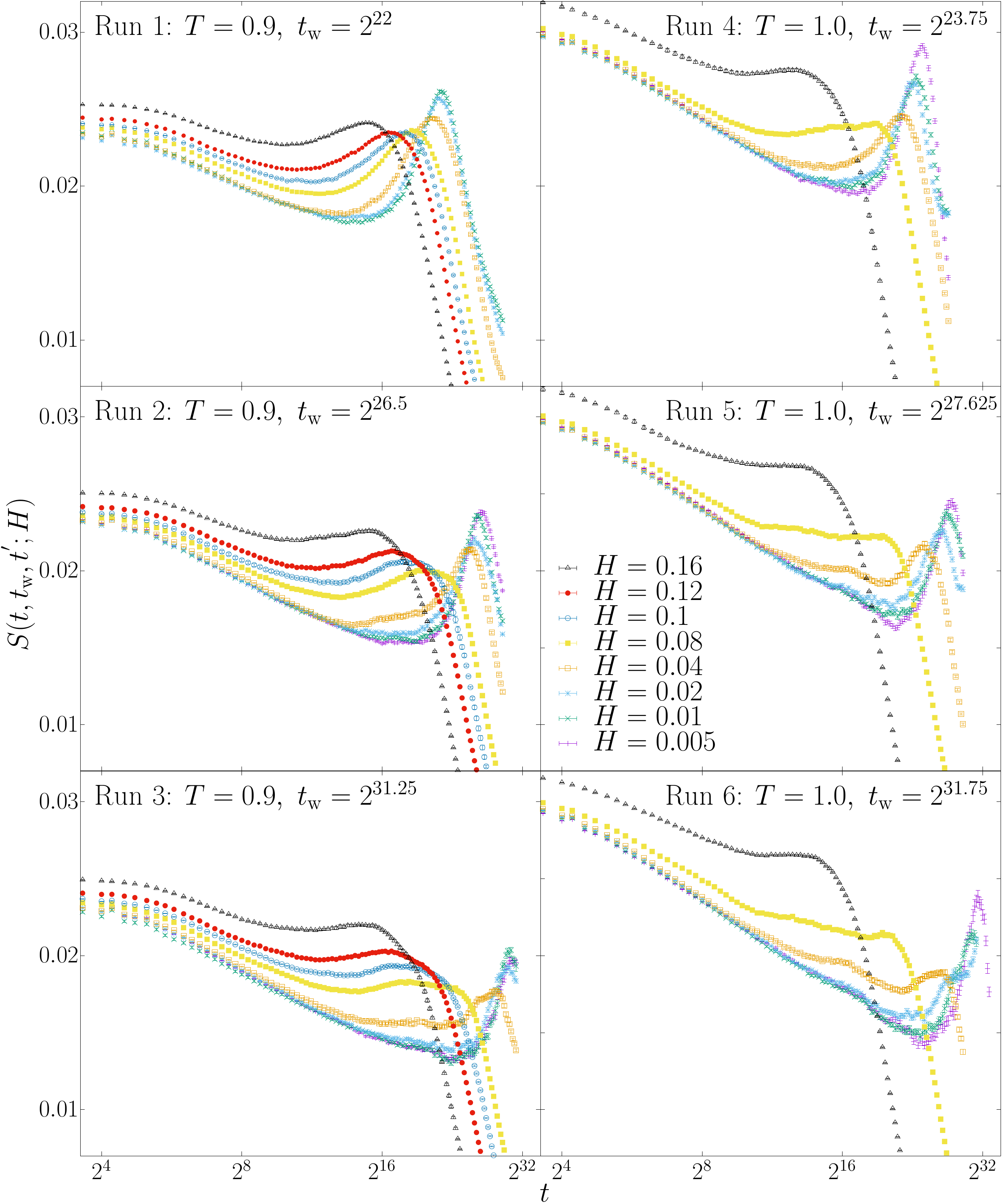}
  \caption{Time evolution of the relaxation rate $S(t,\tw, t';H)$ of Eq.~\eqref{eq:S_of_rescaled_t}
 for the six runs of Table~\ref{tab:details_NUM}. All plots have the time parameter $k=8$ in Eq.~\eqref{eq:tprime}.}
  \label{fig:S_oft_all}
\end{figure}

We employ two tricks to extract the relaxation function $S(t,\tw;H)$ from
simulations.  On the one hand, we perform a de-noising method to regularise the
magnetisation density $M_{\mathrm{ZFC}}(t,\tw;H)$, exploiting the
fluctuation-dissipation relations
(FDR)~\cite{cugliandolo:93,cruz:03,franz:94,franz:98,franz:99}
\begin{equation}
\label{eq:FDR}
  \frac{T}{H} M_\mathrm{ZFC}(t,\tw;H) = {\cal F} (C;H) \, ,
\end{equation}
where ${\cal F}(C;H)$ behaves at large $C(t,\tw;H)$ as ${\cal
F}(C;H)=1-C(t,\tw;H)$. We report the details in Appendix
\ref{Appendix:FDR_smooth}.  On the other hand, we define $S(t,\tw;H)$ as a
finite-time difference
\begin{equation}
\label{eq:S_of_rescaled_t}
S(t, \tw, t' ; H) = \frac{M_\mathrm{ZFC}(t',\tw;H)- M_\mathrm{ZFC}(t,\tw;H)}{\ln\left(t'/t\right)} \; . 
\end{equation}
 In simulations, time is discrete and we store configurations at
times $t_n=$ integer−part−of $2^{n/4}$,  with $n$ an integer. Let us write
explicitly the integer dependence of times $t$ and $t'$ as:
\begin{equation}\label{eq:tprime}
t \equiv t_n, \quad \; \quad t' \equiv t_{n+k},
\end{equation}
where $k$ is an integer time parameter. The reader will note that there is a
trade-off in the choice of $k$. On the one hand, the smaller $k$ is the better
the finite difference in Eq.~\eqref{eq:S_of_rescaled_t} represents the
derivative. On the other hand, when $k$ grows the statistical error  in the
evaluation of Eq.~\eqref{eq:S_of_rescaled_t} decreases significantly. In this
section we report only the case for $k=8$ (more details about time
discretisation are provided in
Appendix~\ref{Appendix:details_St_construction}).  The numerical $S(t,\tw,
t';H)$ are exhibited in Fig.~\ref{fig:S_oft_all}, where a local maximum in the
long-time region can be seen.

\subsection{A different approach for the computation of $t^\mathrm{eff}_H$ in simulations}
\label{Sec:C_peak}

\begin{figure}[t] \centering
 \includegraphics[width=\linewidth]{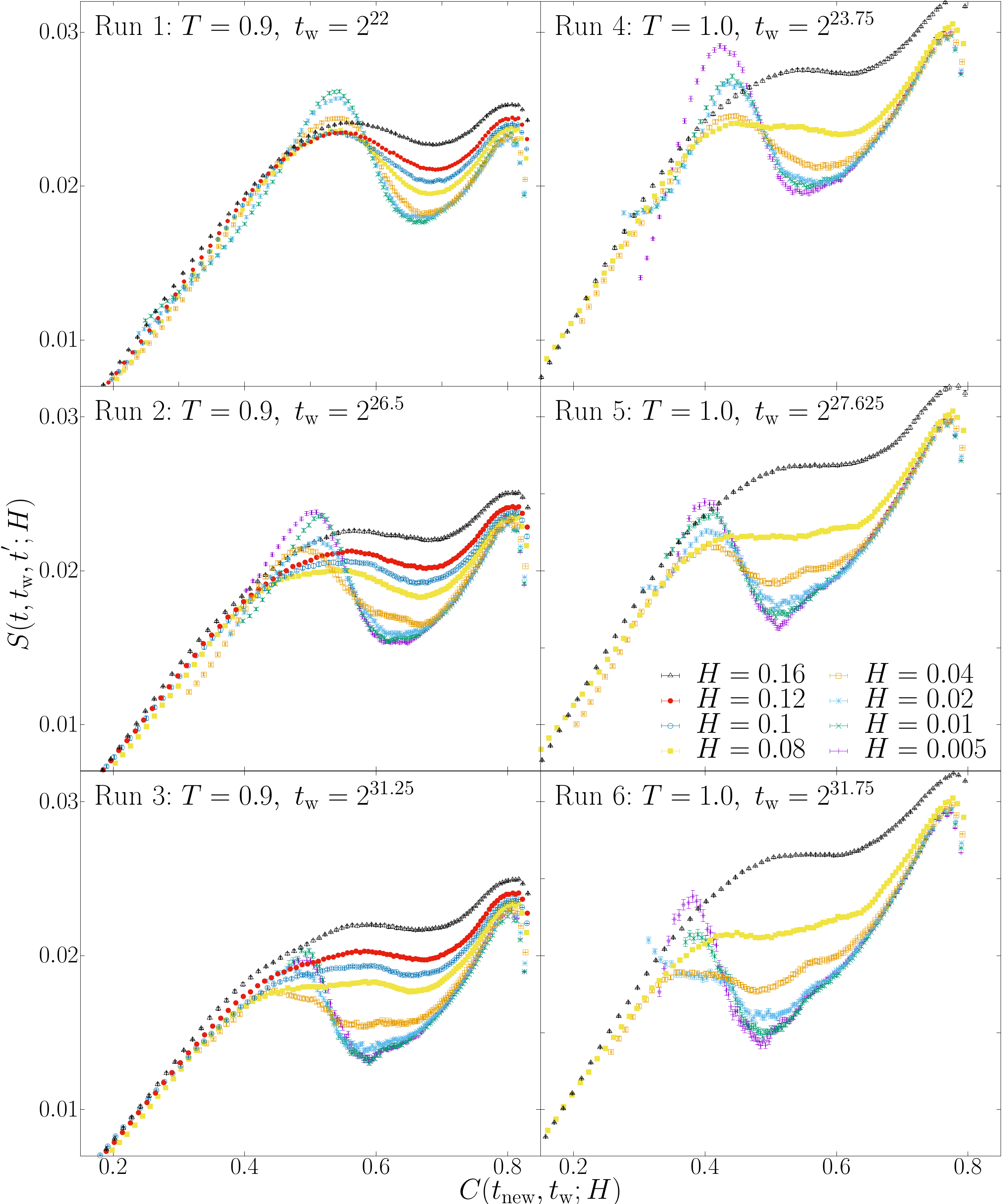}
  \caption{$S(t,\tw, t';H)$ as a function of $C(t,\tw;H)$.
  The peak region is enlarged in Fig.~\ref{fig:zoomS_Ct}.  The physically
relevant peak is the one for small $C$, corresponding to long times. We
consider the reparametrised $t_\text{new}$ with $k=8$ in
Eq.~\eqref{eq:tnew_def}.
}
  \label{fig:SCt_all}
\end{figure}

\begin{figure}[t] \centering
 \includegraphics[width=\linewidth]{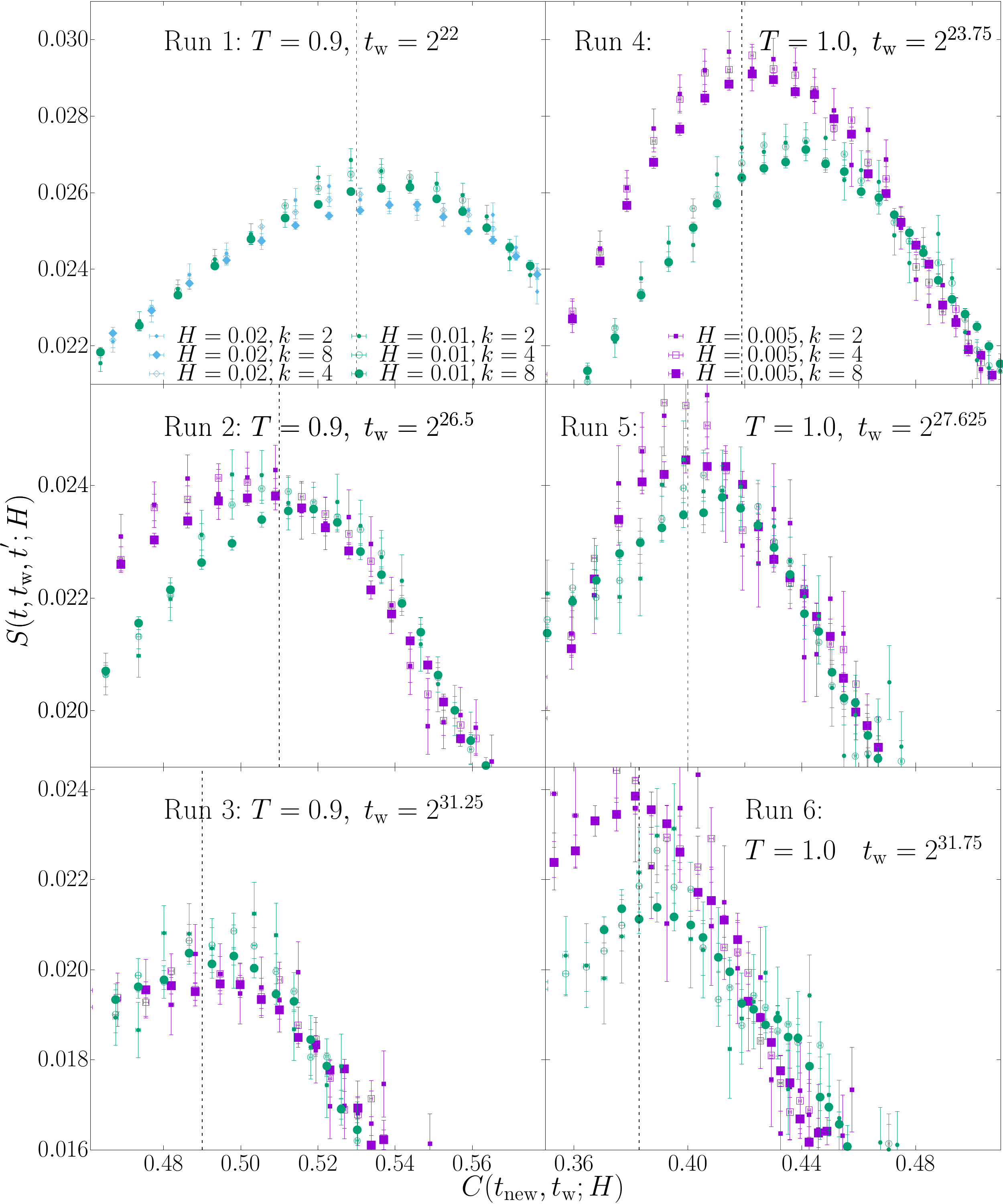}
   \caption{Enlargement of the peak region of $S(t,\tw,t'; H)$ as a function of $C(t,\tw;H)$ for several
    values of time parameter $k$ in Eq.~\eqref{eq:tprime}.
   The dashed black lines indicate the $C_\mathrm{peak}(\tw)$ positions.}
  \label{fig:zoomS_Ct}
\end{figure}
As explained in the Introduction, we are interested in the evaluation of the
time when the relaxation function $S(t,\tw, t';H)$ peaks, namely
$t_H^\mathrm{eff}$. Two problems arise:
\begin{enumerate}
\item The reader will note two separate peaks in the $S(t,\tw,t';H)$ curves of
Fig.~\ref{fig:S_oft_all}: namely the peak at microscopic times $t \sim 4$, and
the peak we are interested in at $t \sim \tw$. Unfortunately, the distinction
between the two is only clear at small $H$. Previous numerical work~\cite{janus:17b} 
did not face this problem, probably because of their smaller
correlation length, $\xi \approx 8$, (non-linear susceptibilities grow very
fast with $\xi$, see the next section).
\item We are most interested in the limit $H \to 0$, which is extremely noisy,
as we have explained above.
\end{enumerate}

An interesting possibility emerges when plotting the relaxation function
$S(t,\tw, t';H)$ in terms of the correlation function $C(t,\tw;H)$, rather than
as a function of time (see Fig.~\ref{fig:SCt_all}). 
$C(t,\tw;H)$ is a decreasing function of time, the long-time region
corresponding to small $C(t,\tw;H)$, and vice versa.
Hence, the \emph{physical} peak in which we are
interested is the one that appears at small $C(t,\tw;H)$ (see Fig.~\ref{fig:SCt_all}).
Analogously to Fig.~\ref{fig:S_oft_all}, we report only the
case for $k=8$ in Fig.~\ref{fig:SCt_all}.

The simulation data strongly suggest that, when $H \to 0$, the correlation
function $C(t,\tw;H)$ approaches a constant value $C_\mathrm{peak}(\tw)$ at the
maximum of the relaxation function. Hence, our proposal is to {define}
$\teff_H$ in simulations as the
time when $C(t,\tw;H)$ reaches the value $C_\mathrm{peak}(\tw)$:
\begin{equation}
\label{eq:C_teff_Cpeak}
C(\teff_H,\tw; H) = C_\mathrm{peak}(\tw) \; .
\end{equation}
As the reader can see, Eq.~\eqref{eq:C_teff_Cpeak} is applicable also at $H=0$,
solving the problem of the vanishing magnetisation in this limit.
\noindent
The crucial point for our new $\teff_H$ definition, see
Eq.~\eqref{eq:C_teff_Cpeak}, is, hence, the computation of
$C_\mathrm{peak}(\tw)$. Two problems arise:
\begin{enumerate}
\item The constant-value $C_\mathrm{peak}(\tw)$ is well defined only for small magnetic field $H$.
\item The relaxation function as a function of the correlation, ${\cal S}(C;H)$, is an implicit function of a reparametrised time,
\begin{equation}
\label{eq:tnew_def}
t_\mathrm{new} = \frac{1}{2} \ln \left( \frac{t_{n+k}}{t_n} \right) \,  ,
\end{equation}
(see Appendix \ref{Appendix:details_St_construction} for details).
\end{enumerate}
Our strategy has been to study, for each run, the behaviour of ${\cal S}(C;H)$
for the two smallest magnetic fields $H$ and for three different time parameters
$k$.  We report in Fig.~\ref{fig:zoomS_Ct} a closeup of the peak region for
${\cal S}(C;H)$, used for the evaluation of $C_\mathrm{peak}(\tw)$.  We report
our estimates for $C_\mathrm{peak}(\tw)$ in Table~\ref{tab:details_NUM}.
 
The relaxation function $S(t_\mathrm{new},\tw;H)$ depends on the correlation
length $\xi(\tw)$, and on the applied magnetic field $H$,
Eq.~\eqref{eq:S_definition}.  We observe, however, that
$S(t_\mathrm{new},\tw;H)$ has a temperature dependence, which we extract by
comparing Runs~4 and~2 in Fig.~\ref{fig:zoomS_Ct}.
These two cases are characterised by
\begin{enumerate}
\item A similar starting correlation length $\xi(\tw;H=0) \approx 11.7$, (see Table~\ref{tab:details_NUM}),
\item The same set of applied magnetic fields, namely $H=0.005$ and $H=0.01$.
\end{enumerate}
\noindent
Yet, there appear to be two different scenarios in the data plotted 
in Fig.~\ref{fig:zoomS_Ct}.  In
Run~2, the peak of  ${\cal S}(C;H)$ is almost the same
for all the rescaled time curves. In Run~4, however,
the peaks separate for different $k$.  As  will be
explained in Section~\ref{Sec:non-linear_scaling}, this difference in behaviour
is caused by increasing non-linear effects in the magnetisation,
$M_\mathrm{ZFC}(t,\tw;H)$.
 
In conclusion,  Eq.~\eqref{eq:C_teff_Cpeak} solves our two problems at once. We
no longer need to resolve the short-time and the long-time peaks in
Fig.~\ref{fig:S_oft_all} and it bypasses the problem of the vanishing
magnetisation as $H$ goes to zero. 

\section{Scaling law}
\label{Sec:scaling_law_all}

We address here three different aspects of the scaling law. The assumptions
that led us to our scaling law are given in Section~\ref{Sec:scaling_law}.
Next, in Section~\ref{Sec:non-linear_scaling_EXP}, we use the scaling law in the
analysis of our experimental data (previous data are also reanalysed in
Section~\ref{Sec:reanalysis_exp_data}), with the corresponding analysis for our
simulations given in Section~\ref{Sec:ratio_time_NUM}. Section~\ref{Sec:non-linear_scaling} 
shows our experimental and numerical
results together, according to the new scaling law. In Section~\ref{Sec:overshoot_phenomena}
we address the nature of the growth of the numerical correlation length
$\xi(\tw)$ in the presence of a magnetic field at temperatures close to the
condensation temperature $\Tg$.

\subsection{Non-linear scaling law}
\label{Sec:scaling_law}

Scaling laws for the spin-glass susceptibility in the vicinity of the
condensation temperature have been proposed and analysed for decades. We first
recall an important early approach, and then develop the scaling law that we
have employed to analyse our experiments and simulations.

Non-linear magnetisation effects, and their scaling properties in spin glasses,
were first introduced by Malozemoff, Barbara, and Imry
\cite{malozemoff:82,malozemoff:82b,chandra:93}, who  introduced the
relation for the singular part of the magnetic susceptibility,
\begin{equation}
\label{eq:chi_singular_def}
\chi_s=H^{2/\delta}f\big(t_r/H^{2/\phi}\big)\,,
\end{equation}
where $f(x)$ is a constant for $x\rightarrow 0$; $f(x)=x^{-\gamma}$ for
$x\rightarrow \infty$; $\phi = \gamma\delta/(\delta - 1) \equiv \beta\delta$;
and $t_r$ is the reduced temperature $T/\Tg$.  This form was used by L\'evy and
Ogielski \cite{levy:86} and by L\'evy \cite{levy:88}, who measured the AC
non-linear susceptibilities of very dilute AgMn alloys above and below $\Tg$ as
a function of frequency, temperature, and magnetic field.  The critical
exponents of Eq.~\eqref{eq:chi_singular_def} were evaluated as $\beta=0.9$,
$\gamma = 2.3$, $\delta=3.3$, $\nu= 1.3$, and $z=5.4$. They differ
substantially from Monte Carlo simulations for short-range Ising systems:
$\beta = 0.782(10)$, $\gamma = 6.13(11)$, $\nu = 2.562(42)$
from~\cite{janus:13}.  The discrepancy in the value of $\gamma$ is very large,
and most probably arises from the lack of an exact value for $\Tg$ in the
experiments.  This illustrates the value of and need for a different approach
for scaling the non-linear magnetisation of spin glasses in the vicinity of
$\Tg$.
 
Our approach is to express the non-linear components of the magnetic
susceptibility in terms of $\xi(t,\tw)$,  the spin-glass correlation length in
a magnetic field $H$.\footnote{The correlation length $\xi(t,\tw)$ is of course
also a function of the temperature $T$, but here we are only interested in the
non-linearity of the magnetisation.} This approach gives the non-linear
magnetisation a direct connection to a measurable quantity and obviates the
need for an accurate value of $\Tg$.
 
The argument goes as follows.  Let $M(t,\tw;H)$ be the magnetisation per spin,
where explicit attention is paid to the waiting (aging) time $\tw$ in the
preparation of the spin-glass state.  The generalised susceptibilities
$\chi_1, \chi_3, \chi _5,\ldots$ are defined through the Taylor expansion
\begin{equation}\label{eq:suscept-defined}
M(H)=\chi_1H+{\frac {\chi_3}{3!}}H^3+{\frac {\chi_5}{5!}}H^5+{\mathcal O}(H^7)\,.
\end{equation}
where, for brevity's sake, we omit arguments $t$ and $\tw$.
 
Under equilibrium conditions, and for a large-enough correlation length
$\xi_\text{eq}$, there is a scaling theory for the magnetic response to an
external field $H$~\cite{parisi:88,amit:05}.  Our main hypothesis in
this work is that this scaling theory holds not only at equilibrium, but also
in the non-equilibrium regime for a spin glass close to $T_\mathrm{g}$ and in
the presence of a small external magnetic field $H$:
\begin{equation}
M(t,\tw;H) = [\xi(t+\tw)]^{y_H-D}  {\cal F} \left( H [ \xi ( t + \tw) ] ^{y_H} , \frac{\xi(t +\tw)}{\xi(\tw)} \right)\,.
\label{eq:m-scaling}
\end{equation}
According to full-aging spin-glass dynamics (see, \emph{e.g.},~\cite{rodriguez:03}),
Eq.~\eqref{eq:m-scaling} tells us that $\xi(t+\tw)/\xi(\tw)$ will be
approximately constant close to the maximum of the relaxation rate (see
Fig.~\ref{fig:exp_St_behavior}), so we shall omit this dependence.  Hence,
combining Eq.~\eqref{eq:suscept-defined} and~\eqref{eq:m-scaling}, one can
express the generalised susceptibilities $\chi_1, \chi_3, \chi _5,\ldots$ in terms
of the spin-glass correlation length $\xi(t,\tw;H)$:
\begin{equation}\label{eq:chi_2n-1prop}
\chi_{2n-1}\propto |\xi(\tw)|^{2n \,y_H -D}\,,
\end{equation}
where we have omitted the arguments $t,H$ for convenience, and
\begin{equation}\label{eq:2yh_def}
2y_H=D-{\frac {\theta(\tilde x)}{2}}\,,
\end{equation}
with $\theta(\tilde x)$ the replicon exponent \cite{janus:17b}.
 
The first term of $M(H)$ in Eq.~\eqref{eq:suscept-defined} is $\chi_1$, which
contains the linear term as well as the first non-linear scaling term
\cite{zhai-janus:20a}, so we write
\begin{equation}\label{eq:chi1_corrections_non-linear}
\chi_1={\frac {\hat S}{T}} + {\frac {a_1(T)}{\xi^{\theta(\tilde x)/2}}}\,.
\end{equation}
where $\hat S$ is the function appearing in the fluctuation-dissipation
relations~\cite{janus:17} and $a_1(T)$ is some unknown constant (hopefully
smoothly varying with temperature).
 
The free-energy variation {per spin} in presence of a magnetic field can
be obtained by integrating the magnetic density, Eq.~\eqref{eq:suscept-defined},
with respect to the magnetic field:
\begin{equation}\label{eq:Ez_zeeman_taylor_chi_expansion}
\Delta F=-\bigg[\frac{\chi_1}{2} H^2+{\frac {\chi_3}{4!}}H^4 + {\frac {\chi_5}{6!}}H^6+ {\mathcal O}\big(H^8\big)\bigg]\,.
\end{equation}
Substituting the scaling behaviour from Eq.~\eqref{eq:chi_2n-1prop} and Eq.~\eqref{eq:chi1_corrections_non-linear}, 
the free energy $\Delta F$ can be written as (we drop the $\tilde x$ dependence of $\theta$ for brevity)
\begin{equation}\label{eq_Ez_zeeman_nonlinar_an_expansion}
\Delta F=-\bigg[{\frac {\hat S}{2 T}}H^2+\frac{a_1(T)}{\xi^{\theta/2}} H^2+ a_3(T)\xi^{D-\theta} H^4 +a_5(T)\xi^{2D-(3\theta/2)} H^6+{\mathcal O}\big(H^8\big)\bigg]\,,
\end{equation}
where again the $a_n(T)$ are unknowns and (again, hopefully) smoothly varying
functions of temperature.  We use the effective response time, $t_H^{\text
{eff}}$, to reflect the total free-energy change at magnetic field $H$ and
$H=0^+$:
\begin{equation}\label{eq:ratio_effective_time_vs_Ncorr}
\ln\bigg[{\frac {t_H^{\text {eff}}}{t_{ H \to 0^+}^{\text {eff}}}}\bigg]= N_\text{corr} \Delta F\,,
\end{equation}
where $N_\text{corr}$ is the number of correlated spins, $N_\text{corr}=
V_{\text {corr}}/a_0^3$, with $V_\mathrm{corr}$ the correlated spins
volume and $a_0$ the lattice constant or average distance between magnetic
moments.  Combining Eq.~\eqref{eq:ratio_effective_time_vs_Ncorr} with Eqs.~\eqref{eq:Ncorr_new_def} 
and~\eqref{eq:Ez_zeeman_taylor_chi_expansion} leads to
\begin{equation}\label{eq:ratio_effective_time_scaling}
\fl
\ln\bigg[{\frac {t_H^{\text {eff}}}{t_{H \to 0^+}^{\text {eff}}}}\bigg]=-b \bigg[ \bigg({\frac {\hat S}{2\;T}}+{\frac {a_1(T)}{\xi^{\theta/2}}}\bigg)\xi^{D-(\theta/2)}H^2\\
+\,a_3(T)\xi^{2D-(3\theta/2)}H^4+a_5(T)\xi^{3D-2\theta}H^6+{\mathcal O}\big(H^8\big)\bigg]\,,
\end{equation}
where  coefficient $b$ is a geometrical factor, see
Eq.~\eqref{eq:Ncorr_new_def}, and  we have absorbed the $k_\text{B} T$ term in the
$a_n(T)$ coefficients.  The correction term $a_1(T)/\xi^{\theta(\tilde x)/2}$ is small
compared to ${\hat S}/T$, so it will be dropped in subsequent expressions.
Eq.~\eqref{eq:ratio_effective_time_scaling} shows that the higher-order terms
have the functional form
\begin{equation}\label{eq:xi_2n-1_expantion}
\chi_{2n-1}{\frac {H^{2n}}{(2n)!}} = a_{2n-1}(T)\xi^{-\theta(\tilde x)/2}\big[\xi^{2y_H}H^2\big]^n\,,
\end{equation}
where 
\begin{equation}
2y_H=D-{\frac {\theta(\tilde x)}{2}}\,.
\end{equation}
This leads to the new scaling relation,
\begin{equation}\label{eq:ratio_effective_time_scaling_final}
\ln\bigg[{\frac {t_H^{\text {eff}}}{t_{H\rightarrow 0^+}^{\text {eff}}}}\bigg]={\frac {\hat S}{2 T}}\,\xi^{D-\theta(\tilde x)/2}H^2
+\,\xi^{-\theta(\tilde x)/2}{\mathcal G}\big(T,\xi^{D-\theta(\tilde x)/2}H^2\big)\,,
\end{equation}
where the geometrical factor $b$ has been absorbed into the scaling function
${\mathcal G}$.  Comparison with the previous, more classical, relation, Eq.~\eqref{eq:chi_singular_def},
evinces  the simplicity and power of our approach
to scale the non-linear magnetisation in the vicinity of $\Tg$.

\subsection{Experimental non-linear magnetisation}
\label{Sec:non-linear_scaling_EXP}

\begin{figure}[t]
        \centering
                \includegraphics[width = 0.6\columnwidth]{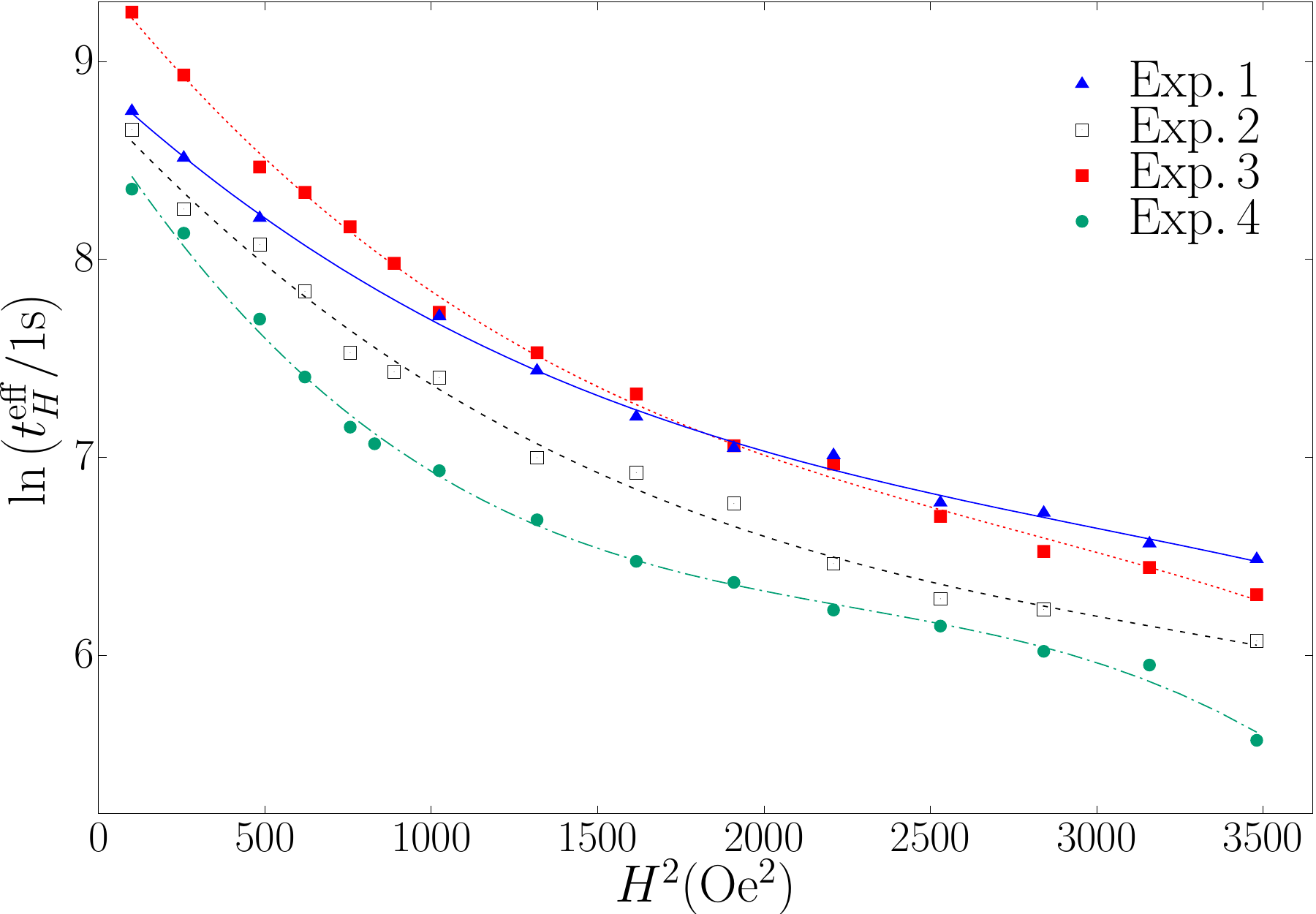}
	\caption{A plot of the peak times $t_H^{\text {eff}}$ for the
single-crystal CuMn 6 at. \% {vs} $H^2$ for the four values of $\Tm$ and $\tw$
listed in Table~\ref{tab:exp_details}.  The slope for small $H^2$ is used to
extract $\xi(\tw)$ in Table~\ref{tab:exp_details}, and the lines come
from the fits to the scaling law introduced in Section~\ref{Sec:scaling_law}.}
        \label{fig:ratio_teff_EXP}
\end{figure}

We extract the effective waiting time $t_H^{\text {eff}}$ in
Eq.~\eqref{eq:Delta_max_definition} from the time at which $S(t,\tw;H)$ is a
maximum, as before.  Our results for all four conditions in
Table~\ref{tab:exp_details} are exhibited as a function of $H^2$ in
Fig.~\ref{fig:ratio_teff_EXP}.
The slope of the data in Fig.~\ref{fig:ratio_teff_EXP} at small values of the
magnetic field $H$ generates the spin-glass correlation length $\xi(\tw;\Tm)$
from Eqs.~\eqref{eq:zeeman_Ez_definition} and \eqref{eq:Ncorr_definition},
see Table~\ref{tab:exp_details}, which also lists the employed
values for the replicon exponent $\theta ({\tilde x})$. 
These results will allow us to
express the non-linear susceptibility in terms of $\xi(\tw)$.

An example of the measured relaxation function $S(t, \tw; H)$ is plotted for
$\Tm= 28.5$~K and $\tw=10\,000$~s in Fig.~\ref{fig:exp_St_behavior} for five
different magnetic fields, while the effective response times, $\ln t_H^{\text
{eff}}$, are plotted in Fig. \ref{fig:ratio_teff_EXP} for all four experiments
listed in Table \ref{tab:exp_details}.  Note the remarkable similarity in shape
of the original experimental results for $\ln t_H^{\text {eff}}$ in
Fig.~\ref{fig:joh99_data} with our results in Fig.~\ref{fig:ratio_teff_EXP}.
Also, note the fits of all four of our results for $\ln \,t_H^{\text {eff}}$ in
Fig.~\ref{fig:ratio_teff_EXP} to the scaling relationship for the non-linear
magnetisation, Eq.~\eqref{eq:ratio_effective_time_scaling}, which will be
described in more detail below.
 
Because the scaling relationship,
Eq.~\eqref{eq:ratio_effective_time_scaling_final}, depends upon the magnitude
of the waiting time in $\xi(t,\tw;\Tm)$, two different values of $\tw$ were used
at the same intermediate temperature $\Tm = 28.75$ K, among the three
temperatures (28.5~K, 28.75~K, and 29.0~K) listed in
Section~\ref{Sec:exp_details} and in Table~\ref{tab:exp_details}, to test 
Eq.~\eqref{eq:ratio_effective_time_scaling_final} at a given
temperature.  This allows us to discriminate between the influence of
temperature and of waiting time on $\xi(t,\tw;\Tm)$.  In this way, we are able to
demonstrate explicitly that $\xi(t,\tw;\Tm)$ is the control parameter.
 
\begin{figure}[t] \centering 
 \centering
 \includegraphics[width=\textwidth]{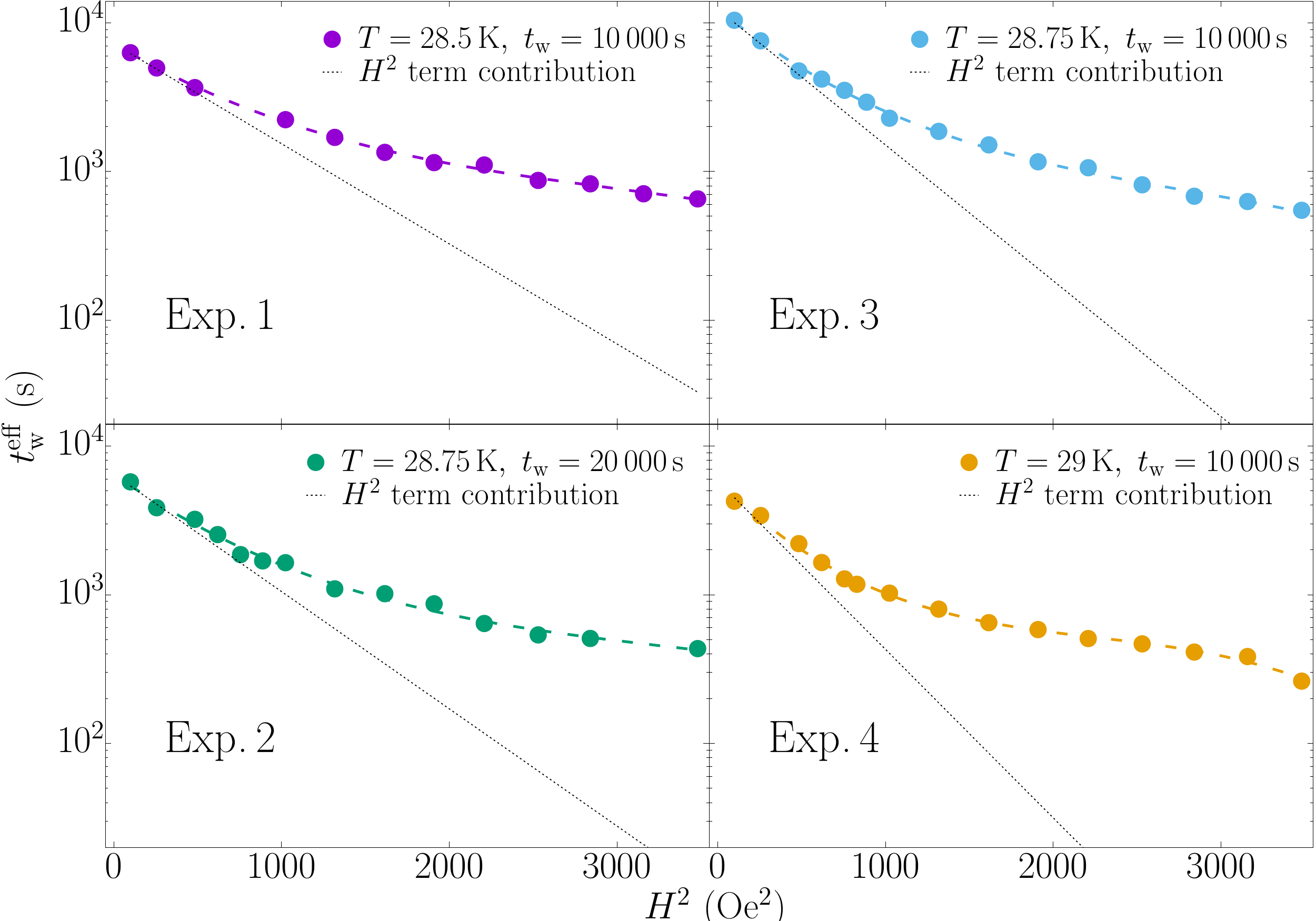}
  \caption{Plots of the peak time $t_H^{\text {eff}}$ for the single-crystal
CuMn 6 at. \% vs $H^2$ for the four experimental regimes of Table~\ref{tab:exp_details}.
The straight lines are extrapolations of the linear
term in the magnetisation, and the dashed lines are fits to Eq.~\eqref{eq:ratio_effective_time_scaling}.}
  \label{fig:teff_vs_H2_EXP}
\end{figure}

It is useful to display $t_H^{\text {eff}}$ against $H^2$ individually for
each of the four values of $\Tm$ and $\tw$.  They are exhibited above in
Fig.~\ref{fig:teff_vs_H2_EXP}.  The data for $\ln \teff_H$ is fitted to the
function
\begin{equation}
\label{eq:fitting_technique_exp}
f(x)= c_0 + c_2 \, x + c_4 \, x^2 + c_6 \,x^3 + \mathcal{O}(x^4)   \; ,
\end{equation}
where $x\equiv H^2$ and the $c_n$ coefficients, according to Eq.~\eqref{eq:ratio_effective_time_scaling}, correspond to:
\begin{eqnarray}
c_0 &=& \ln \left( \teff_{H \to 0^+} \right)\,, \\
c_2 &=& \left[\frac{\hat{S}}{2 \;\Tm}\right] \xi^{D-\theta(\tilde x)/2}\,, \\
c_4 &=&  a_3(\Tm)\ \xi^{2D-3\theta(\tilde x)/2}\,, \label{eq:scaling-c4}\\
c_6 &=&  a_5(\Tm)\ \xi^{3D-2\theta(\tilde x)} \, .
\end{eqnarray}
Notice that we have absorbed the geometrical prefactor $b$ of
Eq.~\eqref{eq:ratio_effective_time_scaling} in the non-linear coefficients
$a_n(\Tm)$ and in the linear coefficient $\hat{S}$, and we neglect the
sub-leading coefficient $a_1(\Tm)/\xi^{\theta(\tilde x)/2}$. 

The effect of increasing temperature with waiting time held constant can be
seen in the difference between the measured $t_H^{\text {eff}}$ and the
extrapolated value of the linear magnetisation term (quadratic in $H^2$) for
the largest magnetic field ($H=59$ Oe) in Exps.~1, 2 and~4 in
Fig.~\ref{fig:teff_vs_H2_EXP}.  Non-linear effects grow for larger $\tw$, hence
larger $\xi(t,\tw;\Tm)$ at the same temperature, which can be seen by comparing
experiments~2 and~3.  The linear and non-linear coefficients of
Eq.~\eqref{eq:ratio_effective_time_scaling} can be extracted from fits of the
data in Fig.~\ref{fig:teff_vs_H2_EXP} (dashed lines), whose resulting
coefficients are listed in Table~\ref{tab:teff_fits_EXP}.
 
\begin{table}
	\caption{Parameters from our fits to
Eq.~\eqref{eq:fitting_technique_exp} of our experimental data for $\ln
\,t_H^{\mathrm{eff}}$, as a function of $\Tm$ and $\tw$ (data from Fig.~\ref{fig:teff_vs_H2_EXP}). The uninteresting fit
parameter $c_0$ is not included in the table.}
        \label{tab:teff_fits_EXP}
\begin{indented}
\lineup
\item[]\begin{tabular}{@{}c  c  c  l}
                \br
                $\Tm$ (K) & $\tw$(s) & Coefficient & \multicolumn{1}{c}{Numerical value}\\
\mr
                \multirow{3}{2em} {28.5}& \multirow {3}{2em}{10\,000}& $c_2$ & $ -1.55(10) \times 10^{-3}$\\
                & &$c_4$ & $\m4.0(7)\0\0 \times 10^{-7}$\\
                & &$c_6$ & $-4.4(13)\0 \times 10^{-11}$\\
                \mr
                \multirow{3}{2em} {28.75}& \multirow{3}{2em}{10\,000}& $c_2$ &  $-1.82(20) \times 10^{-3}$\\
                & &$c_4$ &$\m 4.6(13)\0 \times 10^{-7}$ \\
                & &$c_6$ & $-4.6(25)\0\times 10^{-11}$\\
                \mr
                \multirow{3}{2em} {28.75}& \multirow{3}{2em}{20\,000}& $c_2$ &  $-2.10(12) \times 10^{-3}$\\
                & &$c_4$ & $\m5.9(8)\0\0\times 10^{-7}$\\
                & &$c_6$ & $-7.0(15)\0\times 10^{-11}$\\
                \mr
                \multirow{3}{2em} {29}& \multirow{3}{2em}{10\,000}& $c_2$ &  $-2.61(13) \times 10^{-3}$\\
                & &$c_4$ & $\m1.02(8)\0 \times 10^{-6}$\\
                & &$c_6$ & $-1.49(16)\times 10^{-10}$\\
                \br
        \end{tabular} 
\end{indented}
\end{table}

To test the scaling relationship of
Eq.~\eqref{eq:ratio_effective_time_scaling_final} we first consider the fits of
the data at $\Tm=28.75$~K for the two waiting times, $\tw=2\times 10^4$~s and
$\tw = 10^4$~s.  The linear term is proportional to $\xi^{D-\theta(\tilde x)/2}$.
The ratio of the two correlation lengths from
Table~\ref{tab:teff_fits_EXP} is, hence,
\begin{equation}\label{eq:ratio_xi_aEXP}
{\frac {\xi(\tw=20\,000\ \text{s})}{\xi(\tw=10\,000\ \text{s})}}=\bigg[{\frac {c_2(\tw=20\,000\ \text{s})}{c_2(\tw=10\,000\ \text{s})}}\bigg]^{1/[D-\theta(\tilde x)/2]}
\approx1.0535\,.
\end{equation}
Should scaling hold according to
Eq.~\eqref{eq:ratio_effective_time_scaling_final}, then consistency requires
that the ratios of the correlation lengths from the non-linear terms be
the same as that for the linear term.  They are:
\begin{equation}\label{eq:ratio_xi_bEXP}
{\frac {\xi(\tw=20\,000\ \text{s})}{\xi(\tw=10\,000\ \text{s})}}=\bigg[{\frac {c_4(\tw=20\,000\ \text{s})}{c_4(\tw=10\,000\ \text{s})}}\bigg]^{1/[2D-3\theta(\tilde x)/2]}
\approx1.0476\,,
\end{equation}
and
\begin{equation}\label{eq:ratio_xi_cEXP}
{\frac {\xi(\tw=20\,000\ \text{s})}{\xi(\tw=10\,000\ \text{s})}}=\bigg[{\frac {c_6(\tw=20\,000\ \text{s})}{c_6(\tw=10\,000\ \text{s})}}\bigg]^{1/[3D-2 \theta(\tilde x)]}
\approx1.0526\,.
\end{equation}
The equality (within experimental error) of
Eqs.~\eqref{eq:ratio_xi_aEXP}--\eqref{eq:ratio_xi_cEXP} is an impressive
experimental verification of scaling
relationship~\eqref{eq:ratio_effective_time_scaling_final}.  Another check is
the growth of the correlation length itself.  At temperature $\Tm=28.75$~K and
for the two waiting times, it is possible to calculate the ratio of the two
values of the correlation length directly, using the expression for power-law
growth~\cite{joh:99,kisker:96},
\begin{equation}\label{eq:xi_evaluation_EXP}
\xi(\tw;\Tm)=a_0\,\hat{C}_1\,\bigg({\frac
{\tw}{\tau_0}}\bigg)^{\hat{C}_2(\Tm/\Tg)}\equiv a_0\,\hat{C}_1\,\bigg({\frac
{\tw}{\tau_0}}\bigg)^{\Tm/(z_\text{c}\Tg)}\,,
 \end{equation}
 where $\hat{C}_1$ and
$\hat{C}_2$ are constants, by definition $\hat{C}_2\equiv 1/z_\text{c}$, and $\tau_0$
is a characteristic \textit{exchange time}, here taken as
$\hbar/k_\text{B}\Tg$.
 
Using the growth-rate parameter $z_\text{c}=12.37(107)$~\cite{janus:18,zhai:19} one finds
\begin{equation}
\label{eq:ratio_xinew}
{\frac {\xi(\tw=20\,000\ \text{s})}{\xi(\tw=10\,000\ \text{s})}}\approx \bigg({\frac {2\times 10^4}{10^4}}\bigg)^{\Tm/(12.37\,\Tg)}
= 2^{28.75/(12.37\times 31.5)}\approx1.0525\,.
\end{equation}
Comparing the ratio of $\xi(\tw;\Tm)$ for the two different waiting times,
Eq.~\eqref{eq:ratio_xinew}, from the growth law,
Eq.~\eqref{eq:xi_evaluation_EXP}, with the ratios from fitting to the scaling
relationship, Eqs.~\eqref{eq:ratio_xi_aEXP}--\eqref{eq:ratio_xi_cEXP}, is
remarkable evidence for the consistency of our physical picture.  It
explicitly demonstrates the power of using the spin-glass correlation length as
the primary factor for evaluating the spin-glass non-linear magnetisation in
the vicinity of the transition temperature $\Tg$.
 
The lingering issue from Eqs.~\eqref{eq:Ncorr_new_def}
and~\eqref{eq:ratio_effective_time_scaling_final}, according
to Zhai \emph{et al.}~\cite{zhai:19} ``is that the replicon
exponent [$\theta(\tilde x)$] (\ldots) depends upon both the temperature and $\xi$ through the
crossover variable [$\tilde x$]'', with
\begin{equation}\label{eq:x_equiv_lJxi_def}
\tilde x={\frac {\ell_\text{J}}{\xi(\tw;\Tm)}}\,.
\end{equation}

From the notation of Table~\ref{tab:teff_fits_EXP} and
Eq.~\eqref{eq:Ncorr_new_def} the number of correlated spins is
\begin{equation}\label{eq:Ncorr_vs_xi}
N_\text{corr}={\frac {k_\text{B}\Tm \,c_2}{\chi_{\text {FC}}}}=\xi^{D-\theta(\tilde x)/2}=\xi^{D- [\theta(\ell_\text{J}/\xi)/2]}.
\end{equation}
The left-hand side is a number, the right-hand side is an implicit function of
$\xi$ and $\theta$.  Using the definition of the Josephson length $\ell_\text{J}$
and  of the replicon $\theta(\tilde x)$,\footnote{The reader will find all the necessary
details in Appendix~\ref{Appendix:josephson_lJ}.} one can solve for $\xi$ and
$\theta$ at each of the four values of $\Tm$ and $\tw$ explored experimentally.
The results are displayed in Table~\ref{tab:exp_details}.
 
The aging rate $z_\text{c}$ varies as a function of $\xi$.  Using the data at
$\Tm=28.75$~K, the approximate aging-rate factor is $z_\text{c}=12.37(107)$ at $\xi\sim
200$ lattice spacings $a_0$~\cite{zhai:19}.  Although the correlation length extracted at
28.75~K is larger than that at 28.5~K, the higher temperature sets the
crossover variable $x=\ell_\text{J}(\Tm)/\xi=0.11$ for $\tw=20\,000$~s and $x=0.12$
for $\tw=10\,000$~s, in the range of the crossover variable obtained by us
previously at $\Tm=28.5$~K~\cite{zhai:20}.  Both measurements have exhibited
slowing down of the spin-glass correlation growth rate near the critical
temperature at large correlation lengths.
 
\begin{figure}[t] \centering
\centering
  \includegraphics[width=.65\textwidth]{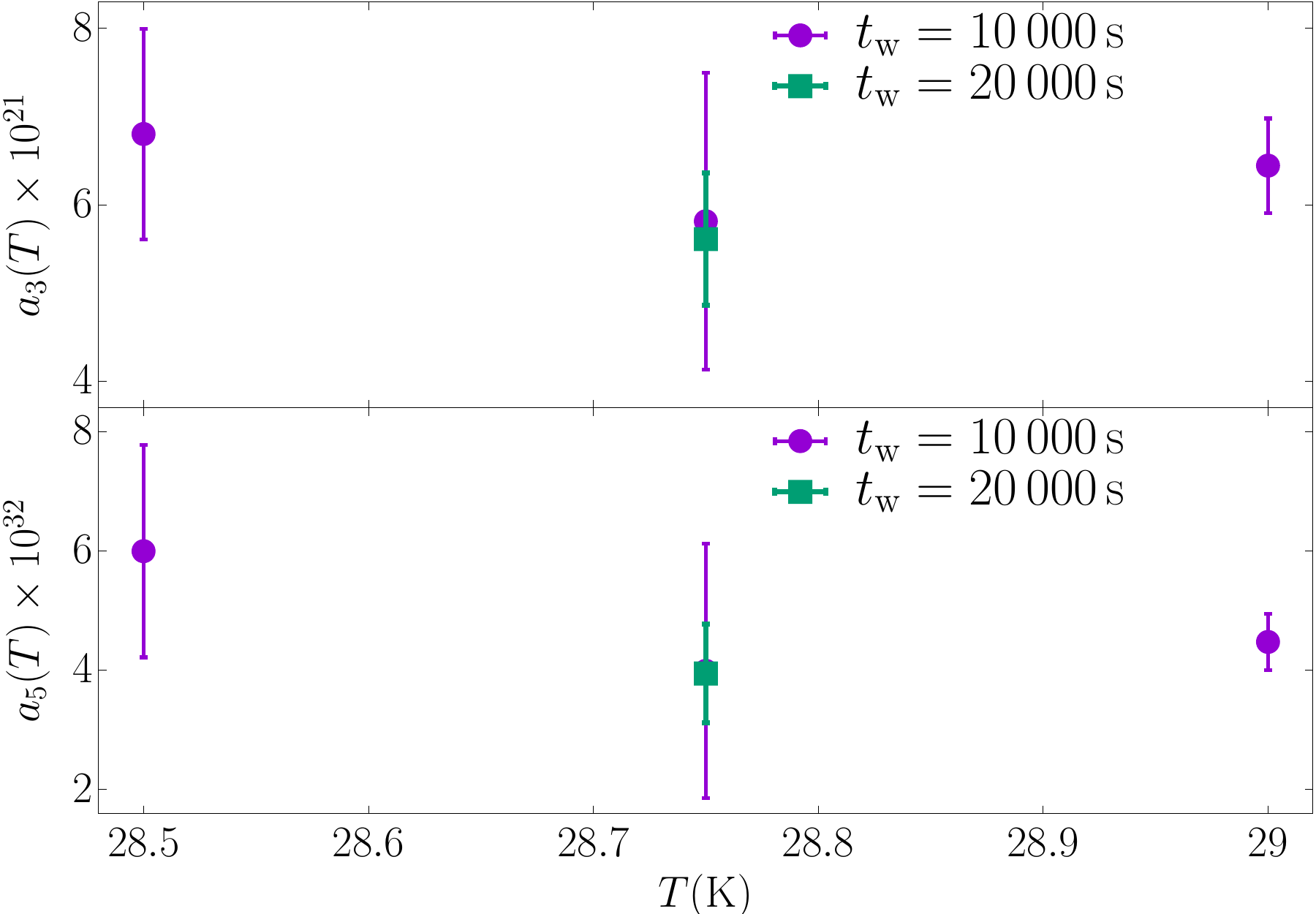}
    \caption{Non-linear coefficients $a_3$ and $a_5$, as defined in Eq.~\eqref{eq:ratio_effective_time_scaling},
 calculated using the extracted values of $\xi$ and $\theta$ for the different measuring temperatures $\Tm$ and waiting times $\tw$ in our experiments.} 
    \label{fig:non-linear_coefficients_a3_a5_EXP}
\end{figure}

Using the average value of $\theta$ from Table~\ref{tab:exp_details}, $\theta =
0.343$, and setting  $z_\text{c}=12.37$, the values exhibited in
Eqs.~\eqref{eq:ratio_xi_aEXP}--\eqref{eq:ratio_xinew} are altered to 1.053,
1.048, 1.052, and 1.051, respectively.  Using the values from
Table~\ref{tab:exp_details}, the temperature-dependent coefficients $a_3(\Tm)$
and $a_5(\Tm)$ of Eq.~\eqref{eq:ratio_effective_time_scaling} can be calculated
for each of the four values of $\Tm$ and $\tw$.  They are displayed in
Fig.~\ref{fig:non-linear_coefficients_a3_a5_EXP}.

From Fig.~\ref{fig:non-linear_coefficients_a3_a5_EXP}, one sees that the
``hope'' expressed after Eq.~\eqref{eq_Ez_zeeman_nonlinar_an_expansion},
\emph{i.e.}, that the temperature dependence of the coefficients $a_n(\Tm)$
appearing in Eq.~\eqref{eq_Ez_zeeman_nonlinar_an_expansion} be weak, is
realised in this set of experiments. For $a_3(\Tm)$, within the experimental
error bars, there is little or no change with temperature. The situation for
$a_5 (\Tm)$ is not as nice, but there appears to be little change with
temperature at the two highest temperatures.
 
It is interesting to test the scaling relationship~\cite{zhai-janus:20a}
\begin{equation}\label{eq:chi2n_1vs_non-linear_an}
\chi_{2n-1} (\tw;\Tm) \propto a_{2n-1} \left[ \xi(\tw;\Tm)\right]^{(n-1)D-n\theta(\tilde{x})/2} \; .
\end{equation}
Thus,
\begin{equation}
\chi_3\propto \xi^{D-\theta(\tilde x)}\,a_3,\qquad 
\chi_5\propto \xi^{2D-{\frac {3\theta(\tilde x)}{2}}}\,a_5.
\label{eq:chi_vs_xi_dependence}
\end{equation}
The measured non-linear susceptibilities are exhibited below for the three temperatures $28.5$~K, 
$28.75$~K and $29.0$~K.

\begin{figure}[t] \centering
\centering
\includegraphics[width=.65\textwidth]{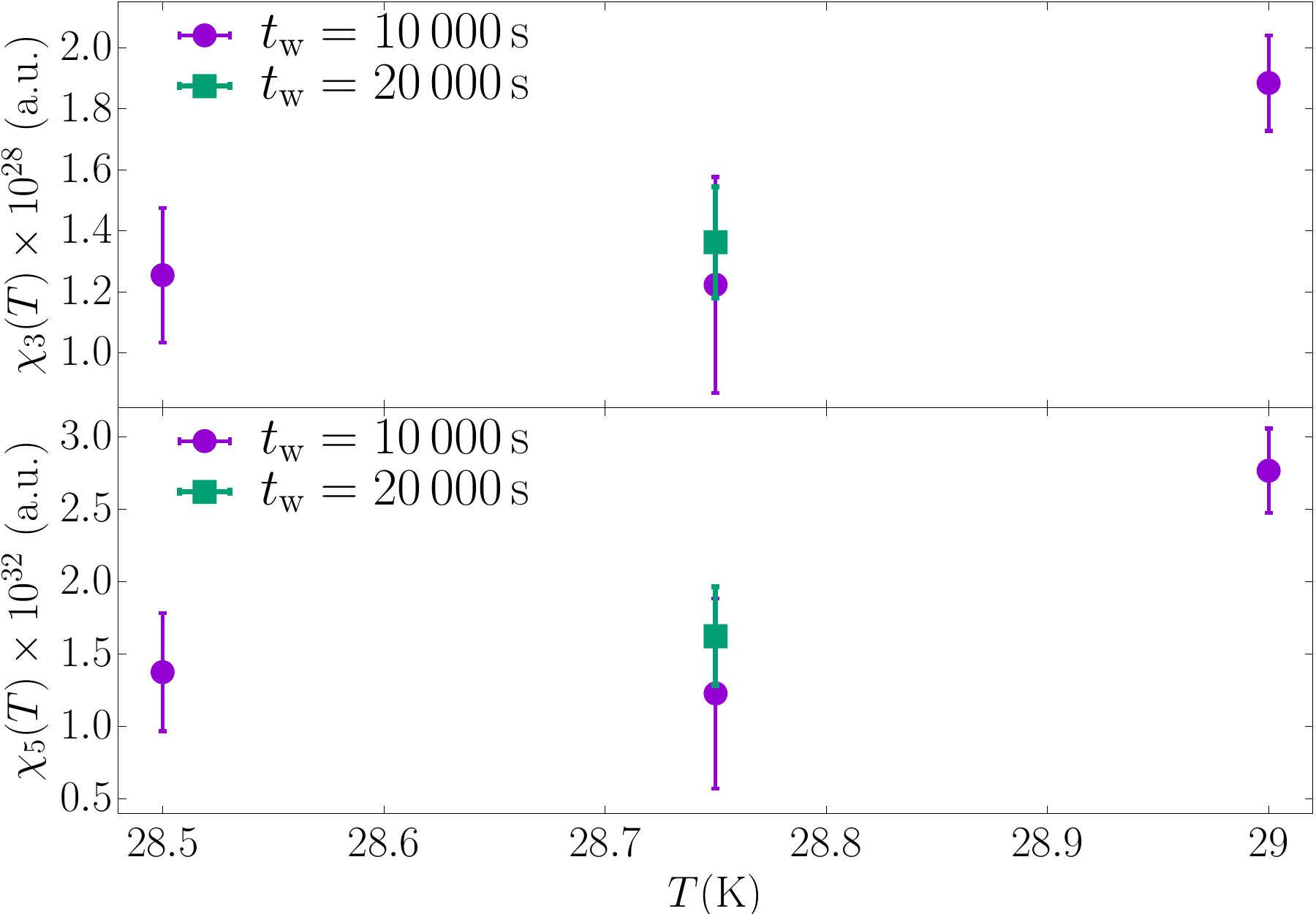}
\caption{Non-linear susceptibilities $\chi_3(\tw;\Tm)$  and $\chi_5(\tw;\Tm)$ from Eq. \eqref{eq:Ncorr_vs_xi}, plotted as a function of temperature 
for the four experimental regimes of Table~\ref{tab:exp_details}.}
    \label{fig:non-linear_chi3_chi5_EXP}
\end{figure}

One can test the scaling relationships~\eqref{eq:chi2n_1vs_non-linear_an}
and~\eqref{eq:chi_vs_xi_dependence} by using the measured values for the
spin-glass correlation length $\xi$, the replicon exponent $\theta(\tilde x)$ from
Table~\ref{tab:exp_details}, and the values of $c_2$ and $c_4$ from
temperatures $28.5$~K and $29$~K and $\tw=10\,000$~s.  For $\Tm=28.5$~K, we have
$\xi= 320.36 \,a_0$, $\theta(\tilde x) = 0.337$ (from Table~\ref{tab:exp_details}), and
$c_4= 4.0\times 10^{-7}$ (from Table~\ref{tab:teff_fits_EXP}, note that we
have ignored the error bars). Similarly, for $\Tm=29.0$ K and $\tw=10\,000$~s we have
$\xi= 391.27 \,a_0$, $\theta(\tilde x) = 0.349$, and $c_4= 10.2 \times
10^{-7}$. Using Eqs.~\eqref{eq:fitting_technique_exp} and~\eqref{eq:scaling-c4}
and the just quoted values of $c_4(\tw;\Tm)$ one finds
\begin{equation}
\frac{\chi_3(\tw=10\,000~\text{s}; \Tm=28.5~\text{K})}{\chi_3(\tw=10\,000~\text{s}; \Tm=29.0~\text{K})} \approx 0.666\,.
\end{equation}
This ratio is well within the error bars of the measured non-linear
susceptibilities in Fig.~\ref{fig:non-linear_chi3_chi5_EXP}. A similar result
is also found for $\chi_5$.

With these scaling observations in hand, it is interesting to wonder about using
them to estimate the condensation temperature $\Tg$.  In principle,
determination of $\Tg$ would require an infinite $\tw$, because $\xi(\Tm)\to\infty$
when $\Tm\to\Tg$.  One expects that any experiment at finite $\tw$ would yield a
maximum for the non-linear susceptibility at a temperature we shall call
$\Tg(\tw)$ because $\tw$ is finite.
 
In principle, then, by measuring $\Tg(\tw)$ for ever larger $\tw$, one could
extrapolate to the true $\tw\rightarrow \infty$ condensation temperature
$\Tg$.  If nothing else, measurements at large values of $\tw$ on laboratory
time scales could establish a lower bound for $\Tg$.
 
The non-linear susceptibility $\chi_3$ diverges as
\begin{equation}
\chi_3(\tw \rightarrow \infty; \Tm)=\chi_0\,{\frac {\Tg(\tw\rightarrow \infty)}{|\Tg(\tw\rightarrow \infty)-\Tm|^\gamma}}\,,
\end{equation}
where $\chi_0$ is a constant independent of temperature, and $\gamma=6.13(11)$~\cite{janus:13}.
For {\it finite} $\tw$, $\chi_3(\tw;\Tm)$ only has a maximum as a function of
temperature.  A way of arriving at this maximum would be to fit the data to the
function
\begin{equation}
\label{eq:chi3_finite_time}
\chi_3(\tw;\Tm)=\chi_0\,{\frac {\Tg(\tw)}{|\Tg(\tw)-\Tm|^\gamma}}\,,
\end{equation}
and then use the data points from just two or three temperatures to extract
$\Tg(\tw)$.  For larger and larger $\tw$, one could in principle extrapolate to
the true $\Tg$.  We emphasise that, though Eq.~\eqref{eq:chi3_finite_time}
suggests $\chi_3(\tw;\Tm)$ diverges at $\Tm=\Tg(\tw)$, it does not, arriving only
at a maximum value for finite $\tw$.  Nevertheless,
Eq.~\eqref{eq:chi3_finite_time} is a way of estimating $\Tg(\tw)$ for use in an
extrapolation procedure.
 
To test whether this trick has any validity, consider the data exhibited in
Fig.~\ref{fig:non-linear_chi3_chi5_EXP}.  Here, $\tw=10~000$~s and, taking $\chi_3(\tw;\Tm)$ at
the centre of the error bars for the two temperatures $28.5$~K and $29$~K, one
finds $\Tg(\tw=10\,000~\text{s})=32$~K.  This value is too high, as magnetisation
measurements suggest $\Tg(\tw\rightarrow \infty)=31.5$~K.  More accurate
determination of the parameters in Table~\ref{tab:teff_fits_EXP} would diminish
the error in $\Tg(\tw)$, but it does suggest a feasible process for taking
laboratory data for finite $\tw$ and extrapolating to find $\Tg(\tw\rightarrow
\infty)$.

\subsection{Reanalysis of previous data}
\label{Sec:reanalysis_exp_data}
Given the above analysis of our recent data, it is convenient to revisit the
work of Joh \textit{et al.}~\cite{joh:99} and of  Bert~\textit{et
al.}~\cite{bert:04}, to examine whether the Zeeman energy is proportional to
$H^2$ or to $H$ (alternatively, to the number of correlated spins or to the
square root of the total number of spins, respectively).  We have already
alluded to the results of these works as displaying the effect of magnetisation
non-linearity.  We now explore this assertion in detail using the analysis of
Subsection~\ref{Sec:non-linear_scaling_EXP}.
 
\begin{figure}[t]
        \centering
        \includegraphics[width = \columnwidth]{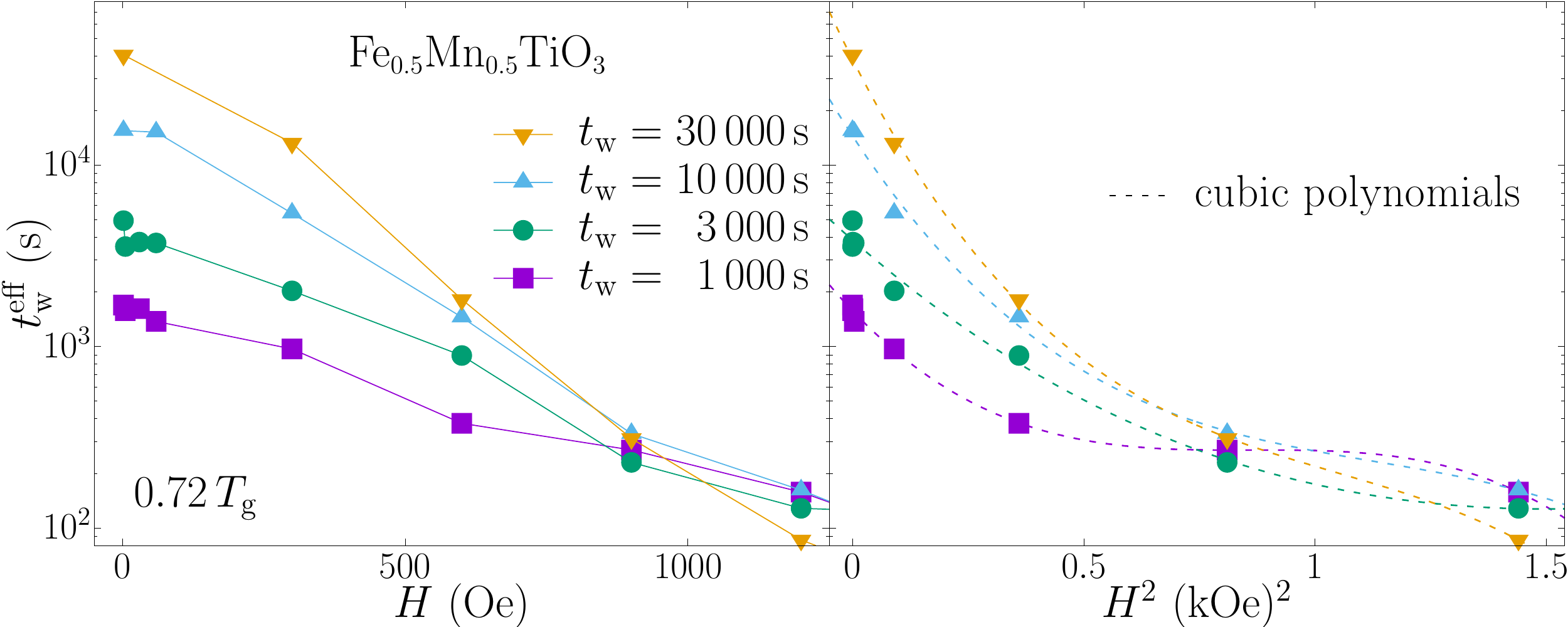}
	\caption{\emph{Left:} Effective waiting times (in log scale) derived from field-change
 experiments on an Ising sample (Fe$_{0.5}$Mn$_{0.5}$TiO$_3$) as a function
of the magnetic field $H$. The plot reproduces Fig.~3 of  Bert \emph{et al.}~\cite{bert:04} (solid lines are linear interpolations to data with same $\tw$).
\emph{Right:} Same data plotted against $H^2$. The dashed lines are fits to Eq.~\eqref{eq:fitting_technique_exp}, with fit parameters
listed in Table~\ref{tab:fit_bert04_data}.  } \label{fig:bert04}
\end{figure}

Fig.~1 of Joh~\textit{et al.} and Fig.~3 of Bert \textit{et al.} are reproduced
in Fig.~\ref{fig:joh99_data} and Fig.~\ref{fig:bert04} in this paper. Both
exhibit significant deviations from an $H^2$ dependence of the $\ln \,
t_H^{\text {eff}}$ with increasing values of $H$.  Bert {\it et
al.}~\cite{bert:04} go on to assert a linear dependence, as exhibited in their
Fig.~3, reproduced here in Fig.~\ref{fig:bert04}.  The magnetic fields
in~\cite{bert:04} are quite large, and the scale of their plot does not cover
the dependence on $H^2$ for small $H$.  Nevertheless, they claim their data
fits a linear dependence of $\ln t_H^{\text {eff}}$ on $H$.  A glance at
the left panel of Fig.~\ref{fig:bert04} suggests how they could rationalise
their conclusion.
 
Yet, as noted by the authors of the experiments in Fig.~\ref{fig:joh99_data}~\cite{joh:99}, a
linear dependence on $H$ is a poor fit to the data at small $H$.  Further,
the argument for the magnetisation's growth with $\sqrt {N_\text{corr}}$ is
supposedly valid at small $H$ \cite{bouchaud:01b}, while the dependence on the
number of correlated spins is argued to be proportional to $\sqrt{N_\text{corr}}$,
rather than linear in $N_\text{corr}$, as from Eq. \eqref{eq:zeeman_Ez_definition}. 
On the other hand, the
data exhibited in Fig.~\ref{fig:bert04} uses magnetic fields
that are substantially larger than those considered in our experiments.
 
We assert that the departure from linearity in $H^2$ as $H$ increases observed
in~\cite{bert:04} is simply the effect of non-linearity.  To prove this, we
apply the scaling relation, Eq.~\eqref{eq:ratio_effective_time_scaling_final},
to their data, doing our best to extract their measured values from their figure.  Our fit to Eq.~\eqref{eq:fitting_technique_exp} is shown in 
the right panel of Fig.~\ref{fig:bert04}
and the resulting $c_n$ are listed in Table \ref{tab:fit_bert04_data}.

\begin{table}[t]
\caption{We report, as a function of $\tw$, the parameters from fits to Eq.~\eqref{eq:fitting_technique_exp} of the data
obtained by Bert \textit{et al.} \cite{bert:04} for
$\ln t_\mathrm{H}^\mathrm{eff}$.  Their data correspond to
$\mathrm{Fe}_{0.5}\mathrm{Mn}_{0.5}\mathrm{TiO}_3$ at $\Tm=0.72~\Tg$ (see Fig.~3
of Ref.~\cite{bert:04}). The fits are shown in our Fig.~\ref{fig:bert04}.
The uninteresting fit parameter, $c_0$, is not included in the table.}
\label{tab:fit_bert04_data}
 \lineup
\begin{indented}\item[]\begin{tabular}{@{}ccc}
\br
$\tw$ (s) & coefficient &  value\\
\mr
\multirow{3}{3em}{1\,000} & $c_2$& $-6.184 \times 10^{-6\0}$\\
& $c_4$& $\m7.193 \times 10^{-12}$\\
&$c_6$& $-2.784 \times 10^{-18}$\\
\mr
\multirow{3}{3em}{3\,000} & $c_2$& $-5.154 \times 10^{-6\0}$\\
& $c_4$& $\m2.322\times10^{-12}$\\
&$c_6$& $-0.269 \times 10^{-18}$\\
\mr
\multirow{3}{3em}{10\,000} & $c_2$& $-8.977\times10^{-6\0}$\\
& $c_4$&$\m7.052 \times 10^{-12}$\\
&$c_6$& $-2.075 \times 10^{-18}$\\
\mr
\multirow{3}{3em}{30\,000} & $c_2$&$-11.600\times10^{-6\0}$\\
& $c_4$&$\m9.440 \times 10^{-12}$\\
&$c_6$&$-3.004 \times 10^{-18}$\\
\br
\end{tabular}
\end{indented}
\end{table}
Although only 1-2 digits are significant in Table~\ref{tab:fit_bert04_data}, we 
write more digits, for the sake of reproducibility.
The fitting quality for $\tw = 10\,000$ s and $\tw =30\,000$ s is better than for the other two waiting times.
Note that the
coefficients $c_n$ listed in Table~\ref{tab:fit_bert04_data} are considerably
smaller than in our Table~\ref{tab:teff_fits_EXP} for our current experiments
on a CuMn 6 at. \% single crystal.  We believe this is because our measurements
are for $\Tm\approx 0.9 \Tg$ whereas Bert {\it et al.} \cite{bert:04} 
worked at $0.72\Tg$, where non-linear terms are
expected to be much smaller.

Using the fitting coefficients from Table \ref{tab:fit_bert04_data} and $\theta(\tilde x) = 0.3$, we obtain
\begin{equation}\label{eq:c2_over_c4_bert}
\frac{\xi(\tw = 30\,000~\text{s})}{\xi(\tw = 10\,000~\text{s})} = \left[\frac{c_2 (30\,000~\text{s})}{c_2 (10\,000~\text{s})}\right]^{1/[D-\theta(\tilde x)/2]}\approx 1.094,
\end{equation} 
\begin{equation}\label{eq:c4_over_c6_bert}
\frac{\xi(\tw = 30\,000~\text{s})}{\xi(\tw = 10\,000~\text{s})} = \left[\frac{c_4 (30\,000~\text{s})}{c_4 (10\,000~\text{s})}\right]^{1/[2D-3\theta(\tilde x)/2]}\approx 1.054,
\end{equation} 
\begin{equation}\label{eq:c6_over_c8_bert}
\frac{\xi(\tw = 30\,000~\text{s})}{\xi(\tw = 10\,000~\text{s})} = \left[\frac{c_6 (30\,000~\text{s})}{c_6 (10\,000~\text{s})}\right]^{1/[3D-2\theta(\tilde x)]}\approx 1.045.
\end{equation}

The three ratios, Eq.~\eqref{eq:c2_over_c4_bert}--\eqref{eq:c6_over_c8_bert} do
not agree with one another perfectly, but again, considering the rawness of the analysis, they are certainly suggestive.  In
summary, we believe that the assessment of Ref. \cite{bert:04} that their data
is evidence for $E_\text{Z}\propto H$ is in error.  Rather, we believe the departure
they observe from linearity in $H^2$ arises from non-linear terms in the
magnetisation as a result of the large magnetic fields utilised in their study.

\subsection{Numerical study of the ratio of the effective times at $H$ and $H=0^+$}
\label{Sec:ratio_time_NUM}

\begin{figure}[t]
\centering
\includegraphics[width = 0.65\columnwidth]{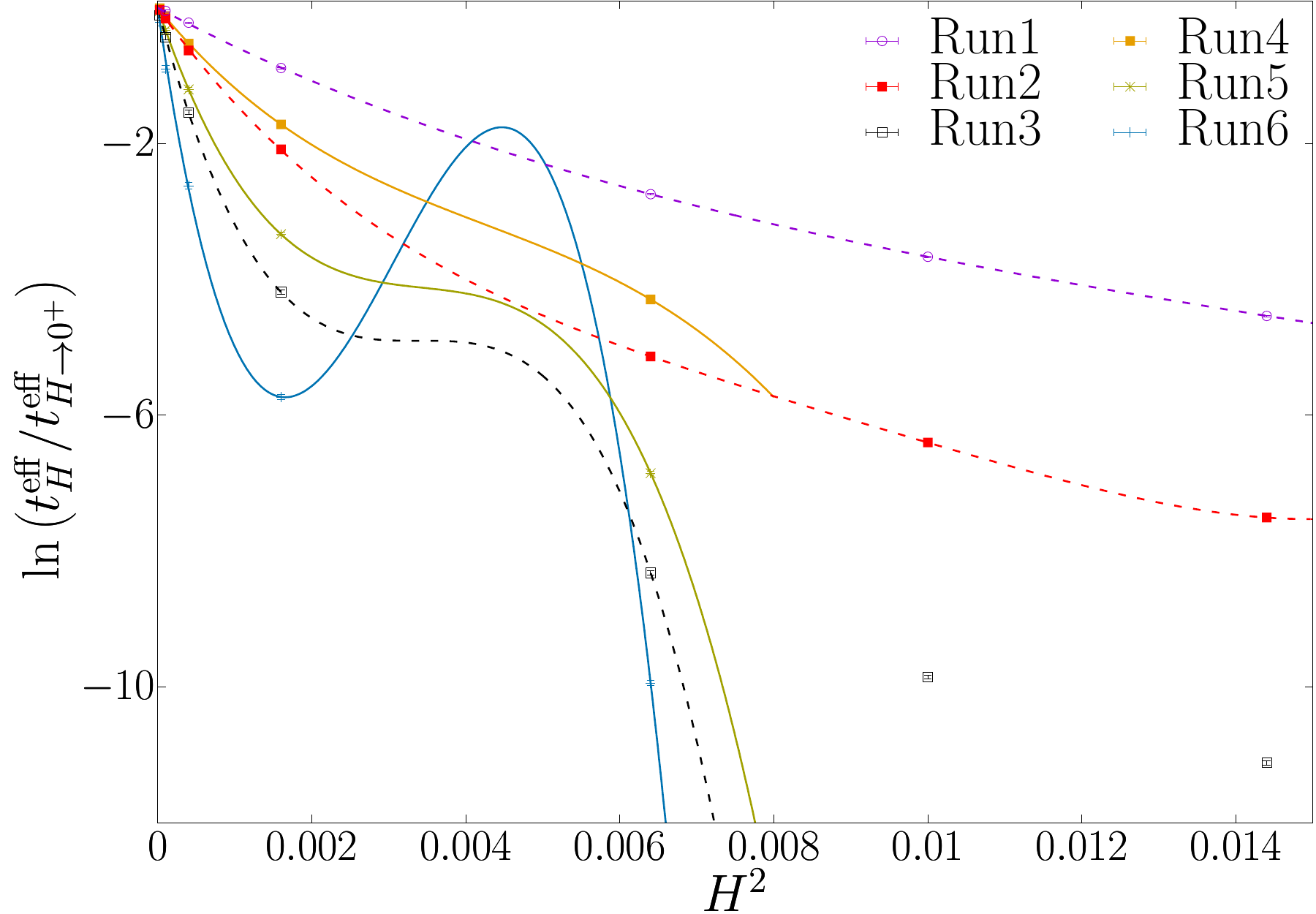}
\caption{The numerical time ratio $\ln (t^{\text{eff}}_H /
t^{\text{eff}}_{H\to 0^+})$ for the six runs of Table~\ref{tab:details_NUM}.
The data were fitted as a polynomial of $H^2$ as reported in
Table~\ref{tab:ratio_times_fits_NUM}.  Continuous lines are fits for data at
$T=1.0$; dashed lines correspond to the data at $T=0.9$.}
\label{fig:ratio_teff_janus}
\end{figure}

We have exploited our proposed
relation~\eqref{eq:C_teff_Cpeak} to extract
effective times $t^\text{eff}_H$, as explained in
Appendix~\ref{Appendix:details_St_construction}.  Our results are displayed in
Fig.~\ref{fig:ratio_teff_janus}. In the subsequent analysis in
Section~\ref{Sec:non-linear_scaling}, we shall need the  derivative of $\ln \left(
t^{\text{eff}}_H / t^{\text{eff}}_{H \to 0+} \right)$ with respect to $H^2$,
evaluated numerically at $H^2=0$. Our main scope here will be evaluating this
derivative.

An obvious strategy would be to fit the numerical data for $\ln \left(
t^{\text{eff}}_H / t^{\text{eff}}_{H \to 0+} \right)$ as we did for the
experimental data in Eq.~\eqref{eq:fitting_technique_exp}. Note that our sought
derivative at $H^2=0$ is just the $c_2$ coefficient in the fit. A welcome
simplification in the analysis of the numerical data is that we can explicitly
put $c_0=0$ in the fit to Eq.~\eqref{eq:fitting_technique_exp} [indeed, we are
able to carry out the fit for  $\ln \left( t^{\text{eff}}_H / t^{\text{eff}}_{H
\to 0+} \right)$ thanks to Eq.~\eqref{eq:C_teff_Cpeak}]. Our fitting parameters
are reported in Table~\ref{tab:ratio_times_fits_NUM}. Unfortunately, as the
reader will note from  Fig.~\ref{fig:ratio_teff_janus}, these fits predict
unphysically wild oscillations that are not observed in the numerical data. A
plausible explanation for these oscillations relies on the very large magnetic
fields used (recall that $H=1$ for the IEA model roughly corresponds to
$5\times 10^4$ Oe in physical units). These huge magnetic fields probably
exceed the radius of convergence of the Taylor expansion of
Eq.~\eqref{eq:ratio_effective_time_scaling_final}. At any rate, the
oscillations cast some doubts on the determination of the derivative at
$H^2=0$. This is why we have turned to a different strategy in order to
validate our computation.
\begin{table}[t]
\caption{Results of the fits to Eq.~\eqref{eq:fitting_technique_exp} of the
numerical data for the time ratio $\ln (t^{\text{eff}}_H / t^{\text{eff}}_{H\to
0^+})$. Note that, in order to stabilise the fits, we needed to include an
extra terms in Eq.~\eqref{eq:fitting_technique_exp} for two cases.  In the
table, $\xi$ stands for $\xi(t=0,\tw; H=0)$ and the fitting range is $0\leq
H^2\leq H^2_\text{max}$.
}
\label{tab:ratio_times_fits_NUM}
\small
\lineup
 \begin{tabular}{@{}c l l l l l l@{\ \ }l} 
 \br
$T$ & \multicolumn{1}{c}{\tw} & \multicolumn{1}{c}{$\xi$} &
\multicolumn{1}{c}{$c_2(t_\mathrm{w};T)$} &
\multicolumn{1}{c}{$c_4(t_\mathrm{w};T)$}  &
\multicolumn{1}{c}{$c_6(t_\mathrm{w};T)$}  & \multicolumn{1}{c}{$c_8(t_\mathrm{w};T)$} & $H^2_\text{max}$\\
\mr
0.9 & $2^{22}$     & \08.294(7)  & $-6.01(6)\0\0\!\times\!10^2   $  & $3.45(13)  \!\times\! 10^4$  & $-1.31(9)\0 \!\times\! 10^6$   & $1.998(19)\!\times\! 10^7$ & $0.025$\\
0.9 & $2^{26.5}$   & 11.72(2)  & $-1.589(19)\!\times\!10^3 $  & $1.98(6)\0 \!\times\! 10^5$  & $-1.41(6)\0\!\times\! 10^7$  & $3.86(18)\0\!\times\! 10^8$ & $0.015$\\ 
0.9 & $2^{31.25}$  & 16.63(5)  & $-4.380(15)\!\times\!10^3 $  & $1.34(8)\0 \!\times\!10^6$  & $-1.33(10)\!\times\! 10^8$  & 0 & $0.010$\\
1.0 & $2^{23.75}$  & 11.79(2)  & $-1.39(6)\0\0\!\times\!10^3 $  & $2.2(4)\0\0  \!\times\!  10^5$  & $-1.8(5)\0\0\!\times\! 10^7$     & 0 & $0.008 $\\
1.0 & $2^{27.625}$ & 16.56(5)  & $-3.36(9)\0\0\!\times\!10^3 $  & $9.4(6)\0\0 \!\times\! 10^5$  & $-9.1(8)\0\0\!\times\! 10^7$     & 0 &$0.008 $ \\
1.0 & $2^{31.75}$  & 23.63(14) & $-7.91(15)\0\!\times\!10^3$  & $3.29(11)\!\times\! 10^6$  & $-3.56(13)\!\times\! 10^8$     & 0 & $0.008 $\\
\br
\end{tabular}
\end{table}
Our starting point, recall \cite{janus:18} and
Eq.~\eqref{eq:ratio_effective_time_scaling}, is the expected scaling behaviour
for the coefficient $c_2( t_\mathrm{w};T)$ listed in
Table~\ref{tab:ratio_times_fits_NUM}.
The non-linear coefficient $c_2(t_\mathrm{w};T)$, reported in
Table~\ref{tab:ratio_times_fits_NUM}, behaves as  \cite{zhai-janus:20a}
\begin{equation}
\label{eq:a1_paramet_dependence_NUM}
c_2(t_\mathrm{w};T)=  \xi^{D-\theta(\tilde{x})/2}  \left( \frac{\hat{S}}{2 T}  + \frac{a_1(T)}{ \xi^{\theta(\tilde{x})/2}} \right)\;,
\end{equation}
using the scaling of the susceptibility $\chi_1$ from
Eq.~\eqref{eq:chi1_corrections_non-linear}. Here, $\hat{S}$ is the function
appearing in the fluctuation-dissipation relation \cite{janus:17} and $a_1(T)$
is a smooth function of temperature, and we have absorbed the geometrical
prefactor $b$ of Eq.~\eqref{eq:ratio_effective_time_scaling} in $a_1(T)$ and
$\hat{S}(T)$. Notice that the $a_1(T) \xi^{-\theta(\tilde x)/2}$ term is
sub-leading compared to $\hat{S}/(2T)$ and it was neglected in the previous
analysis.

\begin{figure}[t] \centering
\includegraphics[width=\linewidth]{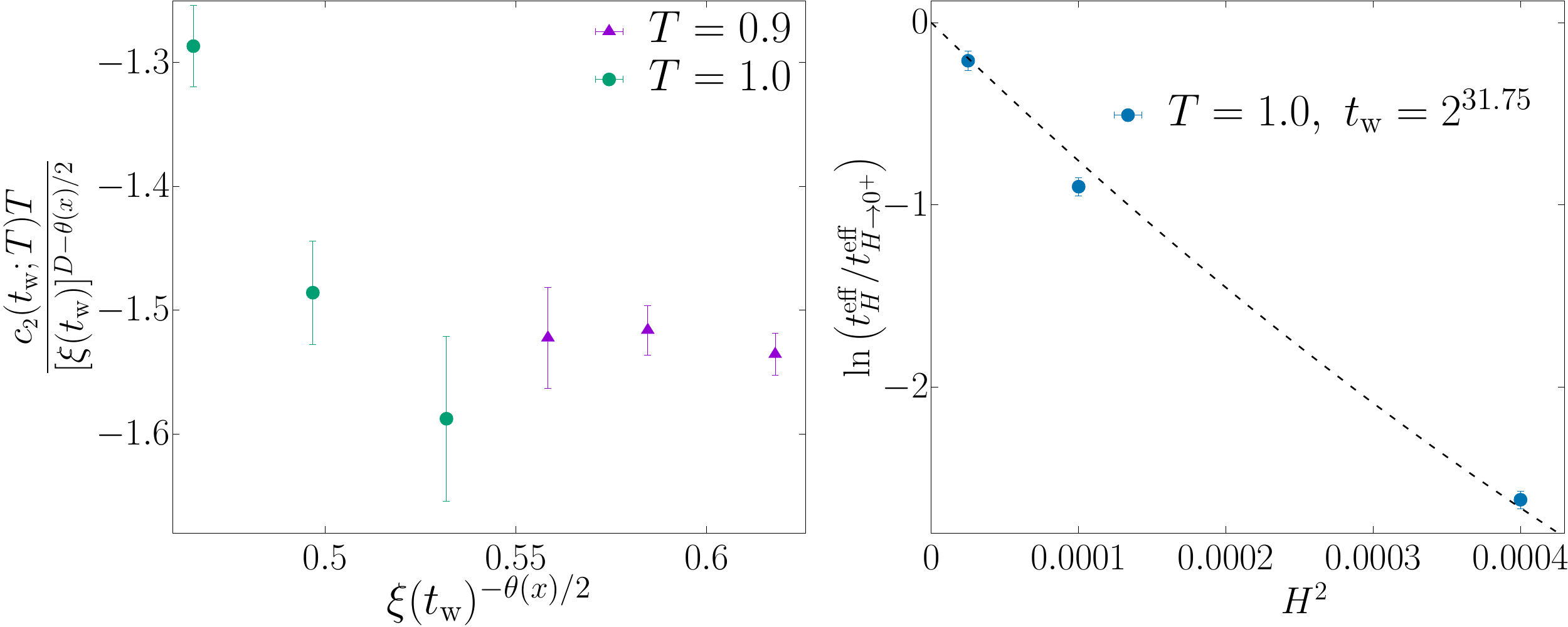}
	\caption{\emph{Left:} Behaviour of the rescaled quantity  $c_2(\tw;T) T
/[\xi(\tw)]^{D-\theta(\tilde x)/2} $ as a function of $\xi(\tw)^{-\theta(\tilde
x)/2}$, see Eq.~\eqref{eq:a2_prediction}.
\emph{Right:} An enlargement of Fig.~\ref{fig:ratio_teff_janus} in the
small-field regime, for the case $\tw=2^{31.75}$ at $T=1.0$ ({Run 6} of
Table~\ref{tab:details_NUM}) and its fit reported in Table
\ref{tab:ratio_times_fits_NUM}.}
   \label{fig:rescaled_c2}
\end{figure}

We rewrite Eq. \eqref{eq:a1_paramet_dependence_NUM} as:
\begin{equation}\label{eq:a2_prediction}
\frac{c_2(t_\mathrm{w};T)}{\xi^{D-\theta(\tilde{x})/2}} = \frac{\hat{S}}{2} + T a_1(T) \xi^{-\theta(\tilde{x})/2} \; .
\end{equation}
and we study this quantity as a function of $
[\xi(\tw)]^{-\theta(\tilde{x})/2}$ in Fig.~\ref{fig:rescaled_c2}.  Note
that in the above expression $\xi$ was \emph{not} obtained from the response
to the magnetic field. Instead, we computed $\xi$ from the correlation
functions at $H =0$ (see Appendix \ref{Appendix:xi_construction} and
Ref.~\cite{janus:08b}).
The data exhibit a constant value, except for the point correspondent to $\tw =
2^{31.75} $ at $T = 1.0$ ({Run 6}). Therefore, we shall accept the
numerical estimation of the derivative at $H^2=0$ through the coefficient $c_2$
for all cases but for our {Run 6}.
In order to clarify what is going on with {Run 6}, we report in the right panel of
Fig.~\ref{fig:rescaled_c2} an enlargement of
Fig.~\ref{fig:ratio_teff_janus} in the small-magnetic-field regime for this
case. As it could be guessed from the left panel, the fitting
procedure clearly underestimates the slope of the curve at $H^2=0$. Therefore,
in order to estimate the derivative for {Run 6}, we have instead relied
on Eq.~\eqref{eq:a2_prediction} by averaging the constant value found in
Fig.~\ref{fig:rescaled_c2} over all other runs 
and by multiplying this averaged constant value by $[\xi(\tw)]^{D-\theta(\tilde{x})/2}$ [see
Eq.~\eqref{eq:a2_prediction}].

\subsection{Non-linear scaling}
\label{Sec:non-linear_scaling}
\begin{figure}[t]
        \centering
        \includegraphics[width=0.67\columnwidth]{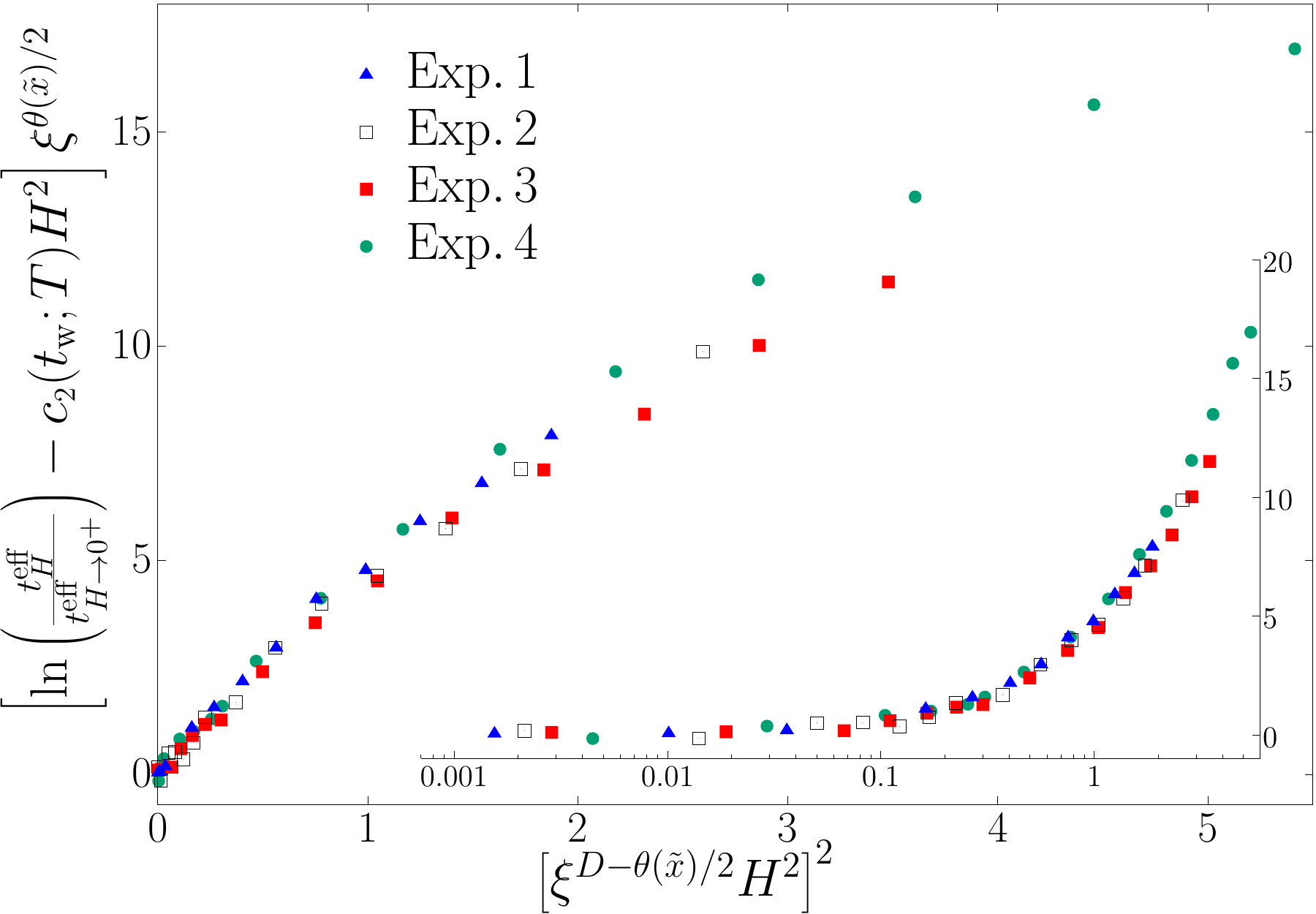}
	\caption{\label{fig:non-linear_scaling_law_EXP} The non-linear parts
from the experimental response time data, $[\ln t^{\text{eff}}_H -c_2(\tw;\Tm)
H^2] \xi^{\theta(\tilde{x})/2}$, plotted against
$(\xi^{D-\theta(\tilde{x})/2}H^2)^2$.  The deviations of the data at $\Tm=29$~K
may be caused by a shift in $\Tg$ as the temperature begins to approach
$\Tg(H)$.
        The small-$x$ range is enlarged in the \emph{inset}.}
\end{figure}

\begin{figure}[t]
        \centering
        \includegraphics[width=0.67\columnwidth]{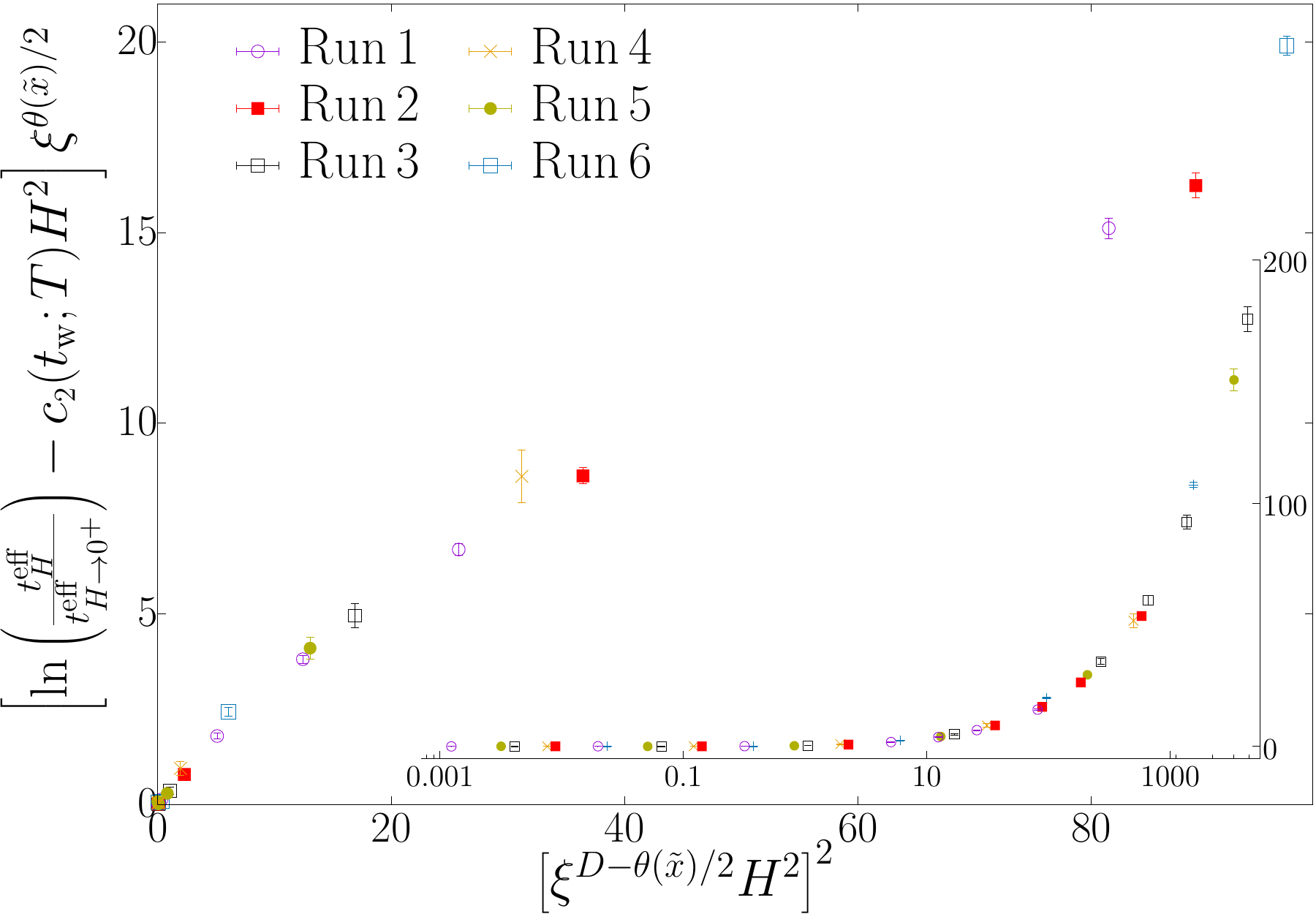}
	\caption{\label{fig:non-linear_scaling_law_NUM} The non-linear parts
from the numerical response time data, $ [ \ln (t^{\text{eff}}_H /
t^{\text{eff}}_{H\to 0^+}) - c_2(\tw;\Tm) H^2] \xi^{\theta(\tilde{x}/2)}$,
plotted against $(\xi^{D-\theta(\tilde{x})/2} H^2)^2$. The abscissa of the
\emph{main panel} is in linear scale and shows a closeup for small values of
$(\xi^{D-\theta(\tilde{x})/2}H^2)^2$. The abscissa of the \emph{insert} is in log
scale in order to report all our numerical data.}
\end{figure}

In order to test the scaling form, Eq.
\eqref{eq:ratio_effective_time_scaling_final}, the data for all of the
non-linear contributions to the magnetisation, experimental and numerical, are
plotted according to the functional form
\begin{equation}
\xi^{-{\frac {\theta(\tilde x)}{2}}}{\mathcal {G}}(T, \xi^{D-{\frac {\theta( \tilde x)}{2}}}H^2)
\end{equation} 
in Fig. \ref{fig:non-linear_scaling_law_EXP}-\ref{fig:non-linear_scaling_law_NUM}.
The fit to scaling relationship~\eqref{eq:ratio_effective_time_scaling_final} is remarkable
and testimony to
the agreement for both the experimental and numerical data.

\subsection{Overshooting phenomena}
\label{Sec:overshoot_phenomena}
We first address the dynamical scaling law for a system in presence of a
magnetic field at temperatures close to $\Tg$ in
Section~\ref{Sec:dynamical_scaling}.  Then, in
Section~\ref{Sec:overshoot_order_system}, we analyse the dynamical scaling for
ferromagnetic systems, either Ising or Heisenberg, in the presence of an
external magnetic field.

\subsubsection{Dynamical scaling close to $\Tg$}
\label{Sec:dynamical_scaling}

\begin{table}[t]
\caption{The aging-rate factors $z(T)$ used in Fig.~\ref{fig:dyn_scaling}.}
\label{tab:z_value}
\begin{indented}
\item[] \begin{tabular}{@{}c c } 
 \br
$T$ & $z(T)$ \\ 
\mr
1.1  & 6.60 \\ 
1.05 & 7.00 \\
1.0  & 7.30 \\ 
0.9  & 8.12 \\
\br
\end{tabular}
\end{indented}
\end{table}
\begin{figure}[t] \centering 
        \includegraphics[width=.7\textwidth]{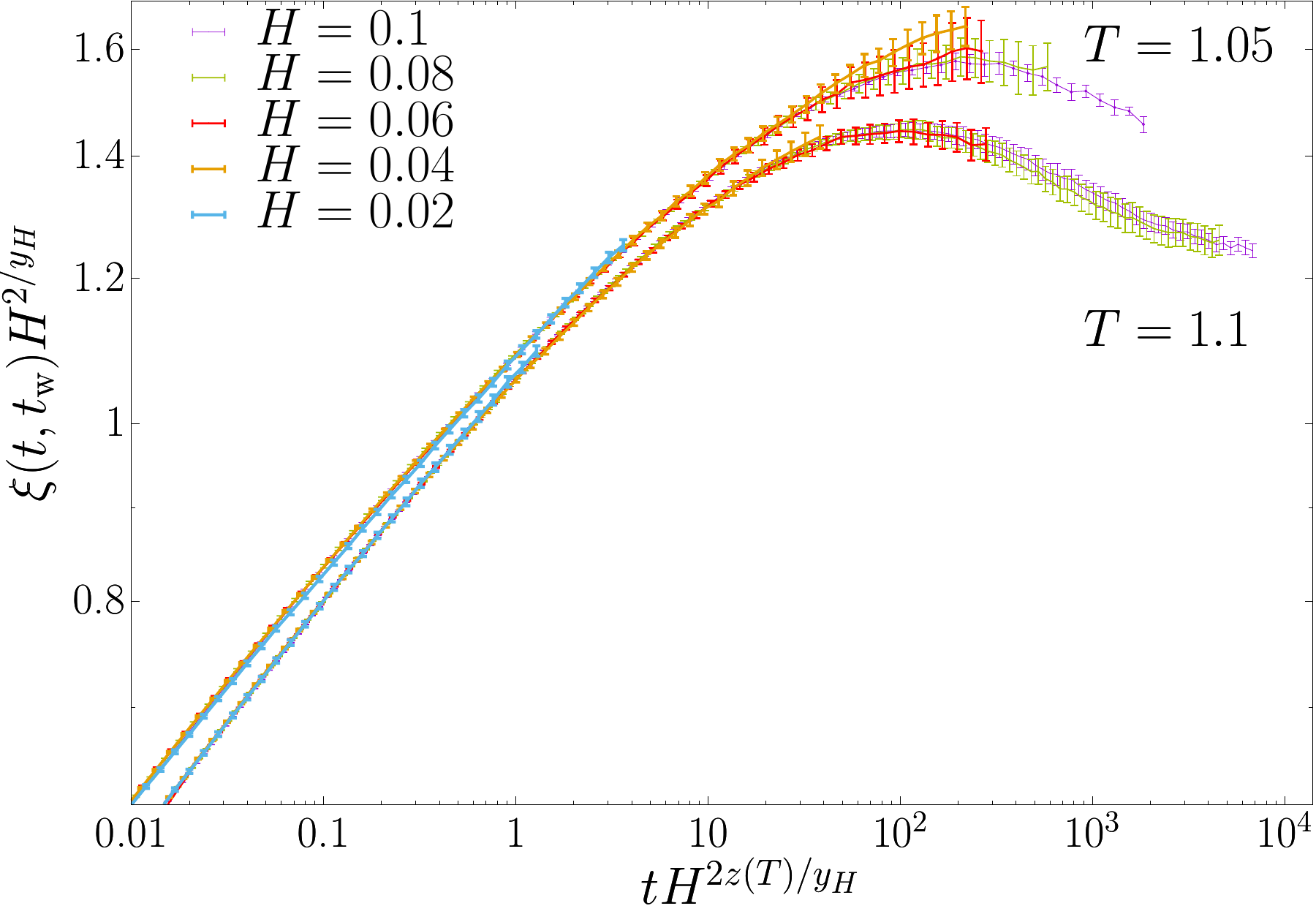}
  \caption{Critical dynamical scaling according
to Eqs.~\eqref{eq:xi_H_dynamical_scaling} and~\eqref{eq:rescaled_time_dynamical_scaling}.
We show data for $T=1.1\approx\Tg$~\cite{janus:13} and for $T=1.05$.}
  \label{fig:dyn_scaling}
\end{figure}
We evaluate the growth of the correlation length $\xi(t,\tw;H)$ in simulations
that mimic the experimental field-cooling protocol (FC), where the
temperature is lowered from above to below $\Tg$ in the presence of a constant
magnetic field $H$.

We performed two independent simulations on Janus II at the critical
temperature $\Tg=1.102(3)$~\cite{janus:13} (in IEA units) and at $T=1.05$ for
several external magnetic fields and $16$ samples.  A protocol equivalent to
FC, but convenient for simulations, is to place a random spin configuration
instantaneously at the working temperature $T$, and turning on the external magnetic
field at the same instant, so that $\tw=0$.

According to Eq.~\eqref{eq:m-scaling}, at the critical temperature $\Tg$, and
for small external magnetic fields $H$, there exists a scaling behaviour that
connects $\xi(t,	\tw ; H)$ with the external magnetic field $H$:
\begin{equation}
\label{eq:xi_H_dynamical_scaling}
[\xi(t, \tw; H) \ H^{2/y_H} ] \quad \propto \text{const.}
\end{equation}
The correlation length $\xi(t,\tw;H)$ grows as
\begin{equation}
\label{eq:xi_vs_t_aging}
\xi(\tw) \propto \tw^{1/z(T)} \, ,
\end{equation}
with an exponent that, in a first approximation, is expected to behave near the critical temperature as \cite{janus:18}:
\begin{equation}
\label{eq:aging_rate_zT}
z(T) \simeq z_\text{c} \,\frac{\Tg}{T} \; , \quad \text{where} \quad z_\text{c} = z(\Tg)= 6.69(6) \, .
\end{equation}
Hence, using  Eq.~\eqref{eq:xi_vs_t_aging} and Eq.~\eqref{eq:aging_rate_zT} in
the scaling argument of Eq.~\eqref{eq:xi_H_dynamical_scaling}, we have
equivalently,
\begin{equation}\label{eq:rescaled_time_dynamical_scaling}
[t \ \times H^{2 z(T)/y_H} ] \quad \propto \text{const.}
\end{equation}
In Table~\ref{tab:z_value} we list the aging-rate factors $z(T)$ used in our
analysis. We plot our rescaled data in Fig.~\ref{fig:dyn_scaling}.
 
The agreement with the scaling prediction, evinced by the data collapse, is
striking.  Our plots also exhibit overshooting, as evidence for the paramagnetic
phase when the magnetic field is turned on.
 
The reader could wonder why we have used Eq.~\eqref{eq:xi_H_dynamical_scaling}
for the scaling analysis at the temperature $T=1.05<\Tg$, and whether this implies
evidence of the absence of the de Almeda-Thouless (dAT) line in finite
dimension. We shall address these questions in Section~\ref{Sec:xi_t_tw_dAT}.

\subsubsection{Overshooting in a ferromagnetic system}
\label{Sec:overshoot_order_system}
By studying two ordered systems we can show that the overshooting phenomenon is, in fact, general.
To demonstrate generality, we have simulated the three-dimensional Ising and
Heisenberg models in a cubic lattice with periodic boundary conditions and size
$L$ at the critical point $\Tc$. The $N=L^D$ ($D=3$) Heisenberg spins interact with
their lattice nearest neighbours through the Hamiltonian
\begin{equation}
  \mathcal{H} =  - \sum_{\langle\boldsymbol{r},\boldsymbol r'\rangle}
  {\boldsymbol{S}}_{{\boldsymbol{r}}}\cdot {\boldsymbol{S}}_{{\boldsymbol{ r'}}} +{\boldsymbol{H}} \cdot \sum_{{\boldsymbol{r}}} {\boldsymbol{S}}_{{\boldsymbol{r}}}\, .
\label{eq:origham}
\end{equation}
where ${\boldsymbol{S}}_{{\boldsymbol{ r}}}$ are unit vector spins and ${\boldsymbol{H}}$ is an external magnetic field.
The connected correlation function is 
\begin{equation}
  \label{eq:cor}
  C(r,t)=\frac{1}{L^3} \sum_{{\boldsymbol{ x}}}
  {\boldsymbol{ S}}_{{\boldsymbol{ x}}}(t) \cdot{{\boldsymbol{ S}}}_{{\boldsymbol{r+x}}}(t) - [{\boldsymbol{m}}(t)]^2\,,
\end{equation}
with 
\begin{equation}\label{eq:m_heisenberg}
{\boldsymbol{ m}}(t) = \frac{1}{L^3}\sum_{{\boldsymbol{ x}}} {\boldsymbol{ S}}_{{\boldsymbol{ x}}}(t)\,.
\end{equation}
We only write equations for the Heisenberg model,
Eqs.~\eqref{eq:origham}--\eqref{eq:m_heisenberg}, but the Ising analogues can be
obtained trivially by just dropping the vector symbol in the
spins. Furthermore, the correlation function $C(r,t)$ will be averaged over
different initial random configurations (runs). We report
the simulation details in Table~\ref{tab:simulation_details_order}.
\begin{table}
\caption{\label{tab:simulation_details_order} Parameters of our ferromagnetic simulations 
for the Ising and Heisenberg models.}
\begin{indented}
\item[]\begin{tabular}{@{}cclc}
\br
System &  $L$ & \multicolumn{1}{c}{$H$} &  Number of runs\\
\mr
\multirow{4}{*}{Ising} & \multirow{4}{*}{256} & 0 & 2\,200\\
& & 0.001 &  2\,600 \\
& & 0.003 &  1\,700 \\
& & 0.005 &  1\,200 \\
\mr
\multirow{4}{*}{Heisenberg} & \multirow{4}{*}{200} & 0 & 2\,660\\
& & 0.003 & 1\,000 \\
& & 0.004 & 1\,600 \\
& & 0.005 & 2\,400 \\
\br
\end{tabular}
\end{indented}
\end{table}

The Ising and Heisenberg model have different symmetry properties, so they belong to two distinct universal classes.
In other words, each model has a distinct value for the critical temperature and exponents.
The Ising model has $\eta=0.036\,297\,8(20)~$\cite{simmons:17},
$z=2.0245(15)$~\cite{hasenbusch:20} and $\beta_\mathrm{c}=0.221\,654\,626(5)$~\cite{ferrenberg:18}. 
Instead, for the Heisenberg ferromagnet
$\eta=0.378(3)$~\cite{campostrini:02,hasenbusch:11},
$z=2.033(5)$~\cite{astillero:19} and
$\beta_\mathrm{c}=0.693001(10)$~\cite{ballesteros:96b} ($\beta_\mathrm{c} \equiv
1/T_\mathrm{c}$).

\begin{figure}[t] \centering 
\includegraphics[width=\linewidth]{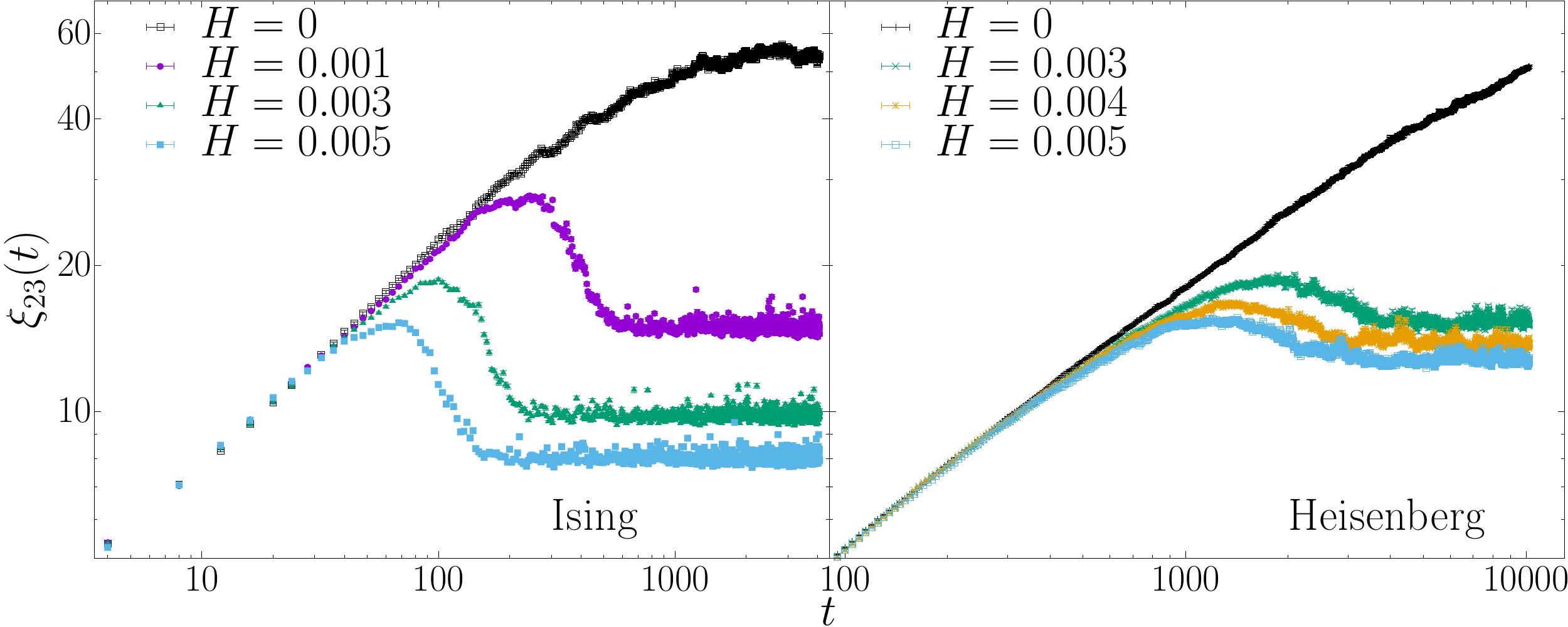}
  \caption{The log-log plots show the behaviour of $\xi_{23}$ versus time for
    the Ising (\emph{left}) and Heisenberg (\emph{right})
    models for three magnetic fields. All 
simulations were performed at the critical temperature $\Tc$ appropriate to each
model. The saturation at long times exhibited by the Ising model at $H=0$ is a
finite-size effect.}
  \label{fig:xi23_ferromagnet}
\end{figure}

As explained in Appendix~\ref{Appendix:xi_construction}, the correlation
length $\xi(t,\tw;H)$ can be calculated with integral estimators~\cite{janus:08b,janus:09b},
\begin{equation}
\label{eq:integral_xi}
I_k (T,\tw) = \int_{0}^{\infty} \mathrm{d} r \, r^k C(r,\tw;T)\, , \quad \xi_{k,k+1}(\tw;T) = \frac{ I_{k+1}(\tw;T)}{I_k(\tw;T)} \, .
\end{equation}
In this Section, we evaluate the correlation length $\xi_{23}(t,\tw;H)$.
As the reader can notice, the growth of $\xi_{23}(t)$ overshoots
before  reaching equilibrium for any external magnetic field for both
ferromagnetic models, see Fig.~\ref{fig:xi23_ferromagnet}.
\begin{figure}[t]
        \includegraphics[width=\textwidth]{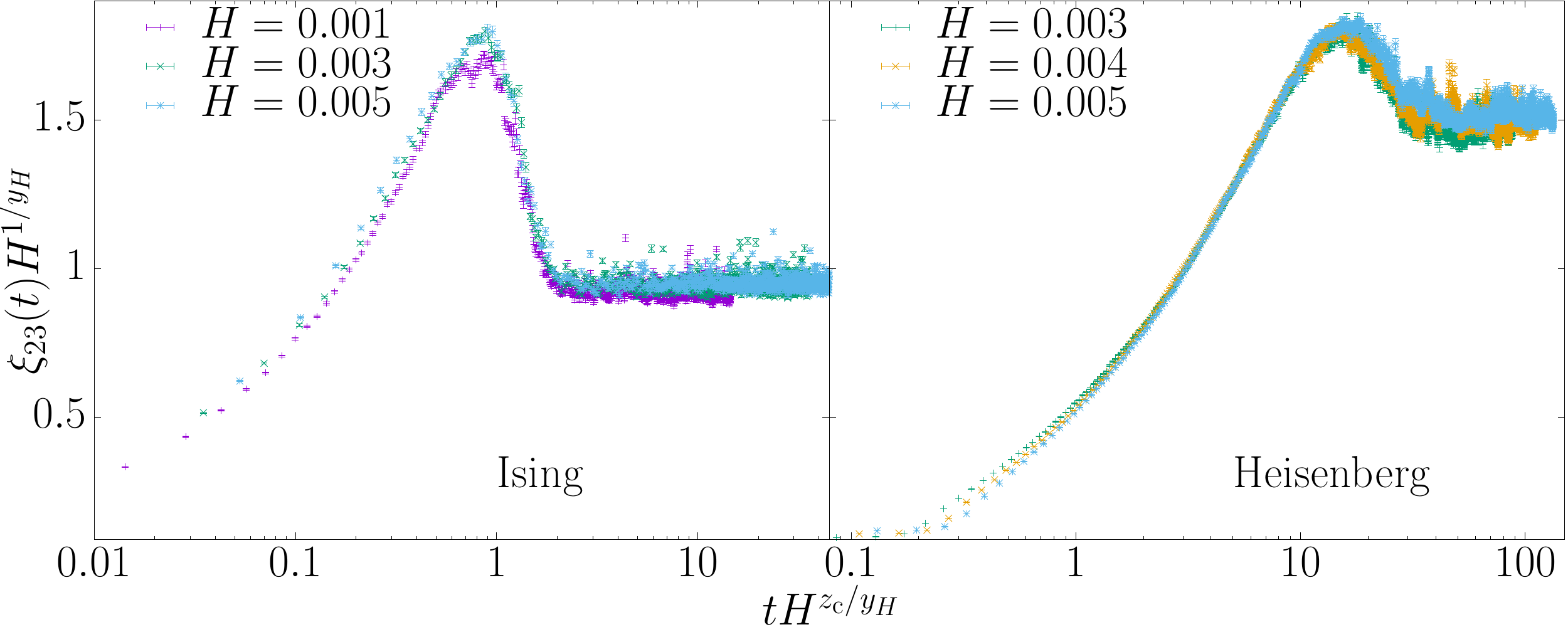}
  \caption{Critical dynamical scaling for ferromagnetic models. The data from
Fig.~\ref{fig:xi23_ferromagnet} for the Ising (\emph{left}) and Heisenberg
(\emph{right}) models for the three non-vanishing magnetic fields are rescaled
following the predictions of the Renormalisation Group [see
Eqs.~\eqref{eq:xi_H_dynamical_scaling_ferromagnetic}--\eqref{eq:rescaled_time_dynamical_scaling_ferromagnetic}].
In this case the relevant variables are $\xi H^{1/y_H}$ and $t
H^{z_\text{c}/y_H}$, with $y_H=(D+2-\eta)/2$ with $D=3$.}
  \label{fig:dyn_scaling_ferromagnet}
\end{figure}

According to Eq.~\eqref{eq:m-scaling}, at the critical temperature $\Tc$, and
for small external magnetic fields $H$, there exists a scaling law that
connects $\xi(t,\tw ; H)$ with the external magnetic field $H$ in
ferromagnetic system:
\begin{equation}
\label{eq:xi_H_dynamical_scaling_ferromagnetic}
[\xi(t,\tw; H) \ H^{1/y_H} ] \quad \propto \text{const.}
\end{equation}
As the reader can notice, Eq.~\eqref{eq:xi_H_dynamical_scaling_ferromagnetic}
differs from Eq.~\eqref{eq:xi_H_dynamical_scaling} in the power of the magnetic
field. In the ferromagnetic system, the relevant external variable is $H$ and
not $H^2$ as it would be for spin glasses \cite{parisi:88, amit:05}.
Analogously to Eq.~\eqref{eq:rescaled_time_dynamical_scaling}, we can rescale time as
\begin{equation}\label{eq:rescaled_time_dynamical_scaling_ferromagnetic}
[t \ \times H^{z(T)/y_H} ] \quad \propto \text{const.}
\end{equation}
We plot our rescaled data in Fig. \ref{fig:dyn_scaling_ferromagnet}.
The agreement with the scaling prediction, both for the Heisenberg and for the
Ising model, is remarkable.

In conclusion, the overshooting phenomenon is general and we have observed it both
in ferromagnetic systems,
Figs.~\ref{fig:xi23_ferromagnet}--\ref{fig:dyn_scaling_ferromagnet}, and in
disordered ones, Fig.~\ref{fig:dyn_scaling}.

\section{Investigation of the dAT line in \boldmath $D=3$}
\label{Sec:xi_t_tw_dAT}

The existence (or not) of the spin-glass condensation in the presence of a
magnetic field remains the subject of some controversy (see,
\emph{e.g.},~\cite{janus:12,larson:13,janus:14c,holler:20}).  In a mean-field
treatment, de Almeida and Thouless~\cite{dealmeida:78} showed that, for the
Sherrington-Kirkpatrick infinite-range mean-field model~\cite{sherrington:78},
there would be a phase transition according to the following relationship for
Ising spin glasses,
\begin{equation}
\label{eq:dAT_line_MF}
\bigg(1-{\frac {\Tg(H)}{\Tg(0)}}\bigg)^3 = {\frac {3}{4}}\,h^2\,,
\end{equation}
with
\begin{equation}
h={\frac {\mu H}{k_\text{B}\Tg(0)}}\,,
\end{equation}
where $\mu$ is the spin magnetic moment.  Conversely, the droplet
model~\cite{fisher:86,bray:86} would predict no phase transition except exactly
at $H=0$.  This dispute was addressed by Lefloch \emph{et al.}~\cite{lefloch:94}.
Their final conclusion bears repetition: ``Thus, even if the spin glass does not
exist in a magnetic field, {\it at least it looks the same as in zero field},
as far as we experimentalists can see''.

In finite dimension and for $T$ very close to the critical temperature
$\Tg(H=0)$, the de Almeida-Thouless (dAT) line, provided it exists, should be
governed by the Fisher-Sompolinsky \cite{fisher:85} relation:
\begin{equation}
\label{eq:dAT_FS_relation_D3}
\left(1-\frac{\Tg(H)}{\Tg(0)} \right) \propto H^{4/\nu(5-\eta)}\,,
\end{equation}
where we have specialised to $D=3$.\footnote{Notice this is the same relation
  used for matching the numerical and experimental scales in
  Section~\ref{Sec:Janus_details}}  Rather than through $\Tg(H)$, we are
interested in describing the dAT line geometrically by the inverse function of
$\Tg(H)$, namely $H_\text{c}(T)$.  Hence, we rewrite Eqs.~\eqref{eq:dAT_line_MF} and
\eqref{eq:dAT_FS_relation_D3} as
\begin{eqnarray}
\label{eq:Hc_dAT_MF}
\fl
\mbox{Mean-Field:}&  \quad \quad& H_\text{c}(T) \propto \left( 1 - \frac{T}{\Tg} \right)^{a_\mathrm{MF}}\,, \quad a_\mathrm{MF}=1.5 \,, \\
\label{eq:Hc_dAT_3D}
\fl
\mathrm{3D:}&  \quad \quad& H_\text{c}(T) \propto \left( 1 - \frac{T}{\Tg} \right)^{a_\mathrm{3D}}\,,  \quad a_\mathrm{3D}=\frac{\nu(5-\eta)}{4} \quad \to\quad  a_\mathrm{3D} = 3.45(5)\,,  
\end{eqnarray}
where we have taken the 3D critical exponents $\nu$ and $\eta$ from Ref.~\cite{janus:13}.
The following considerations, based on
Eqs.~\eqref{eq:Hc_dAT_MF}--\eqref{eq:Hc_dAT_3D}, will be useful:
\begin{itemize}
\item $H_\text{c}(T)$ is a decreasing function of $T$ (remember
  $T\leq \Tg$) and $H_\text{c}(\Tg)=0$. This means that, upon approaching $\Tg$ from below, one
  eventually crosses
  the dAT line for any $H>0$, no matter how small $H$ is.
\item When $H> H_\text{c}(T)$ the system is above the dAT line, in its paramagnetic
  phase: the correlation length, $\xi(t,\tw;H)$, reaches asymptotically its
equilibrium value $\xi_\mathrm{eq}(H)$ for very long time $t$.
\item When $H< H_\text{c}(T)$ we are in the spin-glass phase and one expects to
observe a power-law growth of the correlation length, see
Eq.~\eqref{eq:xi_vs_t_aging}.
\item The $a_\mathrm{3D}$ exponent is much larger than the mean-field (MF)
  one, $a_\mathrm{3D} \approx 2.3 \times a_\mathrm{MF}$. This implies that,
  in $D=3$, the dAT line is \emph{very} flat when one approaches the critical
temperature $T \backsimeq \Tg$.
\end{itemize}

In particular, our last item above suggests an interpretation of the somewhat
surprising results in Fig.~\ref{fig:dyn_scaling}, where data for $T=1.05$
were successfully scaled with the scaling law appropriate for $\Tg$ [recall
that $1.05<\Tg=1.102(3)$ and that at $T=\Tg$ we are in the paramagnetic phase
for any $H>0$].
Assuming that the proportionality coefficient is of order unity,
let us estimate the critical magnetic field at $T=1.05$
exploiting Eq.~\eqref{eq:Hc_dAT_3D}:
\begin{equation}
\label{eq:Hc_T105}
H_\text{c}(T=1.05)\sim 3\times 10^{-5} \, .
\end{equation}
Considering, now, that the smallest magnetic field in
Fig.~\ref{fig:dyn_scaling}, namely $H=0.02$, is larger than
$H_\text{c}(T=1.05)$ by a factor of 1000 or so, there is little surprise in
that a scaling law assuming $H_\text{c}(T=1.05)=0$ works with our data.
\begin{figure}[t] \centering
\centering
   \includegraphics[width=\textwidth]{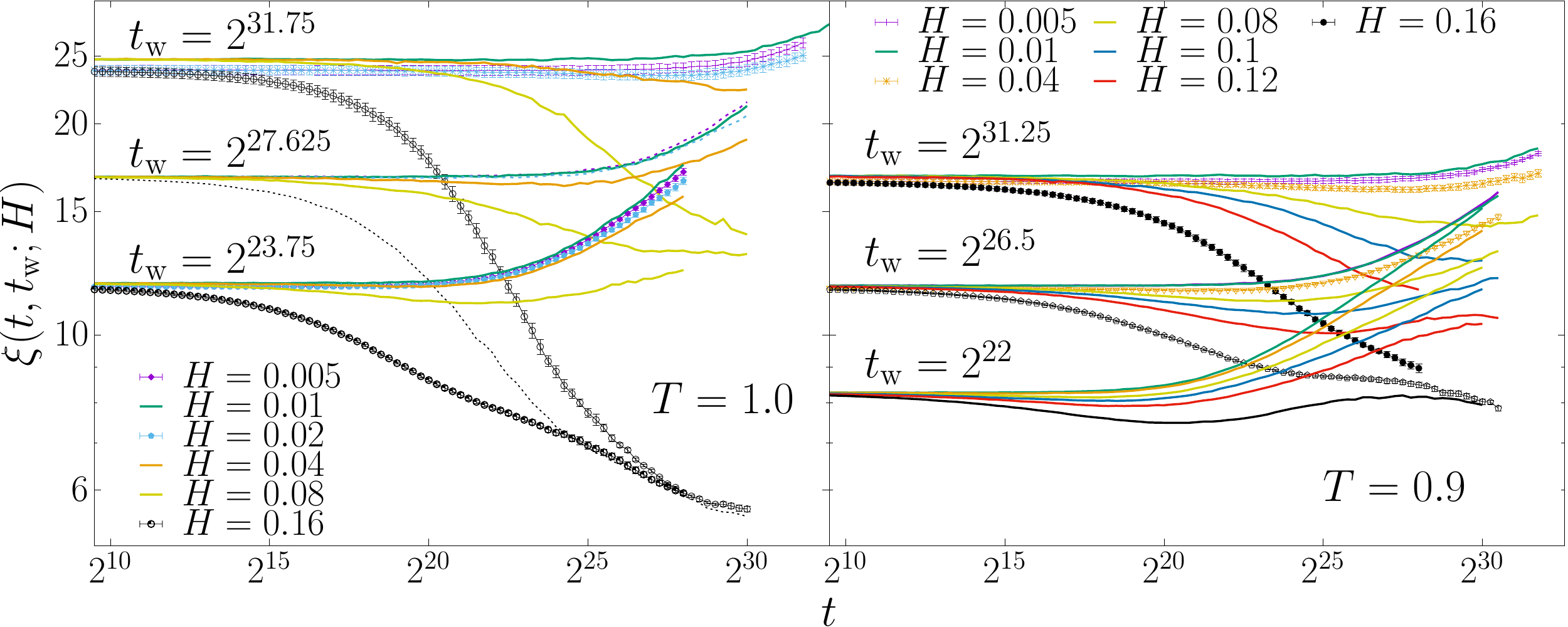}
    \caption{\label{fig:xi_ZFC} Growth of $\xi(t,\tw \neq 0;H)$ in simulations
that mimic the experimental zero-field-cooling protocol. Plots are
in log-log scale.}
\end{figure}

Our focus in this section will be an exploration of the growth of the spin-glass
correlation length, $\xi(t,\tw;H)$, under conditions that mimic the
experimental protocol for measurement of the zero-field-cooled magnetisation,
$M_{\text {ZFC}}(t,\tw;H)$ for $\tw\neq 0$, recall
Section~\ref{Sec:Janus_details}.

In Fig.~\ref{fig:xi_ZFC} we plot $\xi(t,\tw \!\neq\! 0;H)$ as a function of
time for different magnetic fields $H$.
We compute $\xi$ from the microscopic correlation function $C_4(r)$ (see
Appendix~\ref{Appendix:xi_construction}), which requires that we compute error
bars from the sample-to-sample fluctuations. We have simulated different
samples only for some values of $H$ and $\tw$ because of the enormous
computational effort involved.  We show error bars in
Fig.~\ref{fig:xi_ZFC} only in those cases where they can be computed.

The time evolution of the spin-glass correlation length $\xi(t,\tw;H)$ depends
markedly on the interplay between the waiting time $\tw$ and the value of the
magnetic field $H$, see Fig.~\ref{fig:xi_ZFC}.
The system needs several time steps before responding to the switching on of
the magnetic field. Different scenarios appear.

On the one hand, for the largest magnetic fields, namely $H>0.04$ both at $T=0.9$ and $T=1.0$,
the correlation length displays a non-monotonic time behaviour, just as we found in
Section~\ref{Sec:dynamical_scaling} for the dynamics in the paramagnetic phase (recall
that $\tw=0$ in Section~\ref{Sec:dynamical_scaling}).  In particular, for those
cases when the starting correlation length, $\xi(t=0,\tw;H)$, is \emph{larger}
than the equilibrium value $\xi_\mathrm{eq}(H)$, the correlation length
decays. Otherwise, we observe an overshooting phenomenon reminiscent of our
findings in Section~\ref{Sec:dynamical_scaling}, see Fig.~\ref{fig:xi_ZFC}.

On the other hand, for $H<0.08$, we observe that the correlation length
$\xi(t,\tw;H)$ appears to follow the same power-law growth for all the different
waiting times. 
Here we must distinguish between the mean-field $H_\text{c}^{\text {MF}}(T)$
and the Fisher-Sompolinsky scaling $H_\text{c}^{\text {3D}}(T)$, {\it i.e.},
between Eqs.~\eqref{eq:Hc_dAT_MF} and~\eqref{eq:Hc_dAT_3D}.  Using
Eq.~\eqref{eq:dAT_line_MF} for the former, one finds
\begin{equation}\label{eq:Hc_estimation_MF}
H_\text{c}^{\text {MF}}(T=0.9)\approx 0.0675~~ {\text {and}}~~H_\text{c}^{\text {MF}}(T=1.0)\approx 0.02421\,.
\end{equation}
Interestingly, the scaling result, Eq.~\eqref{eq:Hc_dAT_3D}, yields
\begin{equation}\label{eq:Hc_estimation_3D}
H_\text{c}^{\text {3D}}(T=0.9)\approx 0.003~~{\text {and}}~~H_\text{c}^{\text {3D}}(T=1.0)\approx 0.0003\,. 
\end{equation}
For the magnetic fields used in our simulations, therefore, one is presumably in the
condensed state for $0.005\leq H<0.08$ from the perspective of the mean-field
solution of the Sherrington-Kirkpatrick model~\cite{sherrington:78}, while from
the perspective of the Fisher-Sompolinsky scaling~\cite{fisher:85} one is
always in the paramagnetic state since $H>H_\text{c}^{\text {3D}}(T = 0.9, 1.0)$.
Though this latter region is not accessible experimentally through magnetic
measurements, one can argue that the simulation results should be symmetric
around $\Tc(H)$. This is the basis for the comparison between experiment and
simulations contained in Section~\ref{Sec:scaling_law_all} of this paper.

Let us now attempt a scaling analysis similar to the one in
Section~\ref{Sec:dynamical_scaling} for those magnetic field values for which the asymptotic
$\xi_\mathrm{eq}(H)$ can be at least guessed from Fig.~\ref{fig:xi_ZFC}.
We start by modifying scaling relationship~\eqref{eq:xi_H_dynamical_scaling} to
\begin{equation}
\label{eq:xi_scaling_Hc}
\xi(t,\tw;H)\ | H^2-H_\text{c}^2(T) |^{1/y_H}\propto {\text {const.}}
\end{equation}
Next, using Eq.~\eqref{eq:xi_vs_t_aging} in the scaling argument of Eq.~\eqref{eq:xi_scaling_Hc}, we have, equivalently,
\begin{equation}
\label{eq:time_scaling_Hc}
\left( t \times | H^2-H_\text{c}^2(T)|^{ z(T)/y_H} \right) \propto {\text {const.}}
\end{equation}

\begin{figure}[t] 
   \includegraphics[width=\textwidth]{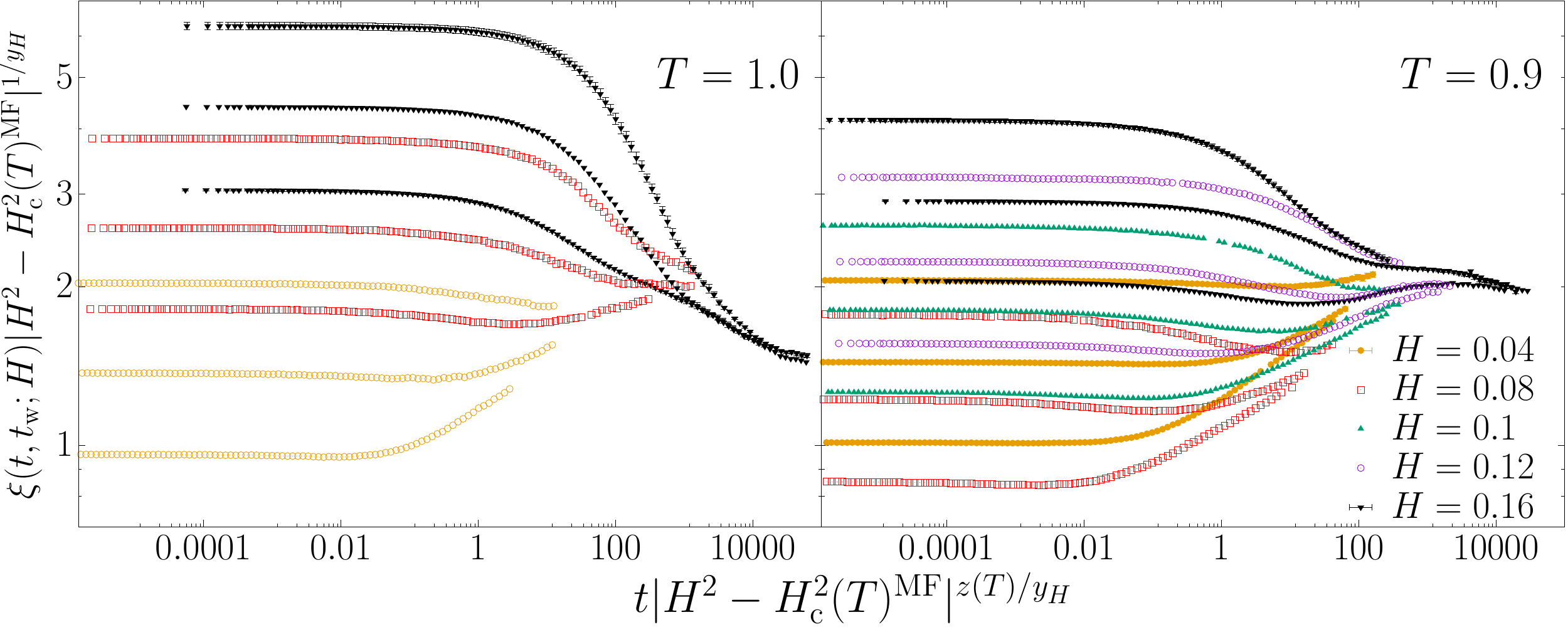}
    \caption{Search for the dAT line in $D=3$ using mean-field scaling. Plots are in
log-log scale and show the behaviour of the rescaled quantities defined in Eqs.~\eqref{eq:xi_scaling_Hc}--\eqref{eq:time_scaling_Hc} for the mean-field
estimators $H_\text{c}(T)^{\mathrm{MF}}$, see
Eq.~\eqref{eq:Hc_estimation_MF}. The aging rates $z(T)$ used in this
scaling are listed in Table~\ref{tab:z_value}, to be found in
Section~\ref{Sec:overshoot_phenomena}.
    \label{fig:dynamical_scaling_xi_coldT_MF}}
\end{figure}

We replot our rescaled data in Fig.~\ref{fig:dynamical_scaling_xi_coldT_MF}
for the mean-field values of $H_\text{c}^{\text {MF}}(T)$, see
Eq.~\eqref{eq:Hc_estimation_MF}. As seen in
the panel for $T=0.9$, there is nearly perfect scaling for
$H\geq 0.08$ but not for $H=0.04$, though the curves do seem to coalesce for
the three different waiting times.

It is tempting to suggest that, for this value of magnetic field, one is in the
condensed phase.  Glancing at Fig.~\ref{fig:xi_ZFC}, however, the growth
of $\xi(t,\tw;H)$ for $H=0.04$ breaks away from the curves for the larger
magnetic fields, so that it is very possible that it would join the
equilibrium curves ({\it i.e.}, the paramagnetic regime) at times longer than
those accessible in our simulations.  This ambiguity softens the interpretation
that we have broached the dAT line in our simulations, as would be predicted
from a mean-field approach.
\begin{figure}[t]
   \includegraphics[width=\textwidth]{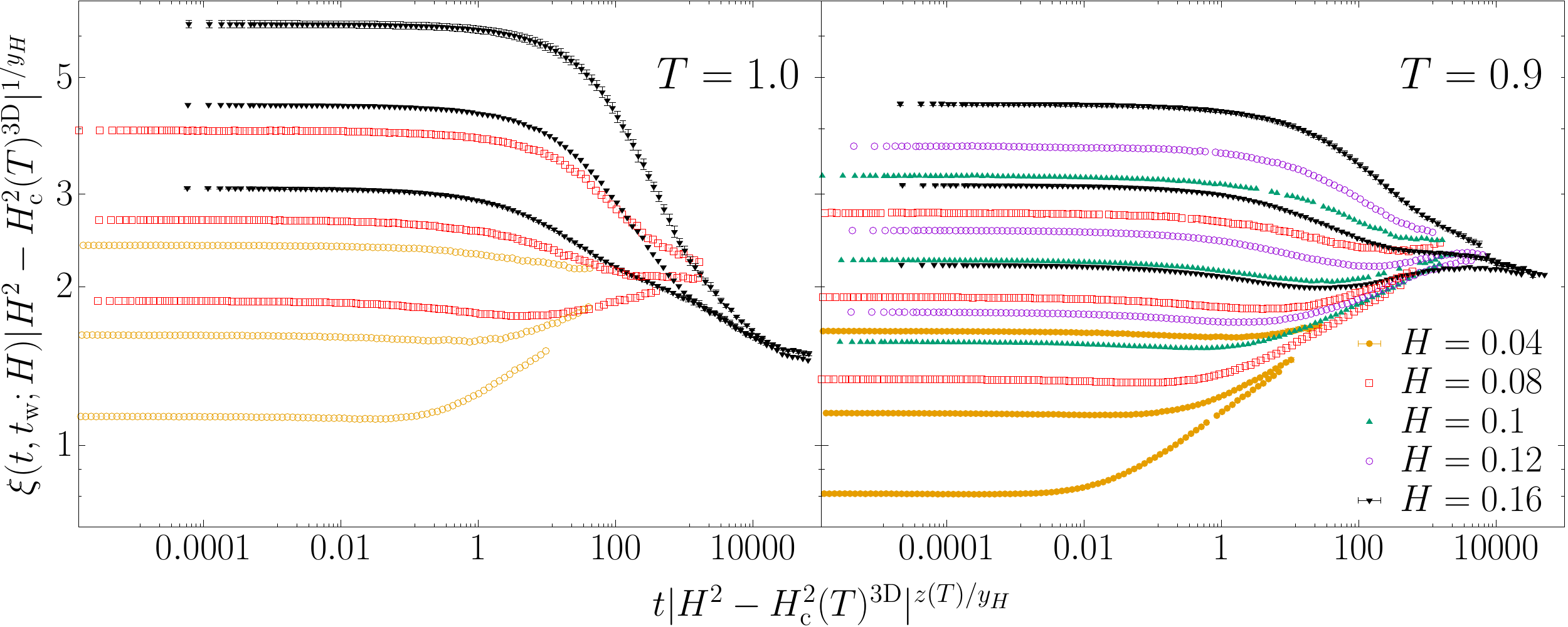}
    \caption{Search for the dAT line in $D=3$ with Fisher-Sompolinsky scaling. Plots are in
log-log scale and show the behaviour of the rescaled quantities defined in 
Eqs.~\eqref{eq:xi_scaling_Hc}--\eqref{eq:time_scaling_Hc} for the Fisher-Sompolinsky
estimators $H_\text{c}(T)^\mathrm{3D}$, see Eq.~\eqref{eq:Hc_estimation_3D}.
We report the aging rates $z(T)$ used in this scaling  in
Table~\ref{tab:z_value} to be found in Section~\ref{Sec:overshoot_phenomena}.
    \label{fig:dynamical_scaling_xi_coldT}}
\end{figure}

However, if we replot our data using the scaling result of
Eq.~\eqref{eq:Hc_estimation_3D}, as exhibited in
Fig.~\ref{fig:dynamical_scaling_xi_coldT}, for the values of $H_\text{c}^{\text
{3D}}(T)$ for $T=0.9$, the data appear to collapse for {\it all} of the
magnetic fields, including $H=0.04$ [$H_\text{c}^{\text {3D}}(T=0.9)\ll 0.04$].
The Fisher-Sompolinsky scaling would, therefore,
support the conjecture that at $T=0.9$, our simulation results for
$H=0.04$ are in the paramagnetic regime.

As to the rescaled $T=1.0$ data in Figs.~\ref{fig:dynamical_scaling_xi_coldT_MF} 
and~\ref{fig:dynamical_scaling_xi_coldT}, they are of low quality,
limiting the magnetic fields to relatively large values. 
The three values ($H=0.04,~0.08,~0.16$) for
which it is feasible to rescale are all above the $H_\text{c}^{\text {MF,
3D}}(T=1.0)$ values given by Eqs.~\eqref{eq:Hc_estimation_MF} and
\eqref{eq:Hc_estimation_3D}.  All are, hence, in the paramagnetic regime, as can
seen from the shape of the curves in both figures.

Thus, though the data of Fig.~\ref{fig:xi_ZFC} suggests that, for the
lowest magnetic fields and $T=0.9$, $\xi(t,\tw;T)$ may be growing as a power
law, and thus be in the condensed phase, our limited time scale for the
simulations is unable to conclude that we have, in fact, straddled the dAT
line.  If one assumes Fisher-Sompolinsky scaling,
Eq.~\eqref{eq:Hc_estimation_3D}, all of our simulation results would be in the
paramagnetic region.  Until much longer times scales become reachable (either
at lower temperatures, or smaller magnetic fields), even our powerful Janus~II
simulations are unable to arrive at a definitive conclusion regarding the
existence, or non-existence, of the dAT line for $D=3$ Ising spin glasses.

\section{Conclusions}
\label{Sec:conlusion}
\noindent
This paper demonstrates the unique and powerful combination of experiment,
theory, and simulations addressing complex dynamics.  The use of single
crystals enables experiments to exhibit the consequences of very large
spin-glass correlation lengths.  The power of a special-purpose computer, Janus~II,
in combination with theory, is sufficient to extend simulation time and
length scales to values explored experimentally.  Together, these approaches
unite to develop new and important insights into spin-glass dynamics.

Previous work~\cite{joh:99,janus:17b} explored the reduction of the free-energy
barrier heights responsible for aging in spin glasses by the Zeeman (magnetic
field $H$) energy.  Observations for small magnetic fields, proportional to
$H^2$, were used to extract a quantitative value for the spin-glass
correlation length and its growth rate with time.  As the magnetic
field was increased, however, departures from proportionality to $H^2$ were
observed.  This paper presents detailed experimental observations of this
behaviour and, together with theory~\cite{zhai-janus:20a}, is able to
demonstrate the applicability of a new non-linear scaling law for the
magnetisation in the vicinity of the spin-glass condensation temperature $\Tg$.
Remarkably, Janus~II simulations were able to generate comparable values for
the magnetisation dynamics, with the added value of direct measurement of the
characteristic response time.

The combination of these two approaches has put to rest a decades-old
controversy concerning the nature of the Zeeman energy.  We have shown that the
departures from proportionality to $H^2$ are caused by non-linear terms in the
magnetisation,  and not by fluctuations of the magnetisation that lead to a
Zeeman energy proportional to $H$. Further, the departure from an $H^2$
behaviour that was used to justify the proportionality to $H$ is shown to be a
consequence of non-linear behaviour of the magnetisation in $H$, and fully
accounted for using the new scaling law.  This is an important finding, because
otherwise the extraction of the spin-glass correlation length from the Zeeman-energy
reduction in the barrier height would have been in error.

One of the most interesting findings in the paper is the extraction of the
characteristic response time for spin glasses, $t_H^{{\text {eff}}}$, from
simulations.  It has been made possible by noting that the spin-glass correlation
function reaches a peak {\it at the response time}.  That is,
\begin{equation}
C(t_H^{{\text {eff}}},\tw;H)=C_{{\text {peak}}}(\tw)\,.
\end{equation}
Thus, by extracting $C_{{\text {peak}}}(\tw)$ one can determine the
characteristic response time $t_H^{\text {eff}}$.  It is this observation that
enables simulations to give quantitative values for the non-linear magnetic
susceptibility that can be compared with the new scaling law.

In addition, we have explored the microscopic behaviour of the magnetic states
through the growth of the correlation lengths under two experimental protocols:
zero-field cooling (ZFC) and field cooling (FC).  We have proved that in a system
close enough to the condensation temperature $\Tg$, the Fisher-Sompolinsky
scaling relation holds under out-of-equilibrium conditions (see
Section~\ref{Sec:non-linear_scaling}.)  This will enable us in future
simulations to
compare the magnitude and growth of the spin-glass correlation length under 
two experimental protocols: the dynamics of zero-field-cooled and thermo-remanent
magnetisations. The important point here is that this paper shows
that our analysis will be valid under these non-equilibrium conditions.

We have discovered an overshooting phenomenon that is shown to be general for both
ordered and disordered magnetic systems.  And finally, we have explored the nature
of the spin-glass condensation at $\Tg$ as a function of the external magnetic
field, the so-called de Almeida-Thouless line. We  have presented preliminary
evidence for its existence as a true condensation
transition, but this conclusion should be regarded as provisional.

In conclusion, this paper has explored the nature of the spin-glass state in
the vicinity of its condensation temperature $\Tg$.  We displayed the power of
combining insights from both experiment and simulations, coupled together by
theory.  We look forward to continued investigation of spin-glass dynamics using
this relationship as we examine the microscopic nature of such phenomena as
rejuvenation and memory.  Finally, because spin-glass dynamics has
applications in many diverse fields (ecology, biology, optimisation, and even
social science), our work demonstrates that modelling complex systems is
feasible in finite dimensions.

\section*{Acknowledgements}
\addcontentsline{toc}{section}{Acknowledgements}
 We are grateful for helpful discussions with S. Swinnea about sample characterisation. This work was
partially supported by the U.S. Department of Energy, Office of Basic Energy Sciences, Division of
Materials Science and Engineering, under Award No. DE-SC0013599, and Contract No. DE-AC02-
07CH11358. We were partly funded as well by Ministerio de Econom\'ia, Industria y Competitividad
(MINECO, Spain), Agencia Estatal de Investigaci\'on (AEI, Spain), and Fondo Europeo de Desarrollo
Regional (FEDER, EU) through Grants No. FIS2016-76359-P, No. PID2019-103939RB-I00, No.
PGC2018-094684-B-C21 and PGC2018-094684-B-C22, by the Junta de Extremadura (Spain) and Fondo
Europeo de Desarrollo Regional (FEDER, EU) through Grant No. GRU18079 and IB15013 and by the
DGA-FSE (Diputaci\'on General de Arag\'on -- Fondo Social Europeo). This project has also received
funding from the European Research Council (ERC) under the European Union's Horizon 2020 research
and innovation program (Grant No. 694925-LotglasSy). DY was supported by the Chan Zuckerberg
Biohub and IGAP was supported by the Ministerio de Ciencia, Innovaci\'on y Universidades (MCIU,
Spain) through FPU grant No. FPU18/02665. BS was supported by the Comunidad de Madrid and the
Complutense University of Madrid (Spain) through the Atracci\'on de Talento program (Ref. 2019-
T1/TIC-12776). The simulations of Section~\ref{Sec:overshoot_order_system}
 were carried out at the ICCAEx supercomputer center in Badajoz.
We thank the staff at ICCAEx for their assistance.

\appendix
\renewcommand{\thesection}{\Alph{section}}
\section{Technical details about our simulations}

\subsection{Smoothing and interpolating the data}
\label{Appendix:FDR_smooth}
Our numerical data for the magnetisation at small magnetic fields are rather
noisy, which complicates the process of taking its derivative with respect to
$\ln t$. This derivative is the response function $S(t,\tw;H)$ [recall
Eq.~\eqref{eq:S_oft}]. This is why, before differentiating, we have followed a
de-noising method first proposed in Ref.~\cite{janus:17}. Because we work at
larger correlation lengths (and closer to $\Tg$) than in that work,
however, we have found it preferable to change some
technical details. We explain below the precise de-noising method that we have
followed in this work.
\begin{figure}[t] 
    \centering
     \includegraphics[width=0.7\textwidth]{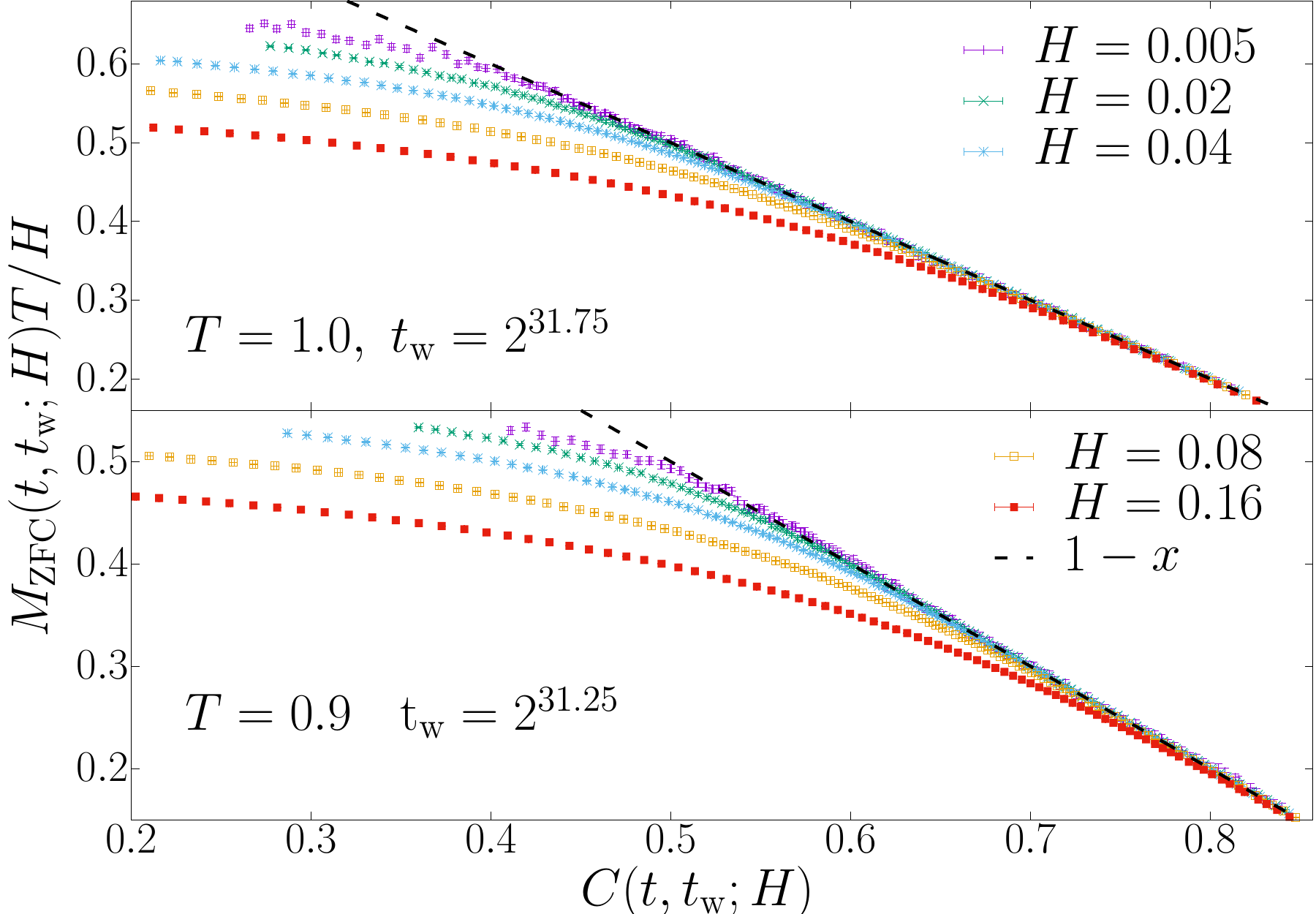}
\caption{\label{fig:FDRT_smooth}The behaviour of $ T
M_\mathrm{ZFC}(t,\tw;H)/H$ is exhibited as a function of $C(t,\tw; H)$.
The \emph{top plot} is for $T=1.0$, $\tw=2^{31.75}$. The
\emph{bottom plot} is for $T=0.9$, $\tw=2^{31.25}$.  We do not report all
the magnetic values for simplicity.}

  \includegraphics[width=0.7\textwidth]{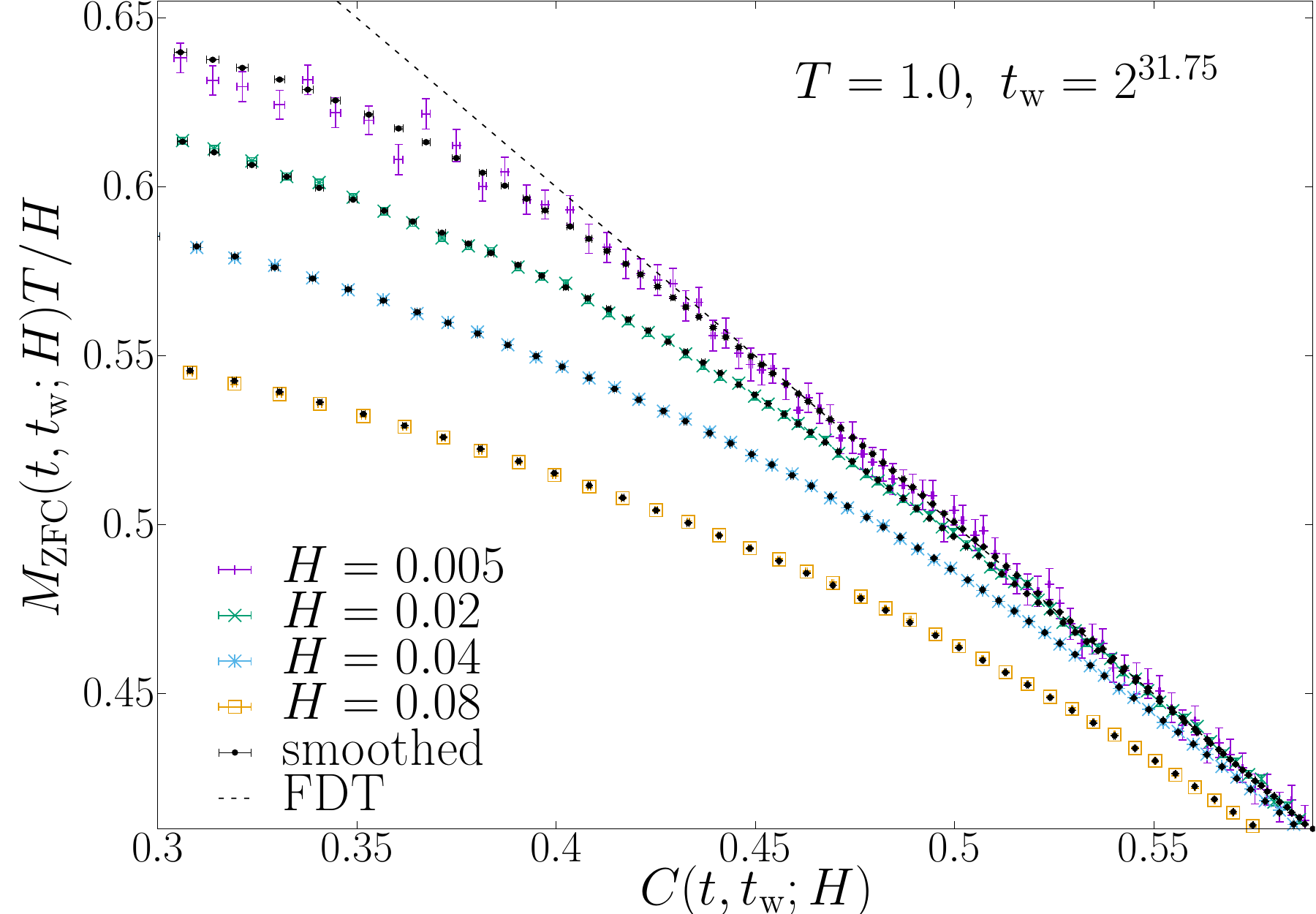}
  \caption{Comparison between the original and smoothed data at $T=1.0$ and waiting time $\tw=2^{31.75}$. One can clearly see the advantage of the de-noising method for the lowest magnetic field. }
  \label{fig:FDR_smooth}
\end{figure}

Our starting observation is that the derivative of both $M_\mathrm{ZFC}(t,\tw;H)$ and $TM_\mathrm{ZFC}
(t,\tw;H)/H$ peak at exactly the same time $\teff_H$. However, $TM_\mathrm{ZFC}(t,\tw;H)/H$
enjoys the advantage of being a very smooth function of the correlation $C(t,\tw;H)$. This smooth
function is named the fluctuation-dissipation
relation~\cite{cugliandolo:93,cruz:03,franz:94,franz:98,franz:99}. The key
point is that, at variance with the magnetisation, $C(t,\tw;H)$ can be
computed with high accuracy for any value of the field $H$, including $H=0$.
Thus, we follow a simple two-step de-noising algorithm:
\begin{enumerate}
\item We fit our data for $TM_\mathrm{ZFC}(t,\tw;H)/H$ as a function of $C(t,\tw;H)$, see Eq.~\eqref{eq:FDR_fit}.
\item We \emph{replace} our data for $TM_\mathrm{ZFC}(t,\tw;H)/H$ by the just-mentioned fitted
function evaluated at $C(t,\tw;H)$.
\end{enumerate}

Our chosen functional form is as follows.
Let the quantity $TM_\mathrm{ZFC}(t,\tw;H)/H$ be approximated by $f(\hat{x})$,
where for notational simplicity we do not write the explicit dependence on
$t,\tw$ and $H$,
\begin{equation}
f(\hat{x}) = f_{\text{L}}(\hat{x}) \frac{1+\tanh[ \mathcal{Q}(\hat{x})]}{2} + f_{\text{s}}(\hat{x}) \frac{1-\tanh[\mathcal{Q}(\hat{x})]}{2}
\label{eq:FDR_fit}
\end{equation}
 with $\mathcal{Q}(\hat{x}) = (\hat{x}-x^*) /w$.
The function $f(\hat{x})$ has distinct behaviour for large and small~$\hat{x}$.
The crossover between the two functional forms is smoothed by the $\tanh[
\mathcal{Q}( \hat{x})]$ functional term, where $x^*$ is the crossover point and
$w$ is the crossover rate. 
The functional form for small $\hat{x}$ is
\begin{equation}
f_{\text{s}}( \hat{x})=  a_0 + \sum_{k=1}^{N} a_k \, \frac{(\hat{x} - \hat{x}_{min})^k}{k!}\,.
\label{eq:fs(x)}
\end{equation}
For the large-$\hat{x}$ region, we choose a polynomial expansion in terms of $(1-\hat{x})$:
\begin{equation}
\label{eq:fL(x)}
f_{\text{L}}( \hat{x})= (1- \hat{x}) + \sum_{k=2}^{N'+1} b_k \, \frac{(1- \hat{x})^k}{k!}\,.
\end{equation}
The polynomial expansion in $1-\hat{x}$ is quite natural in the large-$\hat{x}$
region~\cite{janus:17}, as a deviation from the
fluctuation-dissipation theorem. This theorem, which holds only under
equilibrium conditions, predicts $N'=0$ for Eq.~\eqref{eq:fL(x)} and $x^*=w=0$
for Eq.~\eqref{eq:FDR_fit} [so that, in equilibrium, one would have
$f(\hat x)=(1-\hat x)$ in Eq.~\eqref{eq:FDR_fit}].  In the small-$\hat{x}$
region, there is not a strong justification (other than convenience) for our
choice of $f_{\text{s}}(\hat{x})$. In fact, the choice of Ref.~\cite{janus:17}
for $f_{\text{s}}(\hat{x})$ was a Padé approximant.  The quantity
$ T M_{\mathrm{ZFC}}(t,\tw;H)/H$ turns out to be affected by strong non-linear
effects that increase with increasing external magnetic field $H$ and upon
approaching the glass temperature $T_\mathrm{g}$ (see
Fig.~\ref{fig:FDRT_smooth}).
\begin{table}[p]
\caption{For each of our fits to Eq.~\eqref{eq:FDR_fit} we report: the order of the polynomial in
Eq.~\eqref{eq:fs(x)} $N$, the number of fitted parameters in Eq.~\eqref{eq:fL(x)} $N'$, [$N'=0$ means
$f_L(\hat x)= (1-\hat x)$ ], the crossover parameters $x^*$ and $w$, and the fit's figure of merit
$\chi^2/\text{d.o.f.}$ (d.o.f. stands for degrees of freedom).
Note that we can
only compute the so-called diagonal $\chi^2$, which takes into account only the
diagonal elements of the covariance matrix.  Because of this limitation, we
find a value of $\chi^2$ significantly smaller than the number of degrees of
freedom for many of our fits.}
\label{tab:FDT_fit}
\begin{indented}
\lineup
\item[] \begin{tabular}{@{}c c c c c c c l l l }
 \br
& & $\tw$ & & $H$ & $N$ & $N'$ & \multicolumn{1}{c}{$x^*$} & \multicolumn{1}{c}{$w$} & \multicolumn{1}{c}{$\chi^2/\text{d.o.f.}$} \\
\mr
\multirow{23}{*}{} & & \multirow{7}{*}{} & & 0.010  & 1 & 1 & 0.583(11)  & 0.128(1)  & $\045.721/110$ \\ 
                                            & &                   & &  0.020  & 1 & 1 & 0.604(6)  & 0.1264(6) & $\049.874/110$ \\ 
                                            & &                   & &  0.040  & 3 & 2 & 0.589(4)  & 0.099(4)  & $\027.991/107$ \\
                                            & &     $2^{22}$      & &  0.080  & 3 & 2 & 0.576(7)  & 0.164(11) & $\064.344/107$ \\    
                                            & &                   & &  0.100   & 3 & 2 & 0.681(12) & 0.187(4)  & $\033.133/99$ \\    
                                            & &                   & &  0.120  & 3 & 3 & 0.665(18) & 0.0725(5) & $\031.709/98$ \\    
                                            & &                   & &  0.160  & 4 & 4 & 0.62(4) & 0.052(1)  & $\065.326/104$ \\ \cline{2-10}
                   & & \multirow{8}{*}{} & &  0.005 & 2 & 0 & 0.530(17) & 0.100(7) & $\028.518/112$ \\
                                            & &                   & &  0.010  & 1 & 1 & 0.520(10) & 0.115(4)  & $\037.343/112$ \\ 
                                            & &                   & &  0.020  & 2 & 0 & 0.535(3)  & 0.091(3)  & $\035.535/112$ \\ 
                                            & &     $2^{26.5}$  & &  0.040  & 2 & 0 & 0.653(2)  & 0.033(2)  & $\059.325/112$ \\
$T=0.9$            & &                   & &  0.080  & 3 & 2 & 0.585(7)  & 0.161(11) & $\055.006/109$ \\    
                                        & &                   & &  0.100   & 3 & 2 & 0.674(12) & 0.176(4)  & $\032.874/103$ \\    
                                        & &                   & &  0.120  & 3 & 3 & 0.68(6) & 0.078(11) & $\042.782/102$ \\    
                                            & &                   & &  0.160  & 4 & 4 & 0.623(15) & 0.034(8)& $\077.297/106 $ \\   \cline{2-10}  
                   & & \multirow{8}{*}{} & &  0.005 & 1 & 0 & 0.503(8) & 0.074(7) & $\035.145/123$ \\
                                            & &                   & &  0.010  & 1 & 1 & 0.520(12) & 0.139(2)  & $\032.331/118$ \\ 
                                            & &                   & &  0.020  & 1 & 2 & 0.550(7)  & 0.0335(3) & $\031.219/121$ \\ 
                                            & &   $2^{31.25}$     & &  0.040  & 2 & 0 & 0.554(16) & 0.080(2)  & $\075.790/116$ \\
                                            & &                   & &  0.080  & 3 & 2 & 0.598(6)  & 0.152(8)  & $\038.439/115$ \\    
                                            & &                   & &  0.100  & 3 & 2 & 0.688(10) & 0.170(4)  & $\029.904/107$ \\  
                                            & &                   & &  0.120  & 3 & 3 & 0.655(14) & 0.062(4)  & $\032.396/98 $ \\                                             
                                            & &                   & &  0.160  & 4 & 4 & 0.549(22) & 0.077(6)  & $\067.296/112$ \\  
\mr

\multirow{18}{*}{} & & \multirow{6}{*}{} & &  0.005 & 1 & 0 & 0.371(18) & 0.178(12) & $\060.709/103$ \\
                                            & &                   & &  0.010  & 1 & 0 & 0.411(4)  & 0.129(4)  & $\046.709/103$ \\ 
                                            & &                   & &  0.020  & 1 & 0 & 0.397(2)  & 0.138(2)  & $\076.009/103$ \\ 
                                            & &   $2^{23.75}$     & &  0.040  & 2 & 0 & 0.457(2)  & 0.146(2)  & $\049.401/102$ \\
                                            & &                   & &  0.080  & 4 & 0 & 0.589(6)  & 0.066(1)  & $\036.384/100$ \\    
                                            & &                   & &  0.160  & 4 & 3 & 0.639(12) & 0.061(6)  & $\048.338/98 $ \\ \cline{2-10}                        
                                            
                   & & \multirow{6}{*}{} & &  0.005 & 1 & 0  & 0.37(2) & 0.157(1) & $\034.6242/101$ \\
                                            & &                   & &  0.010  & 1  & 0 & 0.400(8)  & 0.128(5)  & $\045.169/111$ \\ 
                                            & &                   & &  0.020  & 1  & 0 & 0.389(2)  & 0.132(2)  & $\053.121/111$ \\ 
$T=1.0$                & &   $2^{27.625}$    & &  0.040  & 3  & 0 & 0.559(11) & 0.049(4)  & $\032.738/109$ \\
                                            & &                   & &  0.080  & 3  & 1 & 0.638(8)  & 0.023(11) & $491.701/108$ \\                     
                                            & &                   & &  0.160  & 3  & 1 & 0.667(2)  & 0.023(3)  & $140.853/108 $ \\   \cline{2-10}         
                   & & \multirow{6}{*}{} & &  0.005 & 2 & 0  & 0.357(9) & 0.170(12) & $\041.854/127$ \\
                                            & &                   & &  0.010  & 1  & 0 & 0.114(10)  & 0.114(10)  & $\039.202/123$ \\ 
                                            & &                   & &  0.020  & 2 & 0 & 0.488(8)  & 0.116(7)  & $\040.968/118$ \\ 
                                            & &   $2^{31.75}$     & &  0.040  & 3 & 0 & 0.534(12)  & 0.070(4)  & $\033.579/109$ \\                                            
                                            & &                   & &  0.080  & 3 & 1 & 0.631(4)  & 0.023(5)  & $271.914/108$ \\    
                                            & &                   & &  0.160  & 4 & 0 & 0.686(9) & 0.080(3)  & $160.490/108 $ \\  
\br
\end{tabular}
\end{indented}
\end{table}

As we discuss in Appendix \ref{Appendix:over-fit}, it is necessary to select the appropriate order for the polynomials
in Eqs.~\eqref{eq:fs(x)} and~\eqref{eq:fL(x)}. Our preferred
choices are given in Table \ref{tab:FDT_fit}.

Errors are computed using a jackknife procedure. We perform an independent fit
for each jackknife block, and compute errors from the jackknife fluctuations
of the fitted $f(\hat{x})$.  In Fig.~\ref{fig:FDR_smooth} we show a comparison
between the original and smoothed data for the case $T=1.0$ and
$\tw=2^{31.75}$.  As expected, the de-noising technique is most important for
the smallest magnetic fields.

\subsection{Over-fitting problem}
\label{Appendix:over-fit}
A difficulty in our fits to Eq.~\eqref{eq:FDR_fit} is that we can use only the diagonal part of the
covariance matrix in the computation of the goodness-of-fit indicator $\chi^2$. This is the reason
underlying the very small values for $\chi^2$ that we show in Table~\ref{tab:FDT_fit}. As a consequence,
we cannot trust the $\chi^2$ test for selecting the appropriate order for the polynomial expansions in
Eqs.~\eqref{eq:fs(x)} and \eqref{eq:fL(x)}. Hence, we follow a different strategy.

\begin{figure}[t] \centering
  \includegraphics[width=0.7\textwidth]{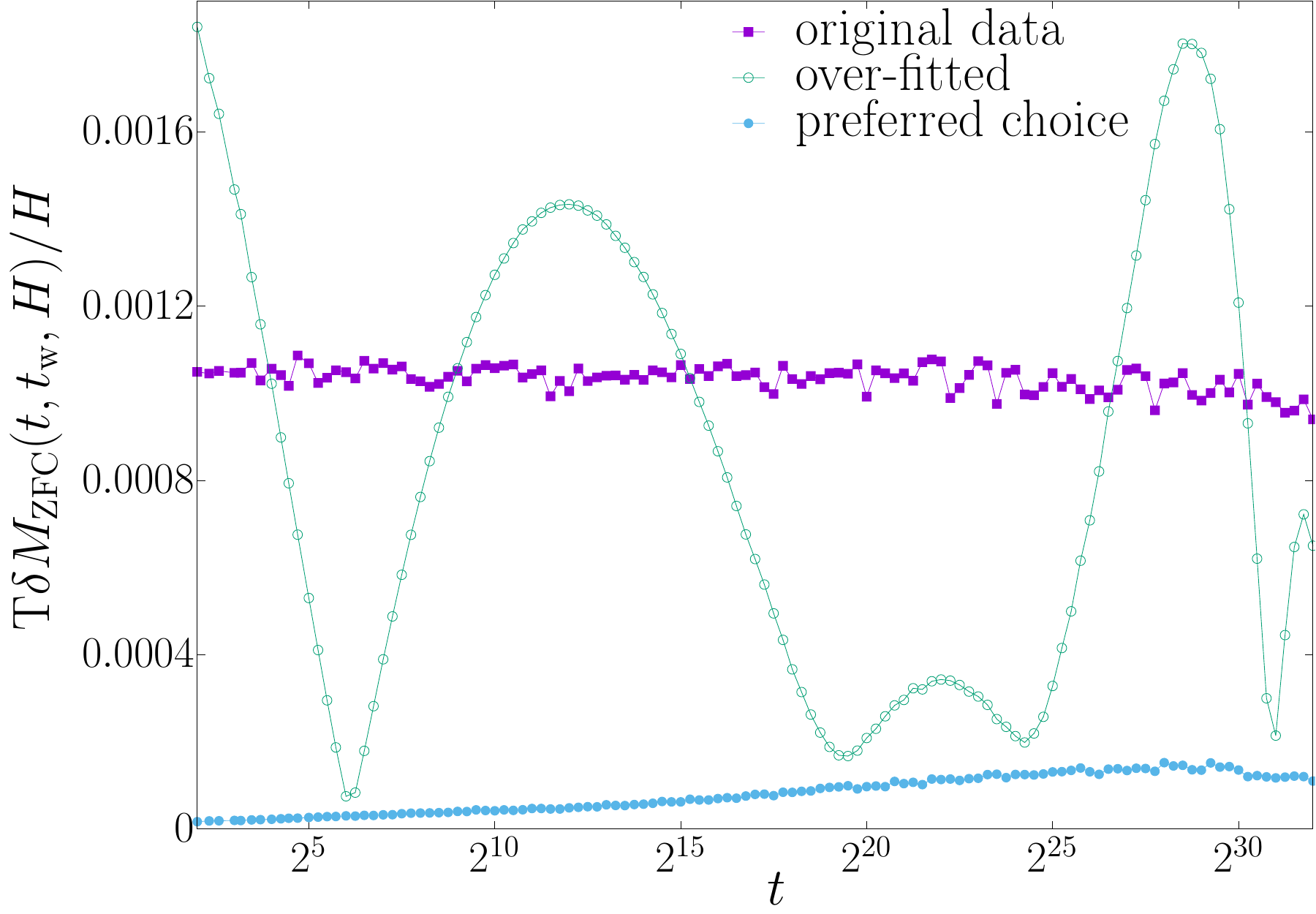}
  \caption{Comparison between the errors for the original and the de-noised
data for $T M_\mathrm{ZFC}(t,\tw;H)/H$  (with $T=0.9$, $\tw=2^{31.75}$ and
$H=0.002$), for the two different  fitting functions $f(\hat{x})$ reported in
Table \ref{tab:FDT_overfit}.}
  \label{fig:FDR_overfit}
\end{figure}

Fortunately, we can also compare the statistical errors that we find for the de-noised $TM_\mathrm{ZFC}
(t,\tw;H)/H$ with different choices of the polynomial expansion (remember that these errors are not
computed from $\chi^2$, but from the jackknife fluctuations).
As an example, consider the case at $T=0.9$
for a waiting time $\tw=2^{31.75}$ and $H=0.002$, which is exhibited in Fig. \ref{fig:FDR_overfit}.
The figure compares the statistical errors of the original, non-de-noised data with the errors found with two
possible choices for the polynomial fits in Eqs.~\eqref{eq:fs(x)} and~\eqref{eq:fL(x)}. Although both
fits are indistinguishable from the point of view of the $\chi^2$ test, see Table~\ref{tab:FDT_overfit}, the
resulting errors are very different. In one case, we find statistical errors that evolve rather smoothly with
$t$. For the second choice, we find wild oscillations in the size of the errors as $t$ varies. When in doubt, we
have always taken the choice that provides the smoother $t$ evolution for the errors. As we said above, our
final choices are reported in Table~\ref{tab:FDT_fit}.

\begin{table}[t]
\lineup
\caption{Details about the two fits shown in Fig.~\ref{fig:FDR_overfit}. We follow the same notational
conventions of Table \ref{tab:FDT_fit}.}
\label{tab:FDT_overfit}
\begin{indented}\item[] \begin{tabular}{@{}c c c c c c  }
 \br
Case & $N$ & $N'$ & $x^*$ & $w$ & $\chi^2/\text{d.o.f.}$ \\
\mr
Over-fitted & 1 & 1 &  0.547(6)  & 0.132(1)\0  & $28.059/122$ \\ 
Our choice  & 2 & 0 &  0.550(7)  & 0.0335(3) & $31.219/121$\\
\br
\end{tabular}
\end{indented}
\end{table}

\subsection{Time discretisation and the calculation of the relaxation function $S(t,\tw;H)$}
\label{Appendix:details_St_construction}

As explained in the main text, the quantity used in experiment
\cite{joh:99} to extract $\teff(H)$ is the relaxation function $S(t,\tw;H)$ of
Eq.~\eqref{eq:S_oft}.
We calculate $S(t,\tw ;H)$ as a finite-time difference:
\begin{equation}
\label{eq:S_tnew_Appendix}
S(t,\tw, t' ; H) = \frac{M_\mathrm{ZFC}(t',\tw ; H)- M_\mathrm{ZFC}(t,\tw; H)}{\ln\left( \frac{t'}{t}\right)} \; .
\end{equation}
In simulations, time is discrete and we have stored configurations at
$t_n=$ integer−part−of
$2^{n/4}$, with $n$ an integer. Let us write the integer dependence of
times $t$ and $t'$ explicitly as:
\begin{equation}
t \equiv t_n, \quad \; \quad t' \equiv t_{n+k},
\end{equation}
where $k$ is an integer time parameter.
Hence, time is rescaled as
\begin{equation}
t_{\mathrm{new}} = \frac{1}{2} \ln\left( \frac{t_{n+k}}{t_n} \right)\,.
\end{equation}
We expressed our observables as functions of $t_\mathrm{new}$:
\begin{eqnarray}
S(t,\tw,t';H) & \quad \to \quad S(t_\mathrm{new}, \tw ; H) \,, \\
C(t,\tw;H) & \quad \to \quad C(t_\mathrm{new}, \tw ; H) \, .
\end{eqnarray}
 
The relaxation function $S(t_\mathrm{new}, \tw; H)$ is trivial to construct, see Eq. \eqref{eq:S_tnew_Appendix}. However, the correlation function $C(t_{\mathrm{new}}, \tw; H)$ needs to be calculated using a linear interpolation.
For any given value of $\tnew$, we looked for our \textit{original} discrete time $t_n$ such that
\begin{equation}
\ln (t_n) < \ln(\tnew) \leq \ln(t_{n+1}) \; .
\end{equation}
Using a linear interpolation, we obtain
\begin{equation}\label{eq:C_tnew_linear_interpolation}
C(\tnew) = \frac{\ln(\tnew) - \ln(t_{n+1})}{\ln(t_{n}) -\ln(t_{n+1})} C(t_{n}) -\frac{\ln(\tnew) - \ln(t_n)}{\ln(t_{n}) -\ln(t_{n+1})} C(t_{n+1}) \, .
\end{equation}

Finally, one can express the relaxation function, $S(\tnew, \tw; H)$, as a
function of the correlation function, ${\cal S} (C;H)$, in much the same manner
as Eq.~\eqref{eq:C_tnew_linear_interpolation}.

\subsection{The $\teff_H$ calculation}
\label{Appendix:teff_details}

As explained in the main text, the extraction of $\teff_H$ from
Eq.~\eqref{eq:C_teff_Cpeak} is delicate because the $C_\mathrm{peak}(\tw)$ are
implicit functions of the rescaled time $t_\mathrm{new}$.
In order to solve Eq.~\eqref{eq:C_teff_Cpeak}, we calculate the $\teff_H$ values through a quadratic interpolation.
First, we calculate the \textit{original} discrete time $t_n$ such that:
\begin{equation}
C(t+t_{n+1}, \tw; H) < C_{\mathrm{peak}} \leq C( t+t_n, \tw; H) \, .
\end{equation}
Then, we solve the three-equation system:
\begin{eqnarray}
C(t+t_{n-1}) &=& \alpha_0 + \alpha_1 x_{n-1} + \alpha_2 x_{n-1}^2\,, \\
C(t+t_{n})   &=& \alpha_0 + \alpha_1 x_n + \alpha_2 x_n^2\,, \\
C(t+t_{n+1}) &=& \alpha_0 + \alpha_1 x_{n+1} + \alpha_2 x_{n+1}^2 \, ,
\end{eqnarray}
where $x_n = \ln t_{n}$ and, for brevity's sake, we omit the arguments $\tw$ and $H$.
The solution generates the $\alpha_i$ coefficients:
\begin{eqnarray}
\alpha_2 &=& \frac{ C(\tw, \tw+ t_{n-1}) - C(\tw,\tw+t_n) }{x_{n-1}-x_{n}}\nonumber\\
&& \quad  -\frac{ C(\tw, \tw+ t_{n+1}) - C(\tw,\tw+t_n) }{(x_{n-1}-x_{n+1})(x_{n+1}-x_n)}  \, , \\
\alpha_1 &=& \frac{ C(\tw, \tw+ t_{n+1}) - C(\tw,\tw+t_n) }{x_{n+1}-x_{n}} - \alpha_2 (x_n + x_{n+1}) \, , \\
\alpha_0 &=&  C(\tw, \tw + t_n) - \alpha_1 x_n - \alpha_2 x_n^2 \, . 
\end{eqnarray}
We can then calculate the $\teff_H$  solving the equation:
\begin{equation}
C_{\mathrm{peak}} = \alpha_0 + \alpha_1 \ln \left( \teff_H \right) + \alpha_2 \left[ \ln \left( \teff_H \right) \right]^2 \, ,
\end{equation}
where only the solution verifying $t_n\leq \teff_H<t_{n+1}$ is physical.

\subsection{The construction of $\xi(t,\tw;H)$}
\label{Appendix:xi_construction}
We shall explain here our computation of the spin-glass correlation length in
the presence of a magnetic field. For the simpler case of $H=0$, we refer the
reader to Refs.~\cite{janus:18,fernandez:19}.

The most informative connected correlator we can construct with $4$ replicas is the replicon propagator \cite{dealmeida:78,dedominicis:06}.
Extending the replicon propagator to the off-equilibrium regime we have:
\begin{equation}
\label{eq:GR_def}
G_\text{R}(\boldsymbol{r},t') = \frac{1}{V} \sum_{\boldsymbol{x}} \overline{( \langle
s_{\boldsymbol{x},t'} \, s_{\boldsymbol{x}+\boldsymbol{r},t'} \, \rangle - \langle
s_{\boldsymbol{x},t'} \rangle \langle s_{\boldsymbol{x}+\boldsymbol{r},t'}
\rangle)^2} \, ,
\end{equation}
where $t'=\tw+t$.
To compute $G_\mathrm{R}(\boldsymbol{r},t')$, we calculate the $4$-replica field
\begin{equation}
\Phi_{\boldsymbol{x},\mathrm{t'}}^{(a,b;c,d)} = \frac{1}{2} \left( s_{\boldsymbol{x},\mathrm{t'}}^{(a)} - s_{\boldsymbol{x},\mathrm{t'}}^{(b)}\right)\left( s_{\boldsymbol{x},\mathrm{t'}}^{(c)} - s_{\boldsymbol{x},\mathrm{t'}}^{(d)} \right) \; ,
\end{equation}
where indices $a,b,c,d$ indicate strictly different replica.
Notice that 
\begin{equation}
\langle \Phi_{\boldsymbol{x},\mathrm{t'}}^{(a,b;c,d)} \Phi_{\boldsymbol{y},\mathrm{t'}}^{(a,b;c,d)} \rangle = ( \langle s_{\boldsymbol{x},\mathrm{t'}} s_{\boldsymbol{y},\mathrm{t'}} \rangle -\langle s_{\boldsymbol{x},\mathrm{t'}} \rangle \langle s_{\boldsymbol{y},\mathrm{t'}}\rangle)^2 \; .
\end{equation}
Therefore, we obtain $G_\mathrm{R}(\boldsymbol{r},{t'})$ by taking the average over the samples
\begin{equation}
\mathrm{E} \left( \Phi_{\boldsymbol{x},\mathrm{t'}}^{(a,b;cd)}
\Phi_{\boldsymbol{y},\mathrm{t'}}^{(a,b;c,d)}\right) =
G_\mathrm{R}(\boldsymbol{x}-\boldsymbol{y},{t'} ) \; .
\end{equation}
With $512$ replicas at our disposal, there are $3 \times 512!/(508!\times 4!)$ ways
of choosing the replica indices $a, b, c$, and $d$. We have found an efficient
way for averaging over (roughly) one third of this astronomic number of
possibilities \cite{paga:20}.
We define the correlation function, $C(\boldsymbol{r},{t'})$ , as the
replicon propagator, $G_\mathrm{R}(\boldsymbol{x}-\boldsymbol{y},{t'})$,
where we consider the difference $\boldsymbol{x}-\boldsymbol{y}$ as the lattice
displacement $\boldsymbol{r} = (r,0,0)$.
Of course, we can align the lattice displacement vector $\boldsymbol{r}$ along
any of three coordinate axes, so we average over these three choices.
The replicon correlator $G_\mathrm{R}$ decays to zero in the large-distance
limit. We, therefore, computed the correlation length $\xi(t,\tw;H)$ exploiting
the integral estimators~\cite{janus:08b,janus:09b}:
\begin{equation}
I_k ({t'};T) = \int_{0}^{\infty} \mathrm{d} r \, r^k C(r,{t'};T), \quad  \quad \xi_{k,k+1}({t'};T) = \frac{ I_{k+1}({t'};T)}{I_k({t'};T)}\,.
\end{equation}
In the main text, we always refer to the $\xi_{12}({t'};T)$ correlation
length except for Section~\ref{Sec:overshoot_order_system} , where we evaluate
$\xi_{23}({t'};T)$.

\section{The Josephson length}
\label{Appendix:josephson_lJ}
For the reader’s convenience, we reproduce here the interpolation proposed in Ref.~\cite{zhai:19} of the
data obtained in Ref.~\cite{janus:18} for the replicon exponent as a function of the Josephson length and
the correlation length.

 The Josephson length, $\ell_\text{J}(T^{(J)})$, scales as
\begin{equation}\label{eq:lJ_def}
\ell_\text{J}\big(T^{(J)}\big)={\frac {b_0+b_1\bigl(\Tc^{(J)}-T^{(J)}\bigr)^{\nu}+b_2\big(\Tc^{(J)}-T^{(J)}\big)^{\omega\nu}}{\big(\Tc^{(J)}-T^{(J)}\big)^{\nu}}},
\end{equation}
where $T^{(J)}$ is the temperature in IEA units,
\begin{equation}
T^{(J)}= \frac{T}{\Tc} T^{(J)}_\text{c} \quad T^{(J)}_\text{c}=1.102(3) \,,
\end{equation}
and we have included analytic ($b_1$) and confluent ($b_2$) scaling corrections
with $\omega=1.12(10)$ and $\nu=2.56(4)$ \cite{janus:13}.  The numerical
coefficients are:
\begin{equation}
\omega = 1.12(10),\\
\nu=2.56(4),~\\
b_0\approx0.1015,~\\
b_1\approx0.3725,~\\
b_2\approx0.1997.~\\
\end{equation}
The replicon exponent $\theta(\xi;T)$ depends on both the temperature and the correlation length $\xi$ through the crossover variable
\begin{equation}
\tilde x= \frac{\ell_\text{J}(T)}{\xi(\tw;T)} \, .
\end{equation}
In fact, $\theta\big(\tilde x(\xi;T)\big)$ can be well interpolated as,
\begin{equation}\label{eq:replicon_def}
\theta(x)=\theta_0+d_1\bigg({\frac {x}{1+e_1x}}\bigg)^{2-\theta_0}+d_2\bigg({\frac {x}{1+e_2x}}\bigg)^{3-\theta_0},
\end{equation}
where
\begin{eqnarray}
\theta_0&\approx&0.30398,\quad
e_1\approx1.38179,\quad
d_1\approx2.72489,\\
e_2&\approx&2.12634,\quad
d_2\approx-9.98389.~\\
\end{eqnarray}

\section{Sample dependence of the non-linear scaling results}
\label{Appendix:Cpeak_samples}
\begin{figure}[t] 
    \centering
        \includegraphics[width=.7\textwidth]{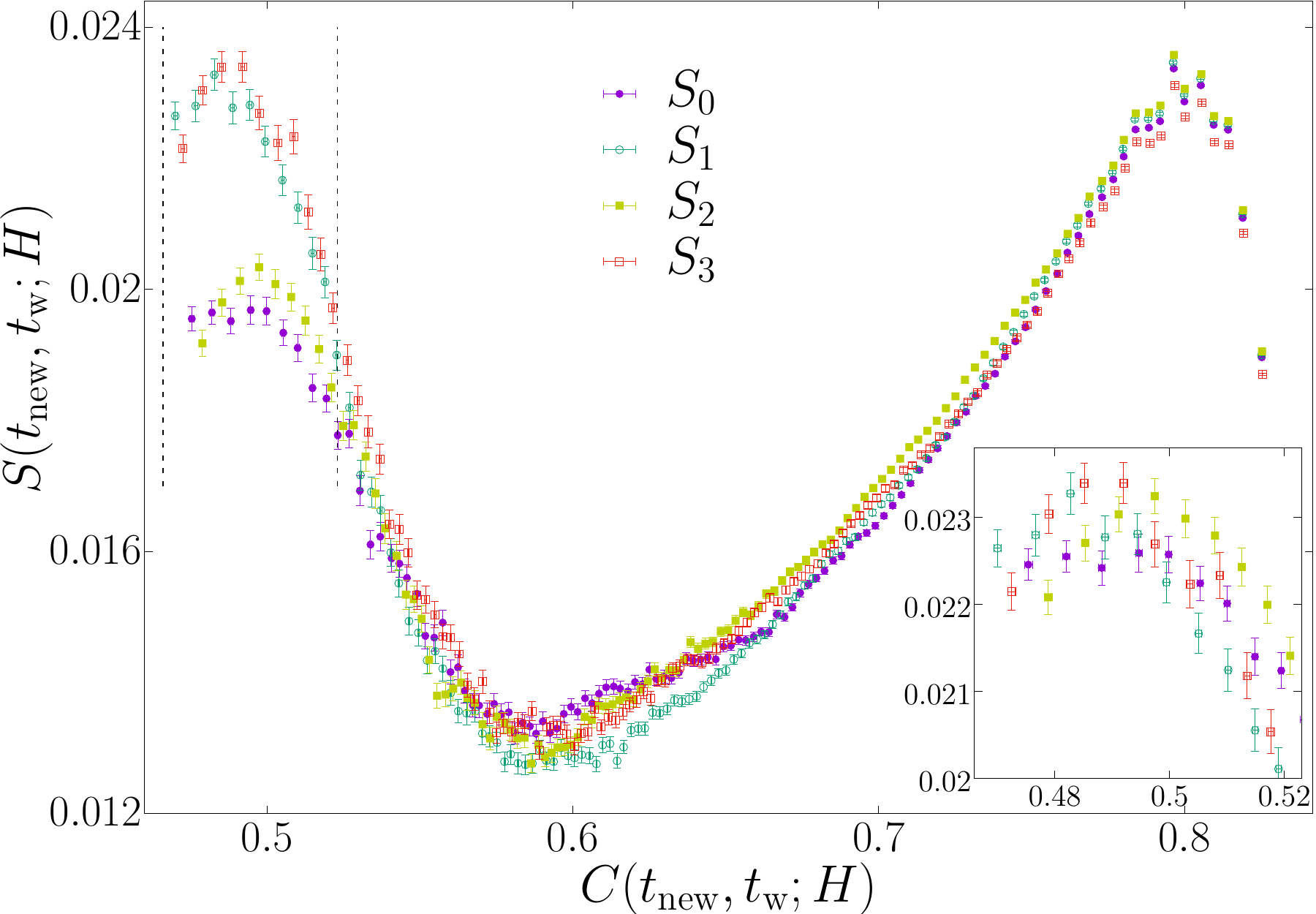}
\caption{Plots show the behaviour of $\mathcal{S}(C;H)$ for four independent
samples for $H=0.005$, $\tw=2^{31.25}$ at $T=0.9$. The 
peak area is enlarged in the \emph{inset} after a vertical shift.}
\label{fig:four_samples}
\end{figure}

\begin{figure}
\centering
 \includegraphics[width=0.7\textwidth]{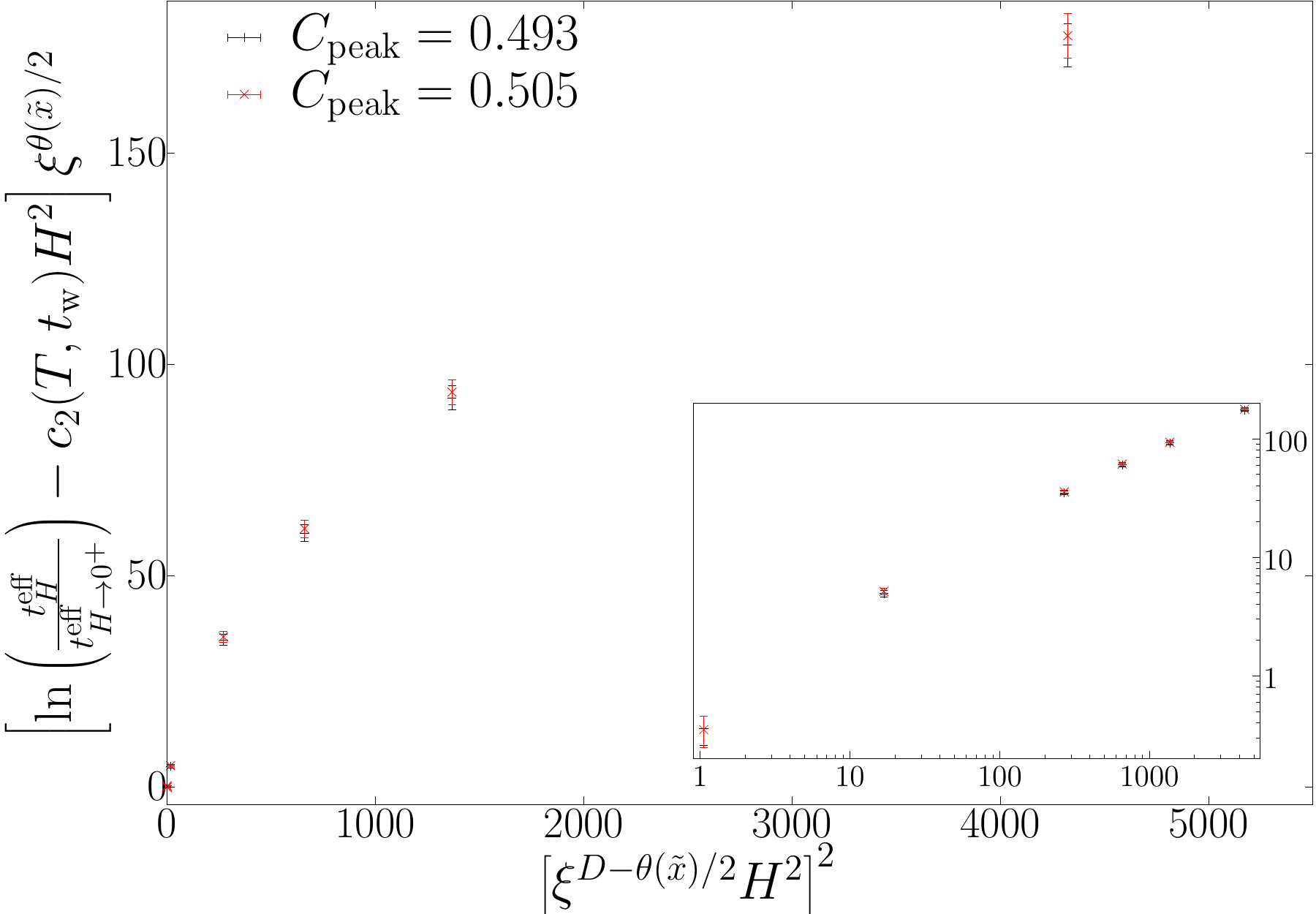}
	\caption{Behaviour of the scaling law for the case $\tw=2^{31.25}$
at $T=0.9$  for the two different $C_{\mathrm{peak}}$ values in
Table~\ref{tab:sample_dependence_Cpeak}. The \emph{main figure} is in
semi-log scale  whereas the \emph{insert} is amplified in a log-log scale. It
can be seen that the data points for the two values of $C_\mathrm{peak}$ are
equivalent within their respective error bars. }
         \label{fig:scaling_law_2Cpeak}
\end{figure}

We demonstrate that the non-linear scaling results are sample independent for
the case $H=0.005$, $\tw=2^{31.25}$, at $T=0.9$.  We simulate four independent
samples and, for each one, we build the relaxation function $\mathcal{S}(C;H)$,
see Section~\ref{Sec:C_peak} and
Appendix~\ref{Appendix:details_St_construction}.  We exhibit them in
Fig.~\ref{fig:four_samples} where we report only the case for $k=8$.  To compare peak
regions of the relaxation function, $\mathcal{S}(C,H)$, we shift the lowest
curves, namely $S_0$ and $S_2$, vertically.  An amplification of the peak
region is shown in the inset of Fig.~\ref{fig:four_samples}.  As one can observe,
there is a sample dependence of the peak position. We report the estimates of
$C_\mathrm{peak}(\tw)$ for each sample in
Table~\ref{tab:sample_dependence_Cpeak}.  Note that the sample $S_0$ is the one
analysed in the main text.  We extract the effective time  $\teff_H$ for each
sample, according to Eq.\eqref{eq:C_teff_Cpeak}.  They are listed in
Table~\ref{tab:sample_dependence_Cpeak}.  The sample dependence found for the
$C_\mathrm{peak}^{S_i}(\tw)$ values is seen in the $\teff_H$ values too.
Accordingly, we repeat the analysis of
Section~\ref{Sec:ratio_time_NUM} using, as input parameter for extracting 
$\teff_H$, the $C_\mathrm{peak}^{S_2}(\tw)$ value shown in
Table~\ref{tab:sample_dependence_Cpeak}.  We analyse the effective time ratio $
\ln \, \frac{t^{\text{eff}}_H}{t^{\text{eff}}_{H\to 0^+}}$ according to Eq.~\eqref{eq:ratio_effective_time_scaling_final}.
We then compare the scaling behaviour for the two values of $C_\mathrm{peak}^{S_i}(\tw)$.
The two sets of data are statistically compatible, see Fig. \ref{fig:scaling_law_2Cpeak}.
This implies that the physical scenario is not affected by the small  uncertainty in the determination of $C_\mathrm{peak}(\tw)$.
We therefore assert that the scaling results are sample independent.

\begin{table}
\caption{Values of $\teff_H$ and $C_\mathrm{peak}(\tw)$ for four independent samples
in the case $H=0.005$, $\tw=2^{31.25}$ and $T=0.9$.}
\label{tab:sample_dependence_Cpeak}
\begin{indented}
\item[] \begin{tabular}{@{}c c c} 
 \br
Sample & $\log_2 \left( \teff_H \right)$  & $C_\mathrm{peak}(\tw)$\\ 
\mr
$S_0$ & 30.434(21) & 0.493 \\ 
$S_1$ & 30.196(23) & 0.493 \\
$S_2$ & 30.546(24) & 0.505  \\ 
$S_3$ & 30.327(21) & 0.505  \\  
\br
\end{tabular}
\end{indented}
\end{table}

\addcontentsline{toc}{section}{References}
\bibliographystyle{iopart-num}
\bibliography{Draft_long_paper_v16.bbl}

\end{document}